\documentclass[reprint,prb,twocolumn,superscriptaddress,longbibliography,showpacs,noeprint]{revtex4-1}
\usepackage{graphicx}
\usepackage[percent]{overpic}
\usepackage{soul}
\usepackage{dcolumn}
\usepackage{bm}
\usepackage{color}
\usepackage{amssymb} 
\usepackage{amsmath}
\usepackage{verbatim}
\usepackage{float}
\usepackage{subfig}
\usepackage{lmodern}
\usepackage[utf8]{inputenc}
\usepackage[T1]{fontenc}
\usepackage[english]{babel}

\usepackage{ragged2e}
\DeclareCaptionJustification{justified}{\justifying}
\renewcommand\Re{\operatorname{Re}}
\renewcommand\Im{\operatorname{Im}}

\newcommand{\la}{\langle}
\newcommand{\ra}{\rangle}
\newcommand{\rt}{\right}
\newcommand{\lf}{\left}

\renewcommand{\k}{\mathbf k}
\newcommand{\s}{\text{s}}
\newcommand{\ka}{\boldsymbol{\kappa}}
\newcommand{\ks}{{\boldsymbol{\kappa}_{\text{s}}}}
\renewcommand{\i}{\text{in}}

\newcommand{\br}{\mathbf r}
\newcommand{\el}{\text{el}}
\newcommand{\eq}[1]{Eq.~(\ref{#1})}
\newcommand{\Nel}{N_{\text{el}}}
\newcommand{\Hel}{\hat H_{\text{el}}}
\newcommand{\Hdel}{\hat H_{\text{el-em}}}
\newcommand{\Hint}{\hat H_{\text{int}}}
\newcommand{\Hem}{\hat H_{\text{em}}}
\newcommand{\Aem}{\hat{\mathbf A}_{\text{em}}}
\newcommand{\bG}{\mathbf G}
\newcommand{\mueven}{\mu_{\text{even}}}
\newcommand{\muodd}{\mu_{\text{odd}}}
\newcommand{\bmj}{\boldsymbol{\mathfrak j}}
\newcommand{\qe}{\text{q.e.}}

\newcommand{\commentout}[1]{\ignorespaces}

\begin{document}
\title{Atomic-scale imaging of laser-driven electron dynamics in solids using subcycle-resolved x-ray-optical wave mixing}
\author{Daria Popova-Gorelova}
\email[]{daria.gorelova@cfel.de}
\affiliation{Center for Free-Electron Laser Science, DESY, Notkestrasse 85, D-22607 Hamburg, Germany}
\affiliation{Department of Physics, Universit\"at Hamburg, Jungiusstrasse 9, D-20355 Hamburg, Germany} 
\affiliation{The Hamburg Centre for Ultrafast Imaging, Universit\"at Hamburg, Luruper Chaussee 149, D-22761 Hamburg, Germany}
\author{Robin Santra}
\email[]{robin.santra@cfel.de}
\affiliation{Center for Free-Electron Laser Science, DESY, Notkestrasse 85, D-22607 Hamburg, Germany}
\affiliation{Department of Physics, Universit\"at Hamburg, Jungiusstrasse 9, D-20355 Hamburg, Germany} 
\affiliation{The Hamburg Centre for Ultrafast Imaging, Universit\"at Hamburg, Luruper Chaussee 149, D-22761 Hamburg, Germany}
\date{\today}

\begin{abstract}

We investigate laser-driven electron dynamics in solids on the atomic scale and in real space within Floquet formalism, and develop a method based on subcycle-resolved x-ray-optical wave mixing to reconstruct those dynamics. We analyze how time-reversal and inversion symmetries influence properties of optically-induced charge distributions and microscopic electron currents. Several examples for the $\mu$th-order microscopic optical response of band-gap crystals are shown and compared for cases when there is either a considerable or a vanishing $\mu$th-order macroscopic response. We then analyze the consequence of crystal symmetries on subcycle-resolved x-ray-optical wave mixing, a process in which an x-ray pulse of a duration shorter than the optical cycle of the driving pulse interacts with a laser-dressed crystal. Based on this analysis, we develop a method to reconstruct amplitudes and phases of Fourier components of optically-induced charge distributions from momentum and delay dependence of x-ray-optical wave-mixing spectra. Subcycle-resolved x-ray-optical wave mixing also reveals phases of temporal oscillations of microscopic optical response and some properties of microscopic laser-driven electron currents.
 
\end{abstract}
\maketitle

\section{Introduction}

Strong-field excitation by light can be used to induce various important mechanisms in solids, such as manipulation of electronic gaps and structure by light \cite{SchiffrinNature12, SchultzeNature12, ChaiPRL18, SederbergNatPhot20, KuehnPRL10, SchubertNature14, SommerNature16, SchlaepferNature18, OkaARCMP19, UzanNatPhot20}, or generation of high harmonics (HHG) \cite{GhimireNature11}. Such processes, on the one hand, have a big potential for the development of petahertz electronics, and, on the other hand, raise many scientific questions about the mechanisms behind them \cite{SchoetzACSPhotonics19, KruchininRMP18}. Access to microscopic properties of laser-driven electron dynamics is necessary for a deeper understanding of strong-field phenomena in solids  \cite{YouNature16, NdabashimiyeNature16, LakhotiaNature20, SchoetzACSPhotonics19}.

X-ray-optical wave mixing, in which an x-ray and an optical pulse simultaneously interact with an electronic system, encodes information about optical response of the system on the atomic scale \cite{FreundPRL70, EisenbergerPRA71, GloverNature12, SchoriPRL17, RouxelPRL18, CohenPRR19, Popova-GorelovaPRB18}. Recent experimental capabilities to generate attosecond x-ray pulses \cite{TeichmannNatComm16, HuangPRL17, ParcApplSci18, LiNatComm17, KaertnerNIMPRSA16, DurisNatPhot20} make subcycle-resolved x-ray-optical wave mixing, a process in which an x-ray pulse with a duration shorter than an optical cycle is used, experimentally feasible. In Ref.~\onlinecite{CitepaperShort}, we present a method to reconstruct microscopic optical response by means of subcycle-resolved x-ray-optical wave mixing. Here, we provide the detailed analysis of subcycle-resolved x-ray-optical wave mixing and show how it leads to the development of that method.





Our approach to analyze subcycle-resolved x-ray-optical wave-mixing signals consists of the description of the interaction of a crystal with an optical pulse beyond the perturbation theory, and, subsequently, the description of the interaction of the laser-dressed system with an ultrashort x-ray pulse. For the first part, we employ the Floquet-Bloch framework, which provides us with a material-specific description of light-matter interaction \cite{HsuPRB06, FaisalPRA97, TzoarPRB75}, and is also a convenient tool to analyze properties of laser-dressed systems, such as selection rules of HHG in periodic structures \cite{HsuPRB06, HiguchiPRL14, MoiseyevPRA15, IkedaPRA18}. The second part is deduced from the general theoretical framework in Ref.~\onlinecite{Popova-GorelovaPRB18} to describe the interaction of general Floquet systems with an x-ray pulse. It is based on the framework of quantum electrodynamics (QED) and the density matrix formalism \cite{Mandel}, which is necessary for a correct description of the interaction of an ultrashort light pulse with a nonstationary electronic system \cite{DixitPNAS12, Popova-GorelovaAppSci18}. Here, employing the Floquet-Bloch formalism and tools developed in Ref.~\onlinecite{Popova-GorelovaPRB18},  we analyze spatial symmetry and temporal behavior of microscopic optically-induced charge and electron-current distributions, and their connection to the delay and momentum dependence of subcycle-resolved x-ray-optical wave-mixing signals. Knowing these connections, we derive a procedure to reconstruct microscopic linear and nonlinear charge rearrangements, and the direction of electron current flow induced by optical excitation.

As we will demonstrate, understanding the information encoded in the subcycle-resolved x-ray-optical wave mixing signal relies on the analysis of the spatial symmetry and temporal dependence of optically-induced charge distributions and the electron current density. Thus, we analyze in Sec.~\ref{Sec_microscopic_within_FloqBloch} properties of the microscopic optical response within the Floquet-Bloch formalism. In Sec.~\ref{Sec_MgO_GaAs}, we calculate the microscopic optical response of two band-gap materials, the semiconductor GaAs and the insulator MgO driven by an intense optical field. By studying these two prototypical materials, we cover two types of crystals, one without inversion symmetry, GaAs, and one with inversion symmetry, MgO. In Sec.~\ref{SectionUXray}, we develop a method to measure the microscopic optical response by means of ultrafast x-ray-optical wave mixing.

\section{Microscopic optically-induced charge and electron-current distributions}

\label{Sec_microscopic_within_FloqBloch}

The Floquet formalism implies that the electric field of the driving field is temporally periodic. It has been shown in Refs.~\onlinecite{Ben-TalJPhB93, FleischerPRA05} that this approximation is already justified for a strong-field optical field with a duration comprising several tens of optical cycles. Throughout this paper, we consider a laser-dressed crystal in a state that is characterized by a single Floquet state. To justify this approximation, it should be assured that the driving field is not too strong to bring the laser-dressed system into a superposition of several Floquet states \cite{BreuerZphD89}. The applicability of these approximations can be verified with the radiation spectrum generated by the system through the driving field. If they do not hold, the radiation spectrum will have additional peaks besides the harmonic ones. Thus, our study applies to the regime in which each generated radiation peak can be clearly assigned to an integer multiple of the driving-laser frequency as, for example, in Refs.~\onlinecite{GhimireNature11, LuuNature15, KuehnPRL10, SchubertNature14, YouNature16}. 


We briefly review the classical limit of the quantized Floquet-Bloch formalism that we use to describe the nonperturbative interaction of an optical electromagnetic field with a band-gap crystal \cite{HsuPRB06, FaisalPRA97, TzoarPRB75, SantraPRA04, ShirleyPR65}. The quantized representation is necessary to introduce the interaction of a laser-dressed crystal with an ultrashort x-ray pulse within the QED framework in the next step \cite{Popova-GorelovaPRB18}. The Hamiltonian of a laser-dressed crystal is given by
\begin{align}
&\Hdel = \Hel+\Hint+\Hem,\label{HamiltonianDrivenSystem}\\
&\Hel = \int d^3r \hat\psi^\dagger(\br)[\mathbf p^2/2+V_c(\br)]\hat\psi(\br),\\
&\Hem = \omega\hat a_{\ka_0,s_0}^\dagger\hat a_{\ka_0,s_0},\\
&\Hint = \overline\alpha\int d^3r\hat \psi^\dagger(\mathbf r)\lf(\Aem(\mathbf r)\cdot\mathbf p\rt)\hat \psi(\mathbf r).
\end{align}
Here, $\Hel$ is the mean-field Hamiltonian of the unperturbed crystal with one-body eigenstates $|\varphi_{m\k}\ra$, where $\k$ is the Bloch wave vector and $m$ is the band and spin index. $V_c(\br) = V_c(\br+\mathbf R)$ is a space-periodic crystal field potential, $\mathbf R$ is a lattice vector. According to the Bloch theorem \cite{Kittel}, the corresponding one-body wave function of $|\varphi_{m\k}\ra$ has the form $\varphi_{m\k}(\br)=e^{i\k\cdot\mathbf r}u_{m\k}(\br)$, where $u_{m\k}(\br) = u_{m\k}(\br+\mathbf R)$ is a space-periodic function. $\mathbf p$ is the canonical momentum of an electron, $\hat \psi^\dagger$ ($\hat \psi$) is the electron creation (annihilation) field operator. $\Hem$ is the Hamiltonian of the electromagnetic field, and $\Hint$ the interaction Hamiltonian between the electromagnetic field and the electronic system, which we describe within the dipole approximation. $\hat a_{\ka,s}^\dagger$ ($\hat a_{\ka,s}$) creates (annihilates) a photon with wave vector $\ka$ and polarization $s$. We assume that only the $\ka_0$, $s_0$ mode with a corresponding polarization vector $\boldsymbol\epsilon_0$ and the energy $\omega = |\ka_0|c$, where $c$ is the speed of light, is occupied in the driving electromagnetic field, and that the state of the field is described by a single-mode coherent state $|\alpha,t\ra$. $\Aem (\mathbf r)$ is the vector potential operator of the electromagnetic field, and $\overline\alpha$ is the fine-structure constant. We neglect the $\Aem^2$ contribution for the optical field. We use atomic units for these and the following expressions.

Since the state of the electromagnetic field $|\alpha,t\ra$ is unaffected by the interaction with the electronic system by assumption, the one-body solution of the time-dependent Schr\"odinger equation $id|\psi_{i,\k},t\ra/dt =  \Hdel|\psi_{i,\k},t\ra$ can be represented as $|\psi_{i,\k},t\ra=|\phi^\el_{i,\k},t\ra|\alpha,t\ra$ with the corresponding electronic one-body Floquet-Bloch wave function  \cite{ShirleyPR65,FaisalPRA97, HsuPRB06}
\begin{align}
\phi^\el_{i,\k}(\br,t)= \sum_{m,\mu}c^i_{ m,\k,\mu}e^{-i\mu\omega t}\varphi_{m\k}(\br)\label{Eq_FB},
\end{align}
where $\mu$ is an integer and the $c^i_{ m,\k,\mu}$ are expansion coefficients. 

\subsection{Electron-density amplitudes}

The electron density of the laser-dressed system evolves in time as \cite{Popova-GorelovaPRB18}
\begin{align}
\rho(\br,t) = \sum_{\mu}e^{i\mu\omega t}\widetilde\rho_\mu(\br)\label{eq_FourierDensity}
\end{align}
with $\mu$th-order density amplitudes
\begin{align}
\widetilde\rho_{\mu}(\br) &= \int\limits_{\text{BZ}}\frac{d^3 k}{V_{\text{uc}}}\sum_{m,m',i,\mu'} c^{i*}_{m',\k,\mu'+\mu}c^i_{m,\k,\mu'}u^\dagger_{m'\k}(\br)u_{m\k}(\br),\label{Eq_ampl_geneq}
\end{align}
where $V_{\text{uc}}$ is the volume of a unit cell, $i$ denotes the index of occupied one-body Floquet-Bloch states and the integration is over the Brilliouin zone. The $\mu$th-order density amplitudes can be connected to properties of the $\mu$th-order macroscopic optical response. For example, the polarization is determined by the time-dependent electron density
\begin{align}
\mathbf P(t) \propto \int d^3 r\br\rho(\br,t),
\end{align}
and the $\mu$th-order component of the polarization is determined by the $\mu$th-order density amplitude
\begin{align}
\widetilde{\mathbf P}^{(\mu)}(\mu\omega) \propto  e^{i\mu\omega t}\int d^3 r\br\widetilde\rho_{\mu}(\br).
\end{align}
Thus, the amplitudes $\widetilde\rho_\mu(\br)$ are optically-induced charge distributions that give rise to a $\mu$th-order macroscopic optical response.

Our connection of the density amplitudes to the macroscopic polarization is consistent with the classical derivation of macroscopic polarization \cite{ShenBook, BloembergenPhRev64}. The conventional expansion of the polarization in orders of $\omega$, $\mathbf P = \widetilde{\mathbf P}^{(1)}(\omega) + \widetilde{\mathbf P}^{(2)}(2\omega)+\widetilde{\mathbf P}^{(3)}(3\omega)+\cdots$, also holds within the Floquet formalism. But in the nonperturbative regime, $\mu$th-order components should not scale as the $\mu$th power of the electric-field amplitude.

The volume integral of a $\mu$th-order density amplitude over a unit cell is given by
\begin{align}
\int d^3 r \widetilde\rho_{\mu}(\br)=\int\limits_{\text{BZ}}\frac{d^3 k}{V_{\text{uc}}}\sum_{m,m',i,\mu'}c^{i*}_{m',\k,\mu'+\mu}c^i_{m,\k,\mu'}\delta_{m,m'}.
\end{align}
We now apply the orthogonality of the expansion coefficients $\sum_{m,\mu'}c^{i*}_{m,\k,\mu'+\mu}c^i_{m,\k,\mu'} = \sum_{m,\mu'}\delta_{\mu'+\mu,\mu'}$ and obtain:
\begin{align}
\int d^3 r \widetilde\rho_{\mu}(\br)=\Nel \delta_{\mu,0}.
\end{align}
As we discuss in Ref.~\onlinecite{CitepaperShort}, this means that optically-induced positively and negatively charged regions cancel each other when volume integrated.


\subsection{Time-reversal symmetry}

\label{SecTimeReversal}

In the following, we will show that the time-reversal symmetry of a crystal determines the temporal behavior of the laser-driven electron oscillations. The derivations below are applied to a general situation, in which the crystal is not necessarily invariant under inversion symmetry in real space. First, let us consider how time-reversal symmetry influences the properties of the Floquet-Bloch functions. For a Bloch function $\varphi_{m\k}(\br)$ of an unperturbed crystal that obeys time-reversal symmetry, it is valid that $\varphi_{m\k}(\br)=\varphi^*_{m-\k}(\br)$ \cite{Kittel}. Similarly to the proof of this property, we show below that for a one-body electronic wave function of a laser-dressed crystal with time-reversal symmetry, it is true that
\begin{align}
\phi^\el_{i,\k}(\br,t)=\phi^{*\el}_{i,-\k}(\br,T/2-t),\label{Eq_tr_FB}
\end{align}
where $T=2\pi/\omega$ is the period of the driving electromagnetic field. 

Applying to the time-dependent Schr\"odinger equation 
\begin{align}
&i\frac{d|\phi^\el_{i,\k},t\ra}{dt}|\alpha,t\ra+i|\phi^\el_{i,\k},t\ra \frac{d|\alpha,t\ra}{dt}\label{Eq_TDSE_phi_ik} \\
&=(\Hel+\Hint+\Hem)|\phi^\el_{i,\k},t\ra|\alpha,t\ra\nonumber
\end{align}
that the coherent state $|\alpha,t\ra$ obeys $i d|\alpha,t\ra/dt =  \Hem|\alpha,t\ra$ and multiplying \eq{Eq_TDSE_phi_ik} by $\la\alpha,t|$, we obtain
\begin{align}
&i\frac{d|\phi^\el_{i,\k},t\ra}{dt} = \Hel|\phi^\el_{i,\k},t\ra+\la\alpha,t |\Hint|\alpha,t\ra|\phi^\el_{i,\k},t\ra.
\end{align}
The matrix element $\la\alpha,t |\Hint|\alpha,t\ra$ gives the interaction Hamiltonian in the classical limit $\Hint^{\text{cl}}(t)=\overline\alpha\int d^3r\hat \psi^\dagger(\mathbf r)\lf(\mathbf A_{\text{em}}(\mathbf r_0,t)\cdot\mathbf p\rt)\hat \psi(\mathbf r)$, where $\mathbf A_{\text{em}}(\mathbf r_0,t) =(c/\omega)\mathbf E_{\text{em}}(\br_0)\cos(\omega t)$ with $\mathbf E_{\text{em}}(\br_0)$ being the amplitude of the electric field. Since we apply the dipole approximation, we ignored the spatial variation of the vector potential and the electric-field amplitude, and substituted the position of the crystal $\br_0$ for $\br$ in $\mathbf A_{\text{em}}(\mathbf r_0,t)$ and $\mathbf E_{\text{em}}(\br_0)$. Thus, the one-body electronic wave function obeys
\begin{align}
&i\frac{\partial\phi^\el_{i,\k}(\br,t)}{\partial t} = \bigl[\Hel+ \Hint^{\text{cl}}(t)\bigr]\phi^\el_{i,\k}(\br,t).\label{Eq_tdSch_cl}
\end{align}

In order to prove \eq{Eq_tr_FB}, we take the complex conjugate of \eq{Eq_tdSch_cl} resulting in
\begin{align}
&-i\frac{\partial\phi^{\el*}_{i,\k}(\br,t)}{\partial t} = \bigl[\Hel- \Hint^{\text{cl}}(t)\bigr]\phi^{\el*}_{i,\k}(\br,t).
\end{align}
Here we applied that $\Hint^{\text{cl}*}(t) = -\Hint^{\text{cl}}(t)$ and $\Hel^*=\Hel$ for crystals with the time-reversal symmetry \cite{Kittel}. We now rewrite the above expression for the time $T/2-t$ taking into account that $\mathbf A_{\text{em}}(\mathbf r_0,T/2-t)=-\mathbf A_{\text{em}}(\mathbf r_0,t)$:
\begin{align}
&i\frac{\partial \phi^{\el*}_{i,\k}(\br,T/2-t)}{\partial t} = \bigl[\Hel+\Hint^{\text{cl}}(t)\bigr]\phi^{\el*}_{i,\k}(\br,T/2-t).
\end{align}
Thus, we obtain that if the wave function $\phi^\el_{i,\k}(\br,t)$ is a solution of the time-dependent Schr\"odinger equation, then $\phi^{\el*}_{i,\k}(\br,T/2-t)$ is also a solution. Since $\phi^\el_{i,\k}(\br+\mathbf R,t) = e^{i\k\cdot\mathbf R}\phi^\el_{i,\k}(\br,t)$, it must be valid that $\phi^{\el*}_{i,\k}(\br+\mathbf R,T/2-t) = e^{-i\k\cdot\mathbf R}\phi^{\el*}_\k(\br,T/2-t)$. Thus, the solution $\phi^{\el*}_{i,\k}(\br,T/2-t)$ is the solution at the Bloch vector $-\k$, and we have proven \eq{Eq_tr_FB}. \eq{Eq_tr_FB} leads to a connection between the corresponding expansion coefficients of Floquet-Bloch functions [cf.~\eq{Eq_FB}],
\begin{align}
c^i_{m\k\mu}=(-1)^\mu c^{i*}_{m-\k\mu},\label{Eq_coef}
\end{align}
which follows from the phase relation $e^{i\mu\omega t} = (-1)^\mu e^{-i\mu\omega (T/2-t)}$.

\subsubsection{Density amplitudes}
\label{SectionDenAmp} 

We now make use of \eq{Eq_coef} and the property of the Bloch functions of crystals with time-reversal symmetry that $u_{m\k}(\br)=u_{m-\k}^*(\br)$ \cite{Kittel} to connect terms with opposite $\k$ in the integral for $\widetilde\rho_{\mu}(\br)$ in \eq{Eq_ampl_geneq}. This allows us to reduce the integration over the Brillioun zone to half of the Brillioun zone (HBZ)
\begin{align}
\widetilde\rho_{\mu}(\br) = &\int\limits_{\text{HBZ}}\frac{d^3 k}{V_{\text{uc}}}\sum_{m,m',i,\mu'}c^{i*}_{m',\k,\mu'+\mu}c^i_{m,\k,\mu'}u^\dagger_{m'\k}(\br)u_{m\k}(\br)\nonumber\\
&+(-1)^{\mu}c.c.\label{Eqdenampcomplex}
\end{align}
It follows from this relation that even-order density amplitudes $\widetilde\rho_{\mueven}(\br)$ are real functions, whereas odd-order density amplitudes $\widetilde\rho_{\muodd}(\br)$ are purely imaginary.


This property has an important consequence for the time dependence of the electron density. We use \eq{Eqdenampcomplex} to combine terms with opposite $\mu$ in the expression for the time-dependent electron density in \eq{eq_FourierDensity}. We also use the relation $\widetilde\rho_{\mu}(\br)=\widetilde\rho_{-\mu}^*(\br)$, which can be easily shown independently of the crystal symmetry. This results in the time dependence of the electron density shown in Ref.~\onlinecite{CitepaperShort}
\begin{widetext}
\begin{align}
\rho(\br,t) = \widetilde\rho_0(\br)-\varrho_1(\br)\sin(\omega t)+\varrho_2(\br)\cos(2\omega t)-\varrho_3(\br)\sin(3\omega t)+\cdots\label{Eq_osc_den}\\
=\widetilde\rho_0(\br)-\sum_{\muodd\ge 1}\varrho_{\muodd}\sin(\muodd\omega t)+\sum_{\mueven\ge 2}\varrho_{\mueven}\cos(\mueven\omega t)
\nonumber
\end{align}
\end{widetext}
for the driving field with the vector potential evolving as $\cos(\omega t)$.  In \eq{Eq_osc_den}, we redefined the density amplitudes as follows:
\begin{align}
&\varrho_{\mueven}(\br)=2\Re[\widetilde\rho_{\mueven}(\br)],\label{Eq_varrho_defe}\\
&\varrho_{\muodd}(\br)=2 \Im[\widetilde\rho_{\muodd}(\br)],\label{Eq_varrho_defo}
\end{align}
which are real functions for both even and odd $\mu$. This representation of electron density amplitudes is more insightful in comparison to the functions $\widetilde\rho_\mu(\br)$, since it demonstrates the actual time dependence of the light-induced charge distributions. Throughout the article, we will mainly refer to the density amplitudes defined by the real-valued functions in Eqs.~(\ref{Eq_varrho_defe}) and (\ref{Eq_varrho_defo}), but will refer to the amplitudes $\widetilde\rho_\mu(\br)$ in Sec.~\ref{SectionUXray}. We, thus, obtain that time-reversal symmetry determines the phases of $\mu$th-order oscillations of optically-induced charge distributions. Broken time-reversal symmetry would lead to a different relation between the phases of induced-charge and electric-field oscillations.


\subsubsection{Electron current density}
\label{SectionElCurrDen}


We now analyze time dependence of the electron current density. The electron current density in the presence of the electromagnetic field is given by \cite{HsuPRB06,KellerBook}
\begin{widetext}
\begin{align}
&\mathbf j(\br,t) = -\rho(\br,t)\mathbf A_{\text{em}}(\br_0,t)+\int\limits_{\text{BZ}}\frac{d^3 k}{V_{\text{uc}}}\Im\lf[\sum_{i}\phi^{\el*}_{i,\k}(\br,t) \boldsymbol \nabla\phi^\el_{i,\k}(\br,t)\rt].\label{Eq_currden_gen}
\end{align}
Using the relation between the expansion coefficients of Floquet-Bloch functions at opposite $\k$ due to the time-reversal symmetry in \eq{Eq_coef}, we obtain the time evolution of the electron current density shown in Ref.~\onlinecite{CitepaperShort}
\begin{align}
\mathbf j(\br,t) = -\sum_{\muodd\ge 1}\bmj_{\muodd}(\br)\cos(\muodd\omega t)-\sum_{\mueven\ge 2}\bmj_{\mueven}(\br)\sin(\mueven\omega t),\label{Eq_osc_current}
\end{align}
where
\begin{align}
\bmj_{\muodd}(\br) =&2\int\limits_{\text{HBZ}}\frac{d^3 k}{V_{\text{uc}}}\Im\widetilde{\mathbf j}_{\k,\muodd} +\frac{\mathbf E_{\text{em}}(\br_0)}{2\omega}[\varrho_{\muodd-1}(\br)+\varrho_{\muodd+1}(\br)],\label{Eq_jodd}\\
\bmj_{\mueven}(\br) = &2\int\limits_{\text{HBZ}}\frac{d^3 k}{V_{\text{uc}}}\Re\widetilde{\mathbf j}_{\k,\mueven}-\frac{\mathbf E_{\text{em}}(\br_0)}{2\omega}[\varrho_{\mueven-1}(\br)+\varrho_{\mueven+1}(\br)]\label{Eq_jeven}
\end{align}
with
\begin{align}
\widetilde{\mathbf j}_{\k,\mu} = \sum_{i,m,m',\mu'}c^{i*}_{m',\k,\mu'+\mu} c^{i}_{m,\k,\mu'} (\varphi_{m'\k}\boldsymbol \nabla \varphi^*_{m\k}-\varphi^*_{m\k}\boldsymbol \nabla \varphi_{m'\k})
\end{align}
\normalsize
are real-valued amplitudes of the electron current density. Thus, the electron current density amplitudes oscillate with a phase shifted by $\pi/2$ with respect to the oscillations of the charge distributions of the same order. When the absolute value of the $\mu$th-order charge oscillation is at a maximum, the $\mu$th-order electron current density is zero and vice versa. 

The expression of the electron current density in \eq{Eq_osc_current} is presented in momentum gauge. As shown in Ref.~\onlinecite{ErnottePRB18}, the electron current density calculated in momentum gauge equals the electron current density in length gauge taking the sum of the interband and intraband contributions into account. The separate contributions of interband and intraband currents, however, must be corrected to be consistent in both gauges. Here, we discuss the total electron current density, not separate contributions.


Applying the continuity equation
\begin{align}
\operatorname{div}\mathbf j(\br,t)& = -\partial\rho(\br,t)/\partial t\label{Eq_den_div_curr}\\
&=\sum_{\muodd\ge 1}\muodd\omega\varrho_{\muodd}(\br)\cos(\muodd\omega t)+\sum_{\mueven\ge 2}\mueven\omega\varrho_{\mueven}(\br)\sin(\mueven\omega t)\nonumber,
\end{align}
\end{widetext}
we obtain the following relation between the amplitudes of the electron current density and the amplitudes of the electron density:
\begin{align}
\operatorname{div}\bmj_\mu(\br) = -\mu\omega\varrho_{\mu}(\br).\label{Eq_den_amp_curr}
\end{align}


Let us now analyze the connection between the dipole moment of the density amplitudes $\int d^3 r \br\varrho_{\mu}(\br)$ and the electron-current-density amplitudes: 
\begin{align}
& \int d^3 r \br\varrho_{\mu}(\br)= -\frac1{\mu\omega}\int d^3 r \br \operatorname{div}\bmj_\mu(\br) \label{Eq_vol_avr_currents} \\
&= -\frac1{\mu\omega}\oint\br[ \bmj_\mu(\br)\cdot d\mathbf S] +\frac1{\mu\omega}\int d^3 r \bmj_\mu(\br).\nonumber
\end{align}
The second line of \eq{Eq_vol_avr_currents} follows from vector-algebra relations for the dipole moment of a divergence. 
Since the volume integral of the optically-induced charge distributions $\varrho_{\mu\neq0}$ is zero, we find
\begin{align}
\mu\omega\int d^3 r \varrho_{\mu} = -\int d^3 r \operatorname{div}\bmj_\mu(\br) = -\oint\bmj_\mu(\br)\cdot d\mathbf S =0.\label{Eq_flux_curr}
\end{align}
Both surface integrals $\oint \bmj_\mu(\br)\cdot d\mathbf S$ and $\oint\br[ \bmj_\mu(\br)\cdot d\mathbf S]$ can be zero only if the electron current density $\bmj_\mu(\br)$ on the boundary of a unit cell is zero. Thus, if this is the case, the $\mu$th-order macroscopic polarization is proportional to the volume integral of the $\mu$th-order electron-current-density amplitudes.





\subsection{Crystal with inversion symmetry}
\label{Sec_InvSym}
In this subsection, we assume that the crystal exposed to periodic driving is invariant under inversion symmetry and, consequently, $\Hel(\br)=\Hel(-\br)$, and derive the symmetry properties of electron density and electron-current density amplitudes shown in Ref.~\onlinecite{CitepaperShort}. The interaction Hamiltonian $\Hint^{\text{cl}}$ is invariant under the transformations $\br\rightarrow-\br$ and $t\rightarrow t-T/2$. Thus, it follows from the time-dependent Schr\"odinger equation in \eq{Eq_tdSch_cl} that \cite{MoiseyevPRA15}
\begin{align}
&i\frac{d\phi^{\el}_{i,\k}(-\br,t-T/2)}{dt} = \bigl[\Hel+\Hint^{\text{cl}}(t)\bigr]\phi^{\el}_{i,\k}(-\br,t-T/2).
\end{align}
Since $\phi^{\el}_{i,\k}(\br+\mathbf R,t) = e^{i\k\cdot\mathbf R}\phi^{\el}_\k(\br,t)$, $\phi^{\el}_{i,\k}(-\br,t-T/2)$ is a solution at $-\k$ and 
\begin{align}
\phi^{\el}_{i,-\k}(-\br,t-T/2) = \phi^{\el}_{i,\k}(\br,t).\label{Eq_FB_invsym}
\end{align}
The Bloch functions of crystals that are invariant under inversion symmetry obey the relation $\varphi_{m\k}(\br) = \varphi_{m-\k}(-\br)= \varphi^*_{m\k}(-\br)$ \cite{Kittel}. Substitution of the relations between Bloch and Floquet-Bloch functions into the expansion of Floquet-Bloch functions in \eq{Eq_FB} gives a relation between the expansion coefficients at opposite $\k$:
\begin{align}
c^i_{ m,\k,\mu}=(-1)^\mu c^i_{ m,-\k,\mu}.
\end{align} 
Comparing it with the relation between the coefficients due to time-reversal symmetry in \eq{Eq_coef}, we find that the coefficients $c^i_{ m,\k,\mu}$ are real.

Substitution of these properties into the expression for the electron density amplitudes via Bloch functions in \eq{Eq_ampl_geneq} leads to the following connection between complex amplitudes at opposite $\br$ 
\begin{align}
\widetilde\rho_{\mu}(-\br) = \widetilde\rho^*_{\mu}(\br).
\end{align}
The property of these amplitudes that either their imaginary or real part is zero depending on the parity of $\mu$ determines how they behave under inversion symmetry. The same holds for the real-valued representation of the density amplitudes in Eqs.~(\ref{Eq_varrho_defe}) and (\ref{Eq_varrho_defo}) that all even-order density amplitudes are invariant under inversion symmetry, whereas all odd-order density amplitudes are opposite under inversion symmetry:
\begin{align}
&\varrho_{\mueven}(\br) = \varrho_{\mueven}(-\br),\\
&\varrho_{\muodd}(\br) = -\varrho_{\muodd}(-\br).
\end{align}
Analogously, using that $\varphi_{m\k}(-\br) = \varphi^*_{m\k}(\br)$ \cite{Kittel}, $\boldsymbol \nabla_{-\br}=-\boldsymbol \nabla_{\br}$ and the coefficients $c^i_{ m,\k,\mu}$ being real, we obtain the following symmetry properties of the current density amplitudes  
\begin{align}
\bmj_{\mueven}(\br) = -\bmj_{\mueven}(-\br),\\
\bmj_{\muodd}(\br) = \bmj_{\muodd}(-\br).
\end{align}
The volume integral of the even-order current density amplitudes is zero
\begin{align}
\int d^3 r\bmj_{\mueven}(\br) = 0
\end{align}
in agreement with the selection rule that even-order harmonics from crystals invariant under inversion symmetry are forbidden \cite{YarivBook}. 

In this Section, we analyzed microscopic properties of optically-induced charge distributions and the electron current density. The temporal dependence of the light-induced oscillations of the electronic state is determined by time-reversal symmetry. We found that components of the electron density oscillate either in phase with the electric field or in phase with the vector potential depending on the parity of the oscillation order. The inversion symmetry of a crystal results in the inversion symmetry of the $\mu$th-order charge distributions and $\mu$th-order amplitudes of the electron current density.  Thereby, their behavior under the transformation $\br\rightarrow-\br$ depends on the parity of the order.  As an outlook, it is interesting to analyze the consequence of other crystal symmetries on the spatial and temporal properties of optically-induced charge distributions. The Floquet-Bloch formalism is a convenient tool to perform such an analysis.

\section{Microscopic optical response in band-gap crystals $\boldsymbol{\mathrm{MgO}}$ and $\boldsymbol{\mathrm{GaAs}}$}
\label{Sec_MgO_GaAs}

\subsection{Computational details}

We diagonalize the Floquet-Bloch Hamiltonian as described in Refs.~\onlinecite{HsuPRB06, Popova-GorelovaPRB18}. We calculate the one-body wave functions $\varphi_{m\k}$ of the field-free Hamiltonian $\hat H_\el$ within the density functional theory using the ABINIT software package \cite{Gonze16,Gonze09,Gonze05} in combination with Troullier-Martins pseudopotentials \cite{Troullier-Martins_Pseudpotentials}. The functions $\varphi_{m\k}$ of valence bands and conduction bands are calculated on a dense grid of $\k$ points in half of the Brillioun zone. The numbers of blocks of the Floquet-Bloch matrix, $\k$ points, and bands are increased in the computations till convergence of the Fourier components of the electron density amplitudes is reached. We apply the scissors approximation \cite{LevinePRL89} to correct a band gap to its experimental value, which is necessary to obtain the correct position of inelastic x-ray scattering in the spectrum as shown in Sec.~\ref{SectionUXray}. 

The conduction bands that are necessary to converge the optical response are actually the bands into which electrons are excited by the electromagnetic field with a nonvanishing probability. The number of conduction bands involved in the interaction with the optical field strongly depends on the intensity of the optical field and crystal properties. This number increases with the intensity of the optical field, and is well above ten in the nonperturbative regime. There are several reasons for such a high required number of conduction bands. The first reason is that the higher the intensity of the optical field, the larger is the probability of an off-resonant transition into energetically high conduction bands. For example, a transition from a valence band into a conduction band with an energy difference detuned by 10 eV from the photon energy of the driving field can contribute to the first harmonic, if the intensity of the optical field is sufficiently high.

The next reason is that the higher the intensity of the optical field, the larger is the probability of a resonant multiphoton transition. As an example, let us consider the seven-photon absorption process induced by a field with a photon energy of $\omega= 1.55$ eV. A transition from a valence band to a conduction band with an energy difference of $7\omega = 10.85$ eV is resonant and should have a dominating contribution to this process. Then, it is crucial to take into account the conduction bands lying at $\approx 11$ eV above the outermost valence band for the calculation of the seventh-order optical response. The number of generated harmonics increases with increasing field intensity, and so should increase the number of conduction bands that are necessary to take into account for the calculation of a high harmonic spectrum. Multiphoton transitions also contribute to the optical response at lower orders in the nonperturbative regime. For example, seven-photon absorption combined with six-photon emission can contribute to the first-order response.


\subsection{Crystal without inversion symmetry: GaAs}

\label{Section_GaAs}

We consider the microscopic optical response of a crystal without inversion symmetry, GaAs, which has a band gap of 1.42 eV \cite{BlakemoreJAppPh82}. We consider optical excitation by an optical field of 1 eV photon energy and an intensity of $4\times10^{11}$ W/cm$^2$. $28\times28\times28$ Monkorst-Pack grid, 4 valence and 56 conduction bands, and 151 blocks of the Floquet-Bloch Hamiltonian are necessary to converge the results. According to the second-order susceptibility tensor of the space group F$\overline{4}$3m \cite{YarivBook}, the second-order macroscopic response of GaAs driven by a field polarized along the $(1,1,1)$ direction is allowed, whereas, for a field polarized along the $(1,0,0)$ direction, it is forbidden. In this subsection, we compare the microscopic optical response of GaAs to the excitation by driving fields polarized along the $(1,1,1)$ direction and along the $(1,0,0)$ direction.

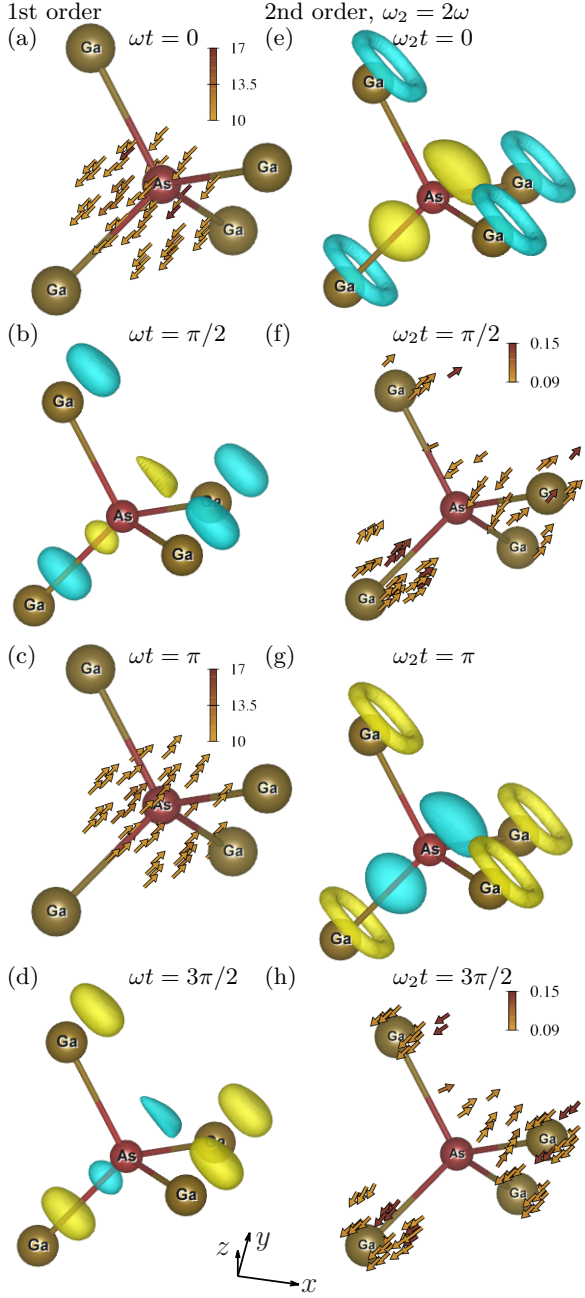
\begin{figure}[H]
\centering
\begin{overpic}[width=0.45\textwidth,tics=10,trim = 0 210 0 0,clip]{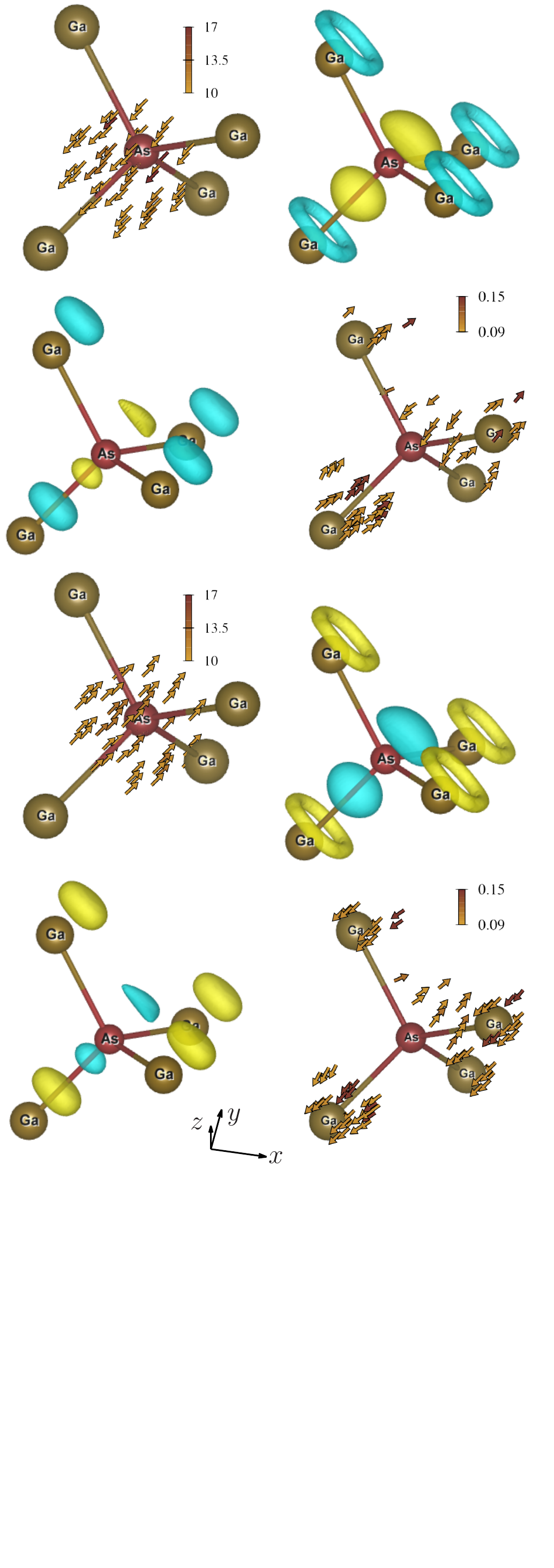}
\put(0,100){1st order}
\put(9.5,98){$\omega t =0$}
\put(9.5,75){$\omega t =\pi/2$}
\put(9.5,50){$\omega t =\pi$}
\put(9.5,25){$\omega t =3\pi/2$}
\put(0,98){(a)}
\put(0,75){(b)}
\put(0,50){(c)}
\put(0,25){(d)}
\put(20,100){2nd order, $\omega_2 = 2\omega$}
\put(30,98){$\omega_2 t =0$}
\put(30,75){$\omega_2 t =\pi/2$}
\put(30,50){$\omega_2 t =\pi$}
\put(30,25){$\omega_2 t =3\pi/2$}
\put(20,98){(e)}
\put(20,75){(f)}
\put(20,50){(g)}
\put(20,25){(h)}
\end{overpic}
\caption{The first- and second-order microscopic optical response of a GaAs crystal at different phases of the driving electromagnetic field polarized along the $(1,1,1)$ direction. A cut of a unit cell of GaAs centered at the As atom is shown. The first column shows the oscillations of the electron density and the electron current density with frequency $\omega$, second column corresponds to the frequency $\omega_2=2\omega$. The yellow and blue colors represent negative and positive charges, respectively. }
\label{Fig_GaAs_Den_Curr_111}
\end{figure}

The first and second columns of Fig.~\ref{Fig_GaAs_Den_Curr_111} show, respectively, the first- and second-order oscillations of the electronic state of GaAs driven by an optical field polarized along the $(1,1,1)$ direction. The first-order oscillations of laser-driven GaAs comprise the oscillations of the electron current density as $-\bmj_{1}(\br)\cos(\omega t)$ and the oscillations of the electron density as $-\varrho_{1}\sin(\omega t)$. The electron densities are represented in terms of an isosurface using VESTA \cite{MommaJAC11}. It is challenging to visualize the $\mu$th-order electron current amplitudes $\bmj_{\mu}(\br)$, since they are three-dimensional vector fields that are nonzero at most points within the unit cell. We plot $\bmj_{\mu}(\br)$, only if its magnitude $|\bmj_{\mu}(\br)|$ is larger than a certain minimum threshold. The magnitudes of $\bmj_{\mu}(\br)$ are color coded and their values are in atomic units. The minimum threshold for $|\bmj_{\mu}(\br)|$ in a given plot is the minimum value of the corresponding color box.

\begin{figure}[t]
\centering
\begin{overpic}[width=0.45\textwidth,tics=10,trim = 0 210 0 0,clip]{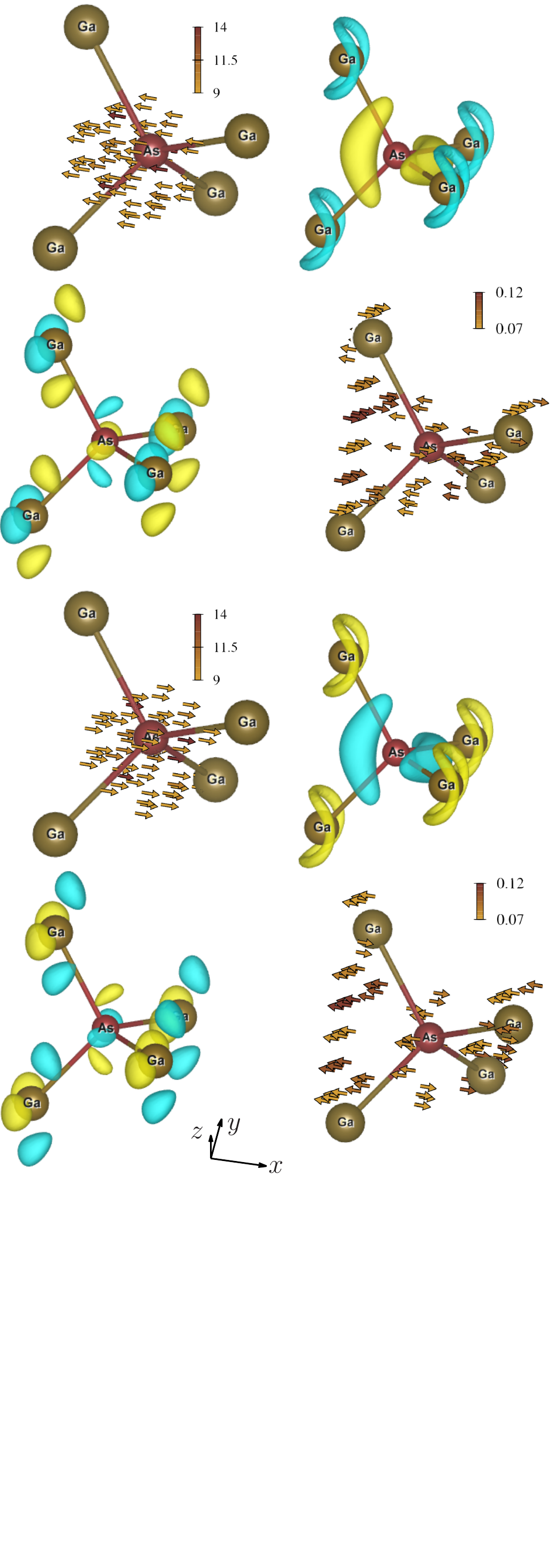}
\put(0,100){1st order}
\put(10,98){$\omega t =0$}
\put(10,75){$\omega t =\pi/2$}
\put(10,50){$\omega t =\pi$}
\put(10,25){$\omega t =3\pi/2$}
\put(0,95){(a)}
\put(0,72){(b)}
\put(0,47){(c)}
\put(0,21){(d)}
\put(22,100){2nd order, $\omega_2 = 2\omega$}
\put(22,98){$\omega_2 t =0$}
\put(22,75){$\omega_2 t =\pi/2$}
\put(22,50){$\omega_2 t =\pi$}
\put(22,25){$\omega_2 t =3\pi/2$}
\put(22,95){(e)}
\put(22,72){(f)}
\put(22,47){(g)}
\put(22,21){(h)}
\end{overpic}
\caption{Same as for Fig.~\ref{Fig_GaAs_Den_Curr_111}, except the driving field is polarized along the $(1,0,0)$ direction.}
\label{Fig_GaAs_Den_Curr_100}
\end{figure}
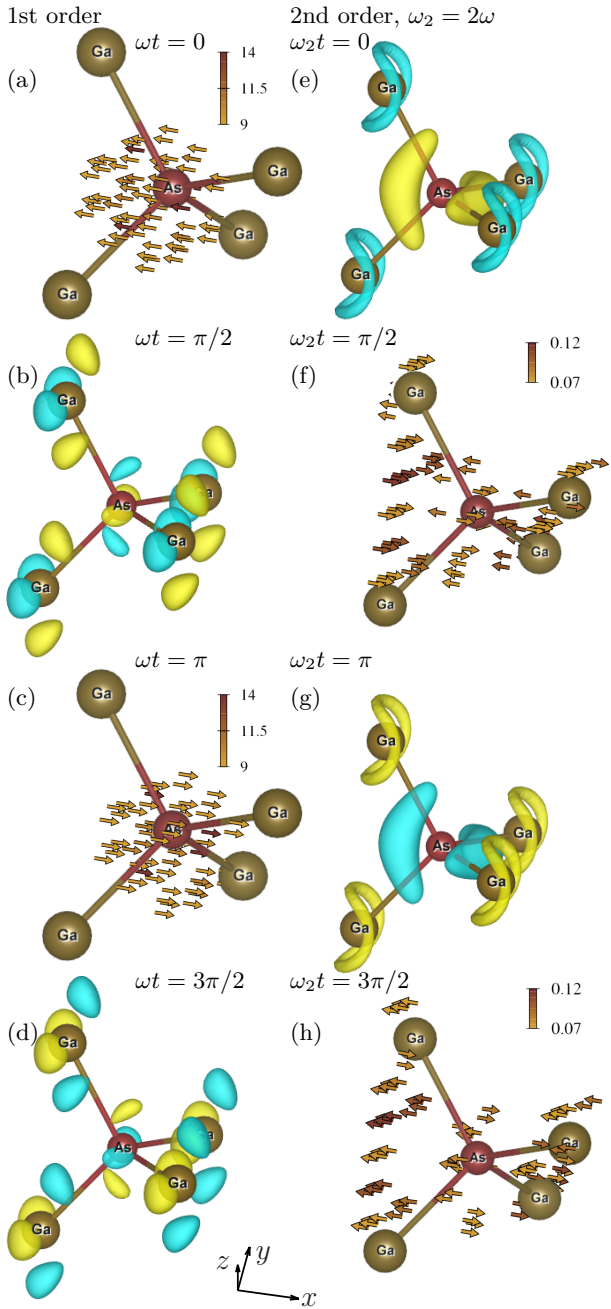

Figure \ref{Fig_GaAs_Den_Curr_111}(a) shows the first-order electron current density at $\omega t = 0$. The vector field $-\bmj_{1}(\br)$ clearly points along the driving-field polarization direction in agreement with the selection rule that the macroscopic first-order polarization of GaAs is aligned with the electric field \cite{YarivBook}. Figure \ref{Fig_GaAs_Den_Curr_111}(b) shows the first-order electron density at $\omega t = 0$. It has a three-fold rotational symmetry with respect to the driving-field polarization direction $(1,1,1)$. The positive charge alternates with the negative charge along the $(1,1,1)$ direction. 


The second column of Fig.~\ref{Fig_GaAs_Den_Curr_111} shows the second-order oscillations of the electronic state that comprise the oscillations of the electron density as $\varrho_2\cos(2\omega t)$ and of the electron current density as $-\bmj_2\sin(2\omega t)$. According to the second-order susceptibility tensor of GaAs \cite{YarivBook}, the second-order macroscopic polarization driven by an electric field polarized along the $(1,1,1)$ direction is also aligned along $(1,1,1)$. Figure \ref{Fig_GaAs_Den_Curr_111}(e) shows the second-order electron density at $\omega t = 0$. It also displays a three-fold rotational symmetry with respect to the driving-field polarization direction $(1,1,1)$. The positive charge alternates with the negative charge along the $(1,1,1)$ direction in agreement with the macroscopic polarization aligned along $(1,1,1)$.

The magnitude of the second-order electron current density reaches the maximum at $\omega t = \pi/4$ and is shown in Fig.~\ref{Fig_GaAs_Den_Curr_111}(f). It has a very complex structure that is difficult to characterize. We calculate the volume integral of $-\bmj_2$ and find that it indeed points in the $(1,1,1)$ direction in agreement with the selection rule for the second-order macroscopic polarization.

The first and second columns of Fig.~\ref{Fig_GaAs_Den_Curr_100} show, respectively, the first- and second-order oscillations of the electronic state of GaAs driven by a field polarized along the $(1,0,0)$ direction. Figure~\ref{Fig_GaAs_Den_Curr_100}(a) shows the first-order electron current density at $\omega t = 0$, when its magnitude is at a maximum. The vector field clearly points along the driving-field polarization direction $(1,0,0)$. This is in agreement with the alignment of the first-order macroscopic polarization of GaAs with the electric field \cite{YarivBook}.

Figure~\ref{Fig_GaAs_Den_Curr_100}(b) shows the first-order electron density at $\omega t = \pi/2$. It has two-fold rotational symmetry with respect to the direction of the driving-field polarization  $(1,0,0)$. It is not obvious how the charge distribution in Fig.~\ref{Fig_GaAs_Den_Curr_100}(b) results in the first-order macroscopic polarization along $(1,0,0)$, but one may notice that negative charges alter with positive charges in the $x$ direction, when looking at charges around the bottom Ga atoms.

According to the second-order susceptibility tensor of GaAs \cite{YarivBook}, its second-order macroscopic polarization is zero for a driving field polarized along the $x$ direction. We obtain that the second-order microscopic optical response is indeed nonzero. Figure~\ref{Fig_GaAs_Den_Curr_100}(e) shows the second-order electron density at $\omega t = 0$. It also has two-fold rotational symmetry about the $x$ axis as $\varrho_1(\br)$ in Fig.~\ref{Fig_GaAs_Den_Curr_100}(b). Figure~\ref{Fig_GaAs_Den_Curr_100}(f) shows the second-order electron current density at $\omega t = \pi/4$. Despite its complex structure, we find in our calculations that its volume integral is indeed zero in agreement with the zero second-order macroscopic polarization.  The magnitudes of the second-order electron current density in Figs.~\ref{Fig_GaAs_Den_Curr_100}(f) and (h) are similar to the magnitudes of the second-order electron current density induced by the field polarized along the $(1,1,1)$ direction in Figs.~\ref{Fig_GaAs_Den_Curr_111}(f) and (h). Thus, microscopic optical response can have comparable magnitudes in cases, when the electric field has a polarization direction resulting in either vanishing or nonzero macroscopic optical response.

In comparison to the first- and second-order microscopic optical response of MgO shown in Ref.~\onlinecite{CitepaperShort}, the microscopic optical response of GaAs is much more complex. The charge distributions in Figs.~\ref{Fig_GaAs_Den_Curr_100}(e) and (g) have no inversion symmetry, and it is not obvious how the corresponding macroscopic polarization becomes zero.

\subsection{Crystal with inversion symmetry, MgO}
\label{Seclaser-drivenMgO}


We show and discuss the first- and second-order microscopic optical response of a crystal with inversion symmetry, MgO, in Ref.~\onlinecite{CitepaperShort}. Here, we compare it to the third- and fourth-order microscopic optical response. We calculate the microscopic optical response to a driving optical field with an intensity of $I_{\text{em}}=2\times 10^{12}$ W/cm$^2$, a photon energy of 1.55 eV, and polarization axis $\boldsymbol\epsilon=(0,0,1)$. An optical field of $2\times 10^{12}$ W/cm$^2$ drives electron dynamics in MgO, which has a band gap of 7.8 eV \cite{RoesslerPhRev67}, nonperturbatively \cite{Popova-GorelovaPRB18}.  The factor $\sqrt {I_{\text{em}}}/\omega$ entering the off-diagonal matrix elements of the Floquet Hamiltonian \cite{HsuPRB06}  is the same as in the calculation of optical response of GaAs. The calculation is performed using a $24\times24\times24$ Monkhorst-Pack grid, four valence and sixteen conduction bands, and 81 blocks of the Floquet Hamiltonian, which are necessary to reach convergence.

\begin{figure}[H]
\centering
\includegraphics[width=0.47\textwidth]{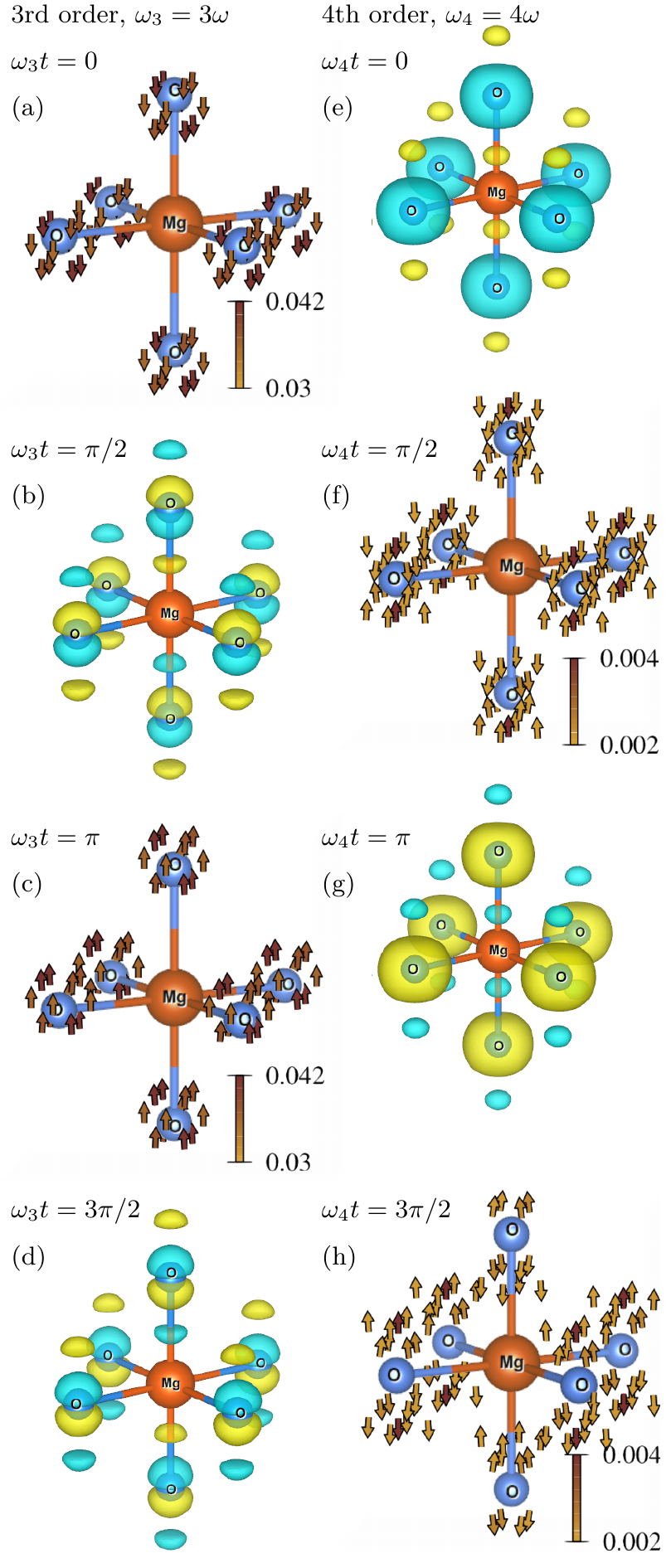}
\caption{The third- and fourth-order microscopic optical response of a MgO crystal at different phases of the driving electromagnetic field polarized along the $z$ direction. A cut of a unit cell centered around the Mg atom is shown. The first and second column shows the oscillations of the electron density and the electron current density with frequency $3\omega$ and $4\omega$, respectively. The yellow and blue colors represent negative and positive charges, respectively.  }
\label{MgO_Den_Curr}
\end{figure}

The first column of Fig.~\ref{MgO_Den_Curr} shows the third-order oscillations of the electronic state of laser-driven MgO crystal, comprising the oscillations of the electron density as $-\varrho_3\sin(3\omega t)$ and of the electron current density as $-\bmj_3\cos(3\omega t)$. Figure \ref{MgO_Den_Curr}(a) shows the third-order electronic state at $\omega t=0$, which is given by the third-order electron current density. Like the first-order electron current density in Ref.~\onlinecite{CitepaperShort}, $\bmj_3(\br)$ points predominantly in the direction of the driving-field polarization. Figure~\ref{MgO_Den_Curr}(b) shows the third-order electron density at $\omega t=\pi/6$. The charge distribution in Fig.~\ref{MgO_Den_Curr}(b) clearly indicates that the third-order macroscopic polarization points along the driving-field polarization direction. The magnitudes of the electron current density $\bmj_3(\br)$ are lower than the magnitudes of $\bmj_2(\br)$ shown in Ref.~\onlinecite{CitepaperShort}. Since the third-order harmonics from laser-dressed MgO has been observed \cite{YouNature16}, we conclude in Ref.~\onlinecite{CitepaperShort} that the second-order microscopic optical response of MgO is considerable.

The second column of Fig.~\ref{MgO_Den_Curr} shows the fourth-order oscillations of the electronic state that comprise the oscillations of the electron density as $\varrho_4\cos(4\omega t)$ and of the electron current density as $-\bmj_4\sin(4\omega t)$.  Since the fourth-order charge distribution is centrosymmetric, it has no dipole moment and leads to zero fourth-order macroscopic polarization. Interestingly, the third-order electron density amplitude of MgO has a very similar structure to its first-order electron density amplitude, and the fourth-order density amplitude has a very similar structure to the second-order electron density amplitude (cf.~Ref.~\onlinecite{CitepaperShort}).

In Ref.~\onlinecite{Popova-GorelovaPRB18}, where we developed the general theoretical framework to describe x-ray diffraction from a laser-driven electronic system, we presented the calculation of a subcycle-unresolved x-ray-optical wave-mixing signal from a laser-driven MgO as an example.  In that study, we found several surprising phenomena that we could not explain.  We can now understand them looking at the figures showing the microscopic optical response of MgO as discussed in Sec.~\ref{Section_xray_MgO}.

\section{Ultrafast x-ray-optical wave mixing}
\label{SectionUXray}

In this Section, we describe an experiment that reveals the complex structure of the optically-induced microscopic charge distribution. We use the general theoretical framework to describe x-ray diffraction from a laser-driven electronic system developed in Ref.~\onlinecite{Popova-GorelovaPRB18}. In that study, we presented a calculation of subcycle-unresolved measurement as an example. Here, we use this framework to develop a method  to reveal insights about laser-driven electron dynamics by means of subcycle-resolved x-ray-optical wave mixing.

We assume that a perfectly coherent x-ray pulse is used to probe the electronic state of a crystal during its interaction with the optical field. The x-ray probe pulse has a Gaussian-shaped electric field amplitude $\mathcal E_{x}(t) = \mathcal E_0\,e^{-2\ln2[(t-t_p)/\tau_p]^2}$, where $\tau_p$ is the x-ray-pulse duration. $t_p$ is the time of x-ray-pulse arrival relative to a reference time $t=0$, when the phase of the optical field $\omega t$ is zero, $\mathcal E_0$ is the peak amplitude. In Ref.~\onlinecite{Popova-GorelovaPRB18}, we showed that the scattering signal from a laser-dressed system is the sum of quasielastic and inelastic contributions,
\begin{align}
P_{\text{tot.}} = P_{\text{q.e.}} + P_{\text{inel.}}.
\end{align}
The quasielastic contribution is due to x-ray scattering causing transitions only within the manifold of initially occupied laser-dressed states. The inelastic contribution is due to x-ray scattering with final states that are different from initially occupied laser-dressed states. Since we use the dipole approximation for the interaction of the optical field and the crystal, the quasielastic part is present only at scattering vectors coinciding with the reciprocal lattice vectors $\bG$. The inelastic contribution is present at all scattering vectors. 

$P_{\qe}(\bG)$ is related to the Fourier transform of the $\mu$th-order density amplitudes:
\begin{align}
P_{\qe}(\omega_\ks,\bG) = P_0\Bigl |\sum_\mu&\widetilde{\mathcal E}_x(\omega_\ks-\omega_{\i}-\mu\omega)\label{Eq_Pqe_denamp}\\
&\times\int d^3 r e^{i\mathbf G\cdot\mathbf r}\widetilde\rho_\mu(\br)\Bigr|^2.\nonumber
\end{align}
Here, $P_0 = \sum_{s_\s}|(\boldsymbol\epsilon_{\i}\cdot \boldsymbol\epsilon^*_{x,\ks,s_\s})|^2\omega_\ks^2/(4\pi^2\omega_{\i}^2c^3)$, where $\boldsymbol\epsilon_{\i}$ is the mean polarization vector of the incoming x-ray beam, $\omega_\ks$ is the energy of a scattered photon with momentum $\ks$, the sum over $s_\s$ refers to the sum over polarization vectors of the scattered photons $\boldsymbol\epsilon^*_{x,\ks s_\s}$ and $\omega_{\i}$ is the mean photon energy of the incoming x-ray beam. $\widetilde{\mathcal E}_x(\omega_\ks-\omega_{\i}-\mu\omega)$ is the Fourier transform of the electric-field amplitude of the x-ray field
\begin{align}
\widetilde{\mathcal E}_x(\omega_\ks-\omega_{\i}-\mu\omega)=&\int_{-\infty}^{\infty}dt \mathcal E_x(t-t_p) e^{i(\omega_\ks-\omega_{\i}-\mu\omega) t}\\
=&\widetilde{\mathcal E}_\mu e^{-i\mu\omega t_p}e^{i(\omega_\ks-\omega_{\i}) t_p},\nonumber
\end{align}
where the
\begin{align}
\widetilde{\mathcal E}_\mu=&\sqrt{\frac{\tau^2_p\pi}{2\ln2}}e^{-(\omega_\ks-\omega_{\i}-\mu\omega)^2\tau_p^2/8\ln 2}
\end{align}
are Gaussian-shaped functions centered at scattered energies $\omega_{\i}+\mu\omega$.

Expanding the modulus squared in \eq{Eq_Pqe_denamp}, we obtain an expression for the quasielastic scattering probability $P_{\qe} (\bG)$ as a function of scattered energy and the time of the probe-pulse arrival, 
\begin{widetext}
\begin{align}
P_{\qe} (\bG)=&P_0\sum_{\mu}\widetilde{\mathcal E}_{\mu}^2\lf|\int d^3 r e^{i\mathbf G\cdot\mathbf r} \widetilde\rho_{\mu}(\br)\rt|^2
+2P_0\sum_{\mu,\Delta\mu>0}\widetilde{\mathcal E}_{\mu+\Delta\mu}\widetilde{\mathcal E}_{\mu}P_{\mu\leftrightarrow\mu+\Delta\mu}(\bG,t_p)\label{PqeInterf},
\end{align}
where
\begin{align}
P_{\mu\leftrightarrow\mu+\Delta\mu}(\bG,t_p)=&\cos(\Delta\mu\omega t_p)\Re\Bigl[\int d^3 r e^{i\mathbf G\cdot\mathbf r}\widetilde\rho_{\mu+\Delta\mu}(\br)\int d^3 r e^{-i\mathbf G\cdot\mathbf r}\widetilde\rho^*_{\mu}(\br)
\Bigr]\label{Eq_Interf_Term}\\
&
-\sin(\Delta\mu\omega t_p)\Im\Bigl[\int d^3 r e^{i\mathbf G\cdot\mathbf r}\widetilde\rho_{\mu+\Delta\mu}(\br)\int d^3 r e^{-i\mathbf G\cdot\mathbf r}\widetilde\rho^*_{\mu}(\br)\Bigr].\nonumber
 \end{align}
\end{widetext}

\normalsize
The first sum over $\mu$ in the expression above is time-independent. It is given by Gaussian-shaped functions centered at scattered energies $\omega_{\i}+\mu\omega$ and describes $\mu$-th order side peaks to the main Bragg peak of a crystal. Their amplitudes are are given by the Fourier transforms of the density amplitudes squared. This term contribute to the quasielastic scattering probability in both subcycle-resolved and subcycle-unresolved measurements.

The second sum in \eq{PqeInterf} is time-dependent and contributes only in a subcycle-resolved measurement. The time-dependent terms in the sum over $\mu$ and $\Delta \mu$ are due to the interference between the side peaks of $\mu$th and $(\mu+\Delta\mu)$th order. Their spectral position and bandwidth are determined by the product of two Gaussian-shaped functions, which is itself a Gaussian-shaped function 
\begin{align}
\widetilde{\mathcal E}_{\mu+\Delta\mu}\widetilde{\mathcal E}_{\mu}=e^{-(\Delta\mu\omega\tau_p)^2/16\ln 2}\widetilde{\mathcal E}^2_{\mu+\Delta\mu/2}\label{Eq_Gauss_middle}
\end{align}
centered at $\omega_{\i}+(\mu+\Delta\mu/2)\omega$. These terms are nonzero as long as the Gaussian functions $\widetilde{\mathcal E}_{\mu+\Delta\mu}$ and $\widetilde{\mathcal E}_{\mu}$ spectrally overlap. We use the criterion that if the factor $e^{-(\Delta\mu\omega\tau_p)^2/16\ln 2}$ is greater than 0.01, the corresponding interference terms cannot be neglected. Then, if the probe-pulse duration $\tau_p$ is less than $1.14\, T/\Delta\mu$, where $T=2\pi/\omega$ is the period of the optical-field cycle, the temporal resolution is sufficient to resolve the oscillations with the frequency $\Delta\mu\omega$. 


For example, if the optical field has a photon energy of 1.55 eV, then the optical period $T$ is 2.67 fs. An x-ray probe pulse with a duration shorter than 3 fs would provide a temporal resolution that is sufficient to resolve oscillations with the frequency $\omega$. The time-dependent part of the spectrum in the spectral interval between $\omega_{\i}$ and $\omega_{\i}+\omega$ is then given by the interference terms between the main peak and the first-order side peak and equals
\begin{align}
&
2\widetilde{\mathcal E}_{1}\widetilde{\mathcal E}_{0}\cos(\omega t_p)\Re\Bigl[\int d^3 r e^{i\mathbf G\cdot\mathbf r}\widetilde\rho_{1}(\br)\int d^3 r e^{-i\mathbf G\cdot\mathbf r}\widetilde\rho^*_{0}(\br)
\Bigr]\label{Eq_P_01}\\
&-2\widetilde{\mathcal E}_{1}\widetilde{\mathcal E}_{0}\sin(\omega t_p)\Im\Bigl[\int d^3 r e^{i\mathbf G\cdot\mathbf r}\widetilde\rho_{1}(\br)\int d^3 r e^{-i\mathbf G\cdot\mathbf r}\widetilde\rho^*_{0}(\br)\Bigr].\nonumber
\end{align}
The time-dependent part of the spectrum in the spectral interval between $\omega_{\i}+\omega$ and $\omega_{\i}+2\omega$ is given by the interference terms between the first- and second-order side peaks and equals to
\begin{align}
&
2\widetilde{\mathcal E}_{2}\widetilde{\mathcal E}_{1}\cos(\omega t_p)\Re\Bigl[\int d^3 r e^{i\mathbf G\cdot\mathbf r}\widetilde\rho_{2}(\br)\int d^3 r e^{-i\mathbf G\cdot\mathbf r}\widetilde\rho^*_{1}(\br)
\Bigr]\label{Eq_P_12}\\
&-2\widetilde{\mathcal E}_{2}\widetilde{\mathcal E}_{1}\sin(\omega t_p)\Im\Bigl[\int d^3 r e^{i\mathbf G\cdot\mathbf r}\widetilde\rho_{2}(\br)\int d^3 r e^{-i\mathbf G\cdot\mathbf r}\widetilde\rho^*_{1}(\br)\Bigr].\nonumber
\end{align}

\subsection{Symmetry of the Fourier transform of the electron density amplitudes}
\label{Sec_Four}

In order to analyze the interference terms in more detail, let us look into the Fourier transform of the density amplitudes. As we have shown in Sec.~\ref{SectionDenAmp}, the even-order density amplitudes of the laser-driven crystal $\widetilde \rho_{\mueven} (\br,t)=\varrho_{\mueven}(\br)/2$ are real functions and the odd-order density amplitudes $\widetilde \rho_{\muodd} (\br,t)=i\varrho_{\muodd}(\br)/2$ are purely imaginary. Thus, we can represent the Fourier transform of an even-order density amplitude as
\begin{align}
\int d^3r e^{i\bG\cdot\br}\widetilde\rho_{\mueven}(\br)= \frac12\mathcal P^g_{\mueven} (\bG)+\frac{i}2\mathcal P^u_{\mueven} (\bG),
\end{align}
and the Fourier transform of an odd-order density amplitude as
\begin{align}
\int d^3r e^{i\bG\cdot\br}\widetilde\rho_{\muodd}(\br)= \frac{i}2\mathcal P^g_{\muodd} (\bG)-\frac{1}2\mathcal P^u_{\muodd} (\bG),
\end{align}
where the functions
\begin{align}
\mathcal P^g_\mu (\bG)=\int d^3 r \cos(\mathbf G\cdot\mathbf r) \varrho_\mu(\br)\label{Eq_Pgmu}
\end{align}
and
\begin{align}
\mathcal P^u_\mu (\bG)=\int d^3 r \sin(\mathbf G\cdot\mathbf r) \varrho_\mu(\br).\label{Eq_Pumu}
\end{align}
are real.
The function $\mathcal P^g_\mu (\bG)$ is an even function of $\bG$ 
\begin{align}
&\mathcal P^g_\mu (\bG) = \mathcal P^g_\mu (-\bG),
\end{align}
whereas $\mathcal P^u_\mu (\bG)$ is an odd function of  $\bG$
\begin{align}
&\mathcal P^u_\mu (\bG) = -\mathcal P^u_\mu (-\bG).
\end{align}

We can now analyze the symmetry of the interference terms $P_{\mu\leftrightarrow\mu+\Delta\mu}(\bG,t_p)$ in \eq{Eq_Interf_Term}. The product of integrals that enters this expression can be represented as
\begin{align}
&\int d^3 r e^{i\mathbf G\cdot\mathbf r}\widetilde\rho_{\mu+\Delta\mu}(\br)\int d^3 r e^{-i\mathbf G\cdot\mathbf r}\widetilde\rho^*_{\mu}(\br)\\
&=\frac{(-i)^{\Delta\mu}}{4}\Bigl [\mathcal P^g_{\mu+\Delta\mu} \mathcal P^g_{\mu} +\mathcal P^u_{\mu+\Delta\mu} \mathcal P^u_{\mu} \nonumber \\
&\quad\quad\quad\quad\quad+i\lf(  \mathcal P^g_{\mu+\Delta\mu}\mathcal P^u_{\mu} -\mathcal P^u_{\mu+\Delta\mu} \mathcal P^g_{\mu}\rt)\Bigr].\nonumber
\end{align}
If $\Delta\mu$ is even, the real part of the above expression is an even function of $\bG$, whereas the imaginary part is an odd function of $\bG$. If $\Delta\mu$ is odd, the dependence is the opposite: the real part of the above expression is an odd function of $\bG$, the imaginary part is an even function of $\bG$. Thus, the centrosymmetric parts of the interference terms oscillate as either $\cos(\Delta\mueven \omega t_p)$ or $\sin(\Delta\muodd \omega t_p)$. The antisymmetric parts oscillate as either $\cos(\Delta\muodd \omega t_p)$ or $\sin(\Delta\mueven \omega t_p)$.

\subsubsection{Crystals with an inversion symmetry}
\label{Symm_x-ray-optical_inversion}

We  now consider the Fourier transform of density amplitudes of crystals with spatial inversion symmetry. We have shown in Sec.~\ref{Sec_InvSym} that such even-order density amplitudes $\varrho_{\mueven}(\br)$ are symmetric with respect to the transformation $\br\rightarrow-\br$. As a result, the integral $\mathcal P^u_{\mueven} (\bG)$ is zero for the even-order amplitudes of a crystal with inversion symmetry. Hence, the Fourier transform of $\widetilde \rho_{\mueven}(\br)$ is a real function
\begin{align}
&\int d^3r e^{i\bG\cdot\br}\widetilde \rho_{\mueven}(\br)= \frac12\mathcal P^g_{\mueven} (\bG).\label{Eq_Prhoeven_inv}
\end{align}

The odd-order density amplitudes of a crystal with inversion symmetry $\varrho_{\muodd}(\br)$ are antisymmetric with respect to the transformation $\br\rightarrow-\br$. Thus, the integrals $\mathcal P^g_{\muodd} (\bG)$ are zero, so that the Fourier transform of an odd-order density amplitude of a crystal with inversion symmetry is also a real function
\begin{align}
&\int d^3r e^{i\bG\cdot\br}\widetilde \rho_{\muodd}(\br)=-\frac{1}2\mathcal P^u_{\muodd} (\bG).\label{Eq_Prhoodd_inv}
\end{align}

Thus, we obtain that the imaginary parts of the products $\int d^3 r e^{i\mathbf G\cdot\mathbf r}\widetilde\rho_{\mu+\Delta\mu}(\br)\int d^3 r e^{-i\mathbf G\cdot\mathbf r}\widetilde\rho^*_{\mu}(\br)$ entering the expression for interference terms in \eq{Eq_Interf_Term} are zero. As the consequence, the time evolution of the quasielastic scattering involves only $\cos(\Delta\mu\omega t_p)$ functions. The time-dependent terms that evolve as $\cos(\Delta\muodd\omega t_p)$ are antisymmetric functions of $\bG$. The terms that evolve as $\cos(\Delta\mueven\omega t_p)$ are centrosymmetric functions of $\bG$.

\subsection{Time dependence of the x-ray-optical wave mixing signal}

\subsubsection{Crystal with broken inversion symmetry}

\begin{figure}
\centering
\includegraphics[width=0.47\textwidth]{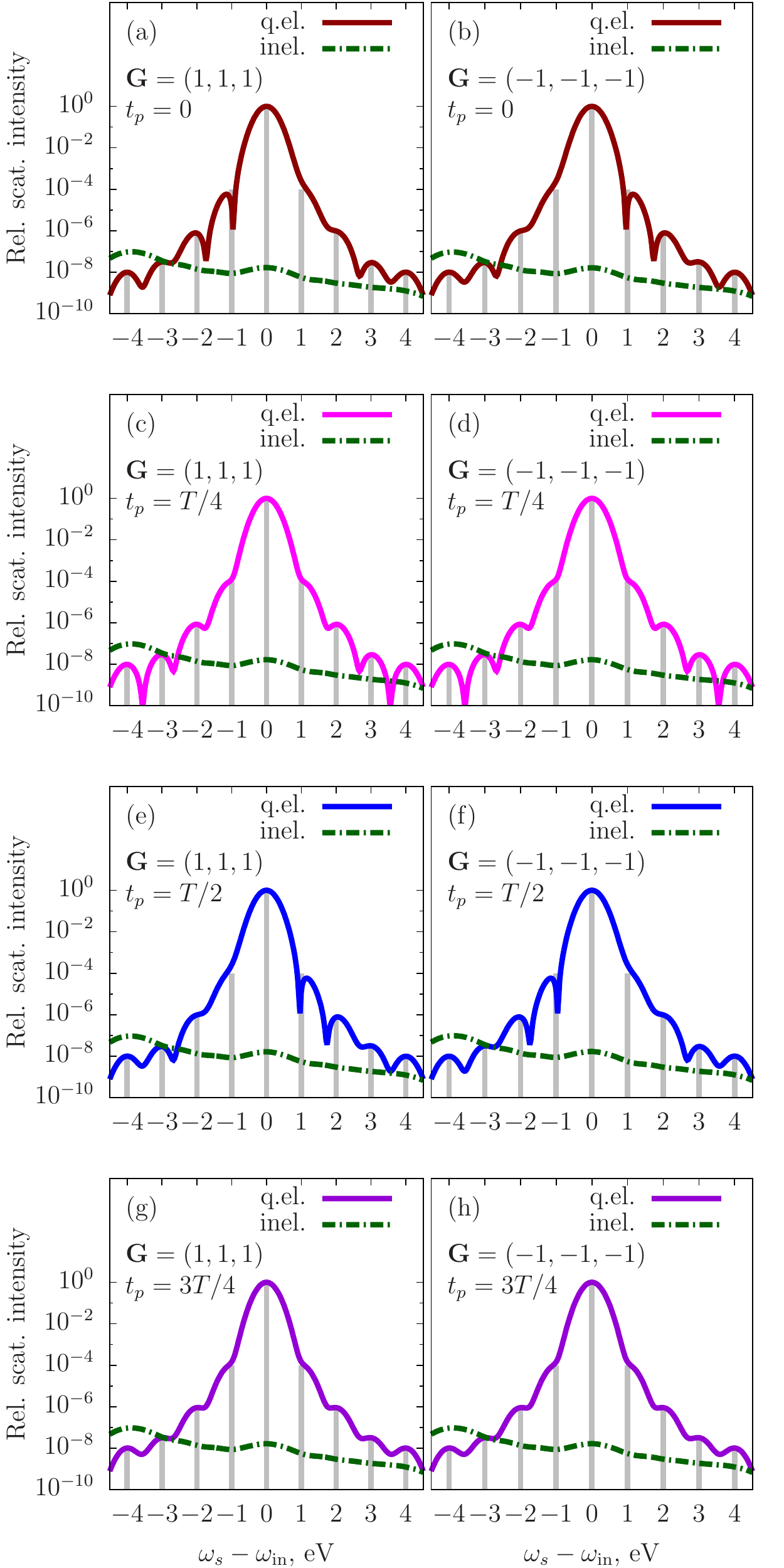}
\caption{Intensities of quasielastic and inelastic x-ray scattering signals at $\mathbf G = (1,1,1)$ and $\mathbf G = (-1,-1,-1)$ from the laser-dressed GaAs crystal at different probe-pulse arrival times as a function of $\omega_\s-\omega_\i$. The intensities are normalized to the intensity of the main Bragg peak of GaAs at $\mathbf G = (1,1,1)$. The gray vertical lines are situated at the positions of the side peaks, $\mu\omega$, and their heights correspond to their relative intensities.}
\label{Fig_Braggs_111GaAs}
\end{figure}

We describe the subcycle-resolved x-ray scattering signal from laser-dressed GaAs, which is a crystal without inversion symmetry. We use the same parameters of the optical field as in Fig.~\ref{Fig_GaAs_Den_Curr_100} in Sec.~\ref{Section_GaAs}, namely, an intensity of $4\times 10^{11}$ W/cm$^2$, a photon energy of $\omega=1$ eV. The polarization of the optical field is along $(1,0,0)$, which leads to zero second-order macroscopic polarization. The optical period of the driving field is 4.14 fs. We assume a probe x-ray pulse duration of 3.5 fs, which provides sufficient temporal resolution to resolve the oscillations of the electronic state of laser-dressed GaAs with frequency $\omega$.

Figure \ref{Fig_Braggs_111GaAs} shows the energy-resolved quasielastic and inelastic scattering signals at the scattering vectors $\bG = (1,1,1)$ and $\bG = (-1,-1,-1)$ at different probe-pulse arrival times. Scattering signals are normalized to the main Bragg peak at $\bG = (1,1,1)$, which is centered at the scattered energy $\omega_\s = \omega_\i$. We have chosen to analyze the signal at the scattering vectors $\bG =\pm (1,1,1)$, because the Fourier transform of the field-free electron density of the GaAs crystal at $\bG=\pm(1,1,1)$ is complex and both its centrosymmetric part $\mathcal P^g_{0} (\bG)$ and antisymmetric part $\mathcal P^u_{0} (\bG)$ are nonzero.

The green dotted lines show inelastic scattering from the laser-dressed GaAs. Inelastic scattering from a field-free GaAs crystal would appear only at scattered energies less than the incoming x-ray photon energy minus the band-gap of 1.42 eV. When GaAs is driven by the optical field, the inelastic signal is modulated. The inelastic scattering is then nonzero at higher scattered energies, but decays with the increasing scattered energy. The inelastic contribution is much higher than the inelastic contribution to the scattering signal from the laser-driven MgO in Ref.~\onlinecite{CitepaperShort}. This is due to the smaller band gap of GaAs, which means that the inelastic signal from a field-free crystal is energetically closer to the main Bragg peak. The inelastic contribution is still much smaller than the intensities of the side peaks at $\omega_\s>\omega_\i$. Thus, we will analyze the probability of quasielastic scattering at $\omega_\s>\omega_\i$, where it dominates in the total x-ray scattering probability from a laser-driven band-gap crystal.


The gray lines in Fig.~\ref{Fig_Braggs_111GaAs} show the intensities of the side peaks relative to the main Bragg peak. As in the case of the laser-dressed MgO in Ref.~\onlinecite{CitepaperShort}, we observe that the intensity of the second-order side peak is nonzero. Thus, x-ray-optical wave mixing reveals the second-order microscopic response of GaAs despite zero second-order macroscopic optical response, when the driving field is polarized along $(1,0,0)$. 

We again find that the quasielastic scattering signal at $t_p=0$ [Figs.~\ref{Fig_Braggs_111GaAs}(a) and (b)] and at $t_p=T/2$ [Figs.~\ref{Fig_Braggs_111GaAs}(e) and (f)] is non-centrosymmetric with respect to $\bG$ at a fixed scattered energy as in the case of MgO in Ref.~\onlinecite{CitepaperShort}. To understand this, let us apply the results of Sec.~\ref{Sec_Four} to the expression for the quasielastic scattering probability in \eq {PqeInterf}:
\begin{widetext}
\begin{align}
P_{\qe} (\bG,\omega_\s>\omega_\i)=&\frac{P_0\widetilde{\mathcal E}_{0}^2}4\Bigl\{ \lf[  \mathcal P^u_{0} (\bG)\rt]^2+  \lf[ \mathcal P^g_{0} (\bG)  \rt]^2\Bigr\}
+\frac{P_0\widetilde{\mathcal E}_{1}\widetilde{\mathcal E}_{0}}2\Bigl\{ 
\cos(\omega t_p)\Bigl[\mathcal P^g_{1} (\bG) \mathcal P^u_{0} (\bG) - \mathcal P^u_{1} (\bG)\mathcal P^g_{0} (\bG)\Bigr] \label{Pqe_noinv}\\
&\quad-\sin(\omega t_p)\Bigl[\mathcal P^g_{1} (\bG)\mathcal P^g_{0} (\bG)
+\mathcal P^u_{1}(\bG)\mathcal P^u_{0} (\bG)\Bigr]\Bigr\}
+\frac{P_0\widetilde{\mathcal E}_{1}^2}4\Bigl\{ \lf[  \mathcal P^u_{1} (\bG)\rt]^2+  \lf[ \mathcal P^g_{1} (\bG)  \rt]^2\Bigr\}\nonumber\\
&+\frac{P_0\widetilde{\mathcal E}_{2}\widetilde{\mathcal E}_{1}}2 
\Bigl\{  -\cos(\omega t_p) \Bigl[\mathcal P^g_{2} (\bG)\mathcal P^u_{1} (\bG)-\mathcal P^u_{2} (\bG)\mathcal P^g_{1} (\bG)\Bigr]\nonumber\\
&\quad+\sin(\omega t_p)\Bigl[
\mathcal P^g_{2} (\bG)\mathcal P^g_{1} (\bG)+\mathcal P^u_{2} (\bG)\mathcal P^u_{1} (\bG)\Bigr]\Bigr\}
+\frac{P_0\widetilde{\mathcal E}_{2}^2}4\Bigl\{ \lf[  \mathcal P^u_{2} (\bG)\rt]^2+  \lf[ \mathcal P^g_{2} (\bG)  \rt]^2\Bigr\}+\cdots
.\nonumber
\end{align}
\normalsize
\end{widetext}
Here, we took into account that only nearest-neighbor side peaks can interfere for the chosen probe-pulse duration. The terms in the above expression are ordered according to their position in the spectrum. The quasielastic scattering signal consists of time-independent and time-dependent contributions. The time-independent contribution is due to the main Bragg peak centered at the scattered energy $\omega_\s=\omega_{\i}$ and its side peaks centered at scattered energies $\omega_\i+\mu\omega$. The corresponding terms are centrosymmetric with respect to the transformation $\bG\rightarrow-\bG$. The relative intensities of the side peaks are shown with gray lines in Fig.~\ref{Fig_Braggs_111GaAs}. They are proportional to the absolute values of the Fourier transforms of the corresponding $\mu$th-order optically induced charge distributions. The sum-frequency signal, which was experimentally observed in x-ray diffraction from laser-driven diamond in Ref.~\onlinecite{GloverNature12}, is the first-order side peak in our terminology \cite{Popova-GorelovaPRB18}.

The time-dependent contribution is due to the interference terms between the nearest-neighbor side peaks of the $\mu$-th and $(\mu+1)$-th order that are given by Gaussian-shaped functions centered at scattered energies $\omega_\i+(\mu+1/2)\omega$. It consists of two parts. The first part evolves in time as $\cos(\omega t_p)$ and is an antisymmetric function of $\bG$. The second contribution evolves in time as $\sin(\omega t_p)$ and is a centrosymmetric function of $\bG$. 

\begin{figure}[t]
\centering
\includegraphics[width=0.42\textwidth]{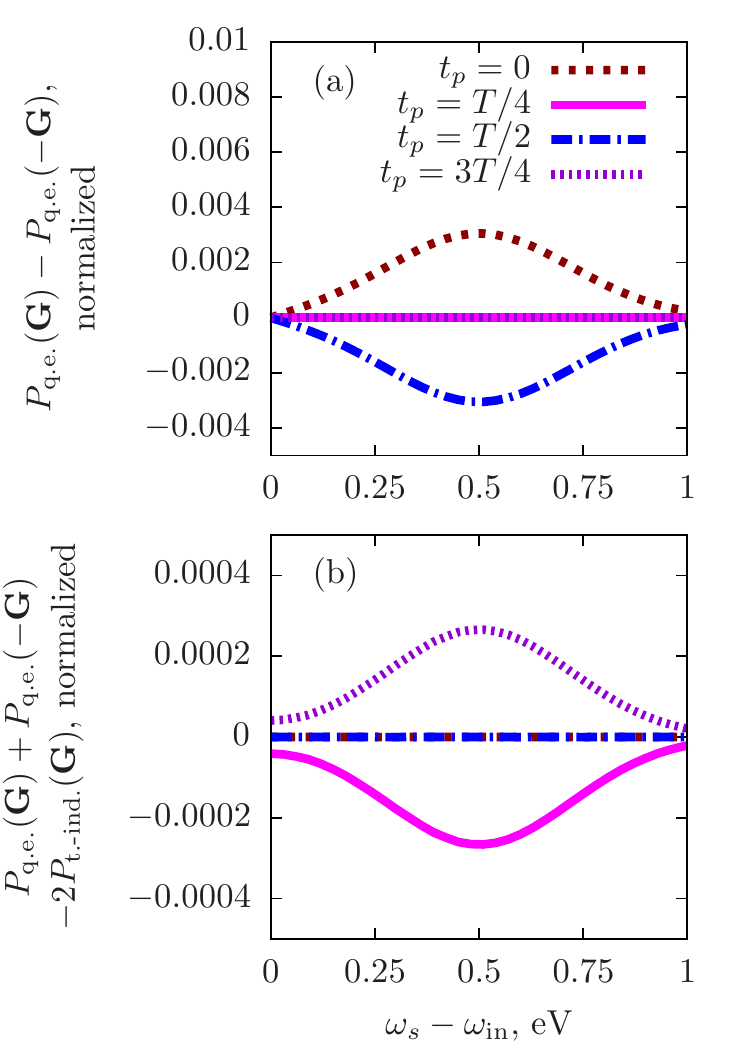}
\caption{(a) Difference and (b) sum of time-dependent contributions to quasielastic scattering at $\mathbf G = (1,1,1)$ and $\mathbf G = (-1,-1,-1)$ from the laser-dressed GaAs crystal at different probe-pulse arrival times as a function of $\omega_\s-\omega_\i$ in the range $[0:\omega]$.}
\label{Fig_DiffSumBraggs_111GaAs}
\end{figure}

Both contributions can be disentangled from the total scattering signal. The first contribution can be disentangled by taking the difference at opposite $\bG$. The difference of the signals at opposite $\bG$ is shown in Fig.~\ref{Fig_DiffSumBraggs_111GaAs}(a) in the spectral range where the main peak and the first-order side peak interfere. Its time dependence coincides with the time dependence of the first-order oscillation of the electron current density $-\bmj_{1}(\br)\cos(\omega t)$. The second contribution can be disentangled by first taking the sum of the signals at opposite $\bG$. The time-independent part $P_{\text{t.-ind.}}$ must be then subtracted. $P_{\text{t.-ind.}}$ can be obtained by taking the sum of signals at opposite $\bG$ at the probe-pulse arrival time $t_p=0$: $2P_{\text{t.-ind.}}=P_{\qe} (\bG,t_p=0)+P_{\qe} (-\bG,t_p=0)$. After the time-independent part is subtracted, the sum of the remaining part of the signals at opposite $\bG$ gives the second contribution. This sum is shown in Fig.~\ref{Fig_DiffSumBraggs_111GaAs}(b) in the spectral range where the main peak and first-order side peak interfere. It evolves in time as $\sin(\omega t_p)$, which coincides with the time evolution of the first-order oscillation of the electron density $-\varrho_1(\br)\sin(\omega t_p)$. Thus, the momentum dependence of subcycle-resolved x-ray-optical wave mixing reveals the time dependence of electron current density and charge density oscillations.

Let us now consider how to reconstruct the electron density amplitudes from the quasielastic scattering signal. The Fourier transform of the electron density amplitudes $\int d^3r e^{i\bG\cdot\br}\varrho_{\mu}(\br)$ are complex functions that can be represented as $\Bigl|  \int d^3r e^{i\bG\cdot\br}\varrho_{\mu}(\br)  \Bigr|e^{i\alpha_\mu(\bG)}$. It follows from the definition of $\mathcal P^g_{\mu} (\bG)$ and $\mathcal P^u_{\mu} (\bG)$ in Sec.~\ref{Sec_Four} that
\begin{align}
&\mathcal P^g_{\mu} (\bG)= \Bigl|  \int d^3r e^{i\bG\cdot\br}\varrho_{\mu}(\br)  \Bigr|\cos[\alpha_\mu(\bG)],\\
&\mathcal P^u_{\mu} (\bG)= \Bigl|  \int d^3r e^{i\bG\cdot\br}\varrho_{\mu}(\br)  \Bigr|\sin[\alpha_\mu(\bG)].
\end{align}
Thus, rewriting the above expression in \eq{Pqe_noinv}, we obtain the delay dependence of the x-ray-optical wave mixing signal in Ref.~\onlinecite{CitepaperShort}:
\begin{widetext}
\begin{align}
P_{\qe} (\bG,\omega_\s>\omega_\i)=&\frac{P_0\widetilde{\mathcal E}_{0}^2}4 \Bigl|  \int d^3r e^{i\bG\cdot\br}\varrho_{0}(\br)  \Bigr|^2
+\frac{P_0\widetilde{\mathcal E}_{1}\widetilde{\mathcal E}_{0}}2\Bigl|  \int d^3r e^{i\bG\cdot\br}\varrho_{0}(\br)  \Bigr|\Bigl|  \int d^3r e^{i\bG\cdot\br}\varrho_{1}(\br)  \Bigr|
\sin(\omega t_p-\alpha_0+\alpha_1) \label{Pqe_noinv_phases}\\
&+\frac{P_0\widetilde{\mathcal E}_{1}^2}4 \Bigl|  \int d^3r e^{i\bG\cdot\br}\varrho_{1}(\br)  \Bigr|^2
-\frac{P_0\widetilde{\mathcal E}_{2}\widetilde{\mathcal E}_{1}}2 \Bigl|  \int d^3r e^{i\bG\cdot\br}\varrho_{1}(\br) \Bigl|  \int d^3r e^{i\bG\cdot\br}\varrho_{2}(\br)  \Bigr| 
\sin(\omega t_p-\alpha_1+\alpha_2)\nonumber\\
&+\frac{P_0\widetilde{\mathcal E}_{2}^2}4 \Bigl|  \int d^3r e^{i\bG\cdot\br}\varrho_{2}(\br)  \Bigr|^2+\cdots\nonumber
.\nonumber
\end{align}
\normalsize
\end{widetext}
The scattering signal evolves in time out of phase with the electric field oscillation of the driving field. The phase shift is determined by the phases of the spatial Fourier transform of the $\mu$th-order optically-induced charge distributions.

The amplitudes $\Bigl|  \int d^3r e^{i\bG\cdot\br}\varrho_{\mu}(\br)  \Bigr|$ can be reconstructed from the intensity of the side peaks centered at scattered energies $\omega_\s=\omega_\i+\mu\omega$. To reconstruct the phases $\alpha_\mu(\bG)$, it is necessary to know the phase of the Fourier transform of the zero-order density amplitude. In our calculations, we obtain $\alpha_0(\bG)=-0.38\pi$ at $\bG = (1,1,1)$ for GaAs. Thus, knowing $\alpha_0(\bG)$ and the time evolution of the scattering signal in the range of $\omega_\s-\omega_\i\in[0,\omega]$ in Fig.~\ref{Fig_DiffSumBraggs_111GaAs}, we obtain the phase $\alpha_1(\bG)=0.1\pi$ of the Fourier transform of the first-order optically-induced charge distribution. Repeating this procedure for the subsequent interference terms and collecting data at various $\bG$, the optically-induced charge distributions can be retrieved.

We apply the relation $\operatorname{div}\bmj_\mu(\br) = -\mu\omega\varrho_{\mu}(\br)$ shown in Sec.~\ref{SectionElCurrDen} to the Fourier transform of the density amplitudes
\begin{align}
\int d^3 r e^{i\mathbf G\cdot\mathbf r} \varrho_\mu(\br)= -\frac1{\mu\omega} \bG\cdot \int d^3 r e^{i\mathbf G\cdot\mathbf r} \bmj_\mu(\br).\label{Eq_Four_curr}
\end{align}
This relation follows from a general relation for a Fourier transform of a divergence of a vector field. This means that by reconstructing Fourier components of the optically-induced charge distributions, one also determines projections of Fourier components of the electron current densities.

\subsubsection{Crystal with inversion symmetry }
\label{Section_xray_MgO}

\begin{figure}[b]
\centering
\includegraphics[width=0.42\textwidth]{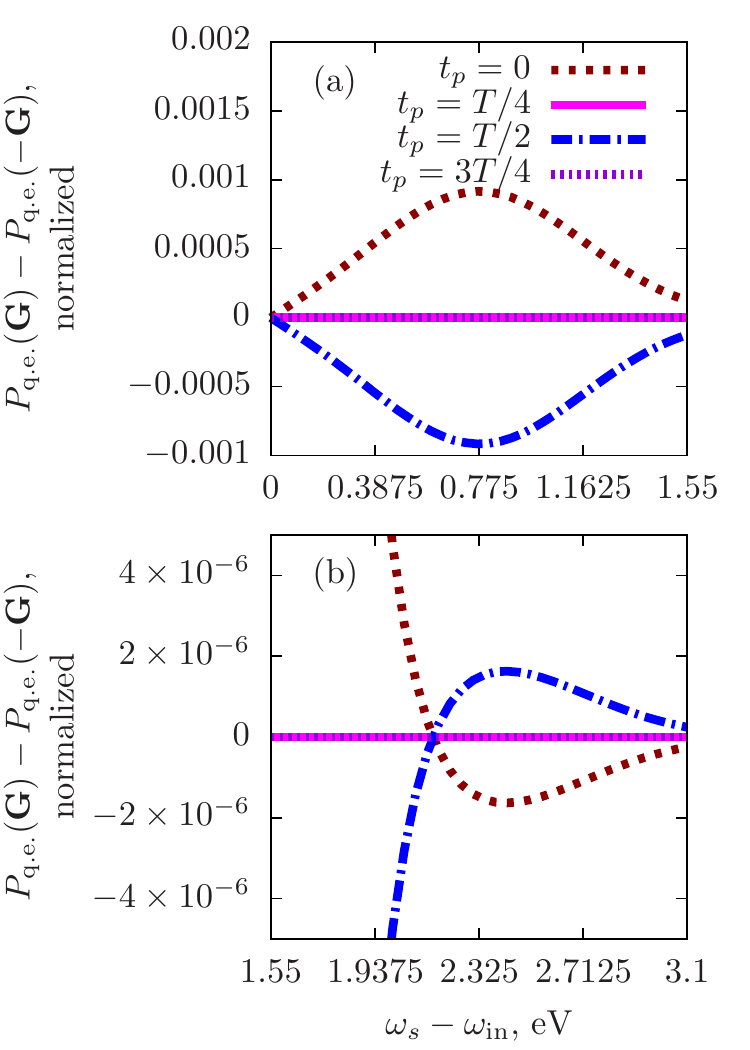}
\caption{Difference of the relative intensities of quasielastic scattering at $\mathbf G = (0,0,2)$ and $\mathbf G = (0,0,-2)$ from the laser-dressed MgO crystal at different probe-pulse arrival times as a function of $\omega_\s-\omega_\i$ (a) in the range $[0:\omega]$ and (b) in the range $[\omega:2\omega]$.}
\label{Fig_DiffBraggs_MgO}
\end{figure}

In Ref.~\onlinecite{CitepaperShort}, we show the ultrafast x-ray scattering signal from the laser-driven MgO crystal, which is a crystal with inversion symmetry. Here, we analyze it in detail. We consider the same parameters of the optical driving field as in Sec.~\ref{Seclaser-drivenMgO}, namely, an intensity of $2\times10^{12}$ W/cm$^2$ and polarization along $(0,0,1)$. The optical field has a photon energy of $\omega = 1.55$ eV, which corresponds to an optical period of $T = 2\pi/\omega = 2.67$ fs. We assume the duration of the probe nonresonant x-ray pulse is 2.0 fs, which is sufficient to resolve first-order oscillations of the electronic state of the laser-driven MgO.

We apply the results of Section \ref{Symm_x-ray-optical_inversion} to the expression of quasielastic scattering probability in \eq {PqeInterf}:
\begin{widetext}
\begin{align}
P_{\qe} (\bG,\omega_\s>\omega_\i)=\frac{P_0}4\Bigl\{&\widetilde{\mathcal E}_{0}^2\lf[\mathcal P^g_{0} (\bG)\rt]^2-2 \widetilde{\mathcal E}_{1}\widetilde{\mathcal E}_{0}\cos(\omega t_p) \mathcal P^u_{1} (\bG) \mathcal P^g_{0} (\bG)
+ \widetilde{\mathcal E}_{1}^2[\mathcal P^u_{1} (\bG)]^2\label{Eq_Pqe_MgO}\\
&-2 \widetilde{\mathcal E}_{2}\widetilde{\mathcal E}_{1}\cos(\omega t_p)  \mathcal P^g_{2} (\bG)\mathcal P^u_{1} (\bG)+\widetilde{\mathcal E}_{2}^2\lf[\mathcal P^g_{2} (\bG)\rt]^2 -2\widetilde{\mathcal E}_{3}\widetilde{\mathcal E}_{2}
\cos(\omega t_p)\mathcal P^u_{3} (\bG)  \mathcal P^g_{2} (\bG)+\cdots\Bigr\}.\nonumber
\end{align}
\end{widetext}
\normalsize
The interference term has only an antisymmetric contribution that oscillates as $\cos(\omega t_p)$ function in the case of a crystal with inversion symmetry. As we discuss in Ref.~\onlinecite{CitepaperShort}, this delay dependence of the x-ray-optical wave-mixing signal indicates that optically-induced charge distributions have inversion symmetry.

In the following, we demonstrate that x-ray-optical wave mixing from a crystal with inversion symmetry is directly connected to a microscopic electron current. Fourier components of the electron density amplitudes are connected with Fourier components of electron current density via \eq{Eq_Four_curr}. The interference between the main Bragg peak and the first-order side peak [the second term in \eq{Eq_Pqe_MgO}] is thus proportional to 
\begin{align}
\mathcal P^g_{0} (\bG)\lf[\bG\cdot \int d^3 r e^{i\mathbf G\cdot\mathbf r} \bmj_1(\br)\rt]\cos(\omega t_p).
\end{align}
Hence, the time evolution of the interference term coincides with the time evolution of the first-order oscillations of the electron current density $-\bmj_1(\br)\cos(\omega t_p)$. As long as the projection of the first-order electron current density on a direction parallel to $\bG$ is nonzero, the interference term is antisymmetric with respect to $\bG$. 





The interference term between the first- and the second-order side peaks [the fourth term in \eq{Eq_Pqe_MgO}] is proportional to
\begin{align}
\lf[\bG\cdot \int d^3 r e^{i\mathbf G\cdot\mathbf r} \bmj_1(\br)\rt]\lf[\bG\cdot \int d^3 r e^{i\mathbf G\cdot\mathbf r} \bmj_2(\br)\rt]\cos(\omega t_p).
\end{align}
The second-order electron current density does not have any distinguished direction [cf.~Ref.~\onlinecite{CitepaperShort}] and the second term is a centrosymmetric function of $\bG$ and $-\bG$. Thus, the interference term is antisymmetric again due to the first-order electron current density. Its temporal dependence also follows the oscillations of the first-order electron current density. 

The interference term between the second- and third-order side peaks [the sixth term in \eq{Eq_Pqe_MgO}] is proportional to  
\begin{align}
\lf[\bG\cdot \int d^3 r e^{i\mathbf G\cdot\mathbf r} \bmj_2(\br)\rt]\lf[\bG\cdot \int d^3 r e^{i\mathbf G\cdot\mathbf r} \bmj_3(\br)\rt]\cos(\omega t_p).
\end{align}
and it is antisymmetric due to the third-order electron current density [cf.~Fig.~\ref{MgO_Den_Curr}(a) and (c)]. The third-order electron current density oscillates as $\cos(3\omega t_p)$, which is faster than the $\cos(\omega t_p)$ oscillations of the antisymmetric term. The discrepancy between the temporal dependence is consistent with the statement that the x-ray-probe pulse of the chosen duration does not provide a sufficient temporal resolution to resolve oscillations with the frequency $3\omega$.

The time-dependent part of the signal can be easily disentangled from the total signal. It is simply the difference between quasielastic scattering signal at opposite $\bG$, $P_{\qe} (\bG)-P_{\qe} (-\bG)$. Figure \ref{Fig_DiffBraggs_MgO}(a) shows $P_{\qe} (\bG)-P_{\qe} (-\bG)$ for $\bG=(0,0,2)$ in the range of scattered energies  $\omega_\s\in[\omega_\i,\omega_\i+\omega]$ at different probe-pulse arrival times. As discussed above, this difference is given by the interference term between the main Bragg peak and the first-order side peak, which is a Gaussian-shaped function centered at $\omega_\i+\omega/2$ with an amplitude proportional to $-4\mathcal P^u_{1} (\bG) \mathcal P^g_{0} (\bG)\cos(\omega t_p)$.

$\mathcal P^g_{0} (\bG)$ is approximately the Fourier transform of the unperturbed density of MgO. Since $\mathcal P^g_{0} (\bG)>0$ and the interference term at $t_p=0$ is positive, $\mathcal P^u_{1} (\bG)$ is negative. Its amplitude can be reconstructed by measuring the intensity of the first-order side peak in a subcycle-unresolved measurement, which is proportional to $|\mathcal P^u_{1} (\bG)|^2$. Alternatively, one can determine $|\mathcal P^u_{1} (\bG)|$ by dividing the maximum intensity of the interference term by $4\mathcal P^g_{0} (\bG) e^{-(\omega\tau_p)^2/16\ln 2}$ [cf.~\eq{Eq_Gauss_middle}]. Thus, we obtain that $\mathcal P^u_{1} (\bG)/\mathcal P^g_{0} (\bG)=-1.7\times 10^{-3}$ for $\bG=(0,0,2)$.

We determined $\mathcal P^u_{1} (\bG)$ and can reconstruct $\mathcal P^g_{2} (\bG)$ from the difference $P_{\qe} (\bG)-P_{\qe} (-\bG)$ for $\bG=(0,0,2)$ in the range of scattered energies $\omega_\s\in[\omega_\i+\omega,\omega_\i+2\omega]$ shown in Fig.~\ref{Fig_DiffBraggs_MgO}(b). The difference is given by the interference term between the first- and the second-order side peaks. It should be a Gaussian-shaped function centered at $\omega_\i+3\omega/2$, but its shape is affected by the Gaussian function centered at $\omega_\i+\omega/2$. The amplitude of the peak is still proportional to $-4\mathcal P^g_{2} (\bG) \mathcal P^u_{1} (\bG)\cos(\omega t_p)$, and we obtain that $\mathcal P^g_{2} (\bG)/\mathcal P^g_{0} (\bG)=-2.0\times10^{-3}$ at $\bG = (2,0,0)$ using the same algorithm. Such a procedure to determine the Fourier transform of the density amplitudes can be repeated as long as the interference terms are detectable. The density amplitudes in real space can be reconstructed if the scattering signal at various $\bG$ is measured. 

\begin{figure}
\centering
\includegraphics[width=0.49\textwidth]{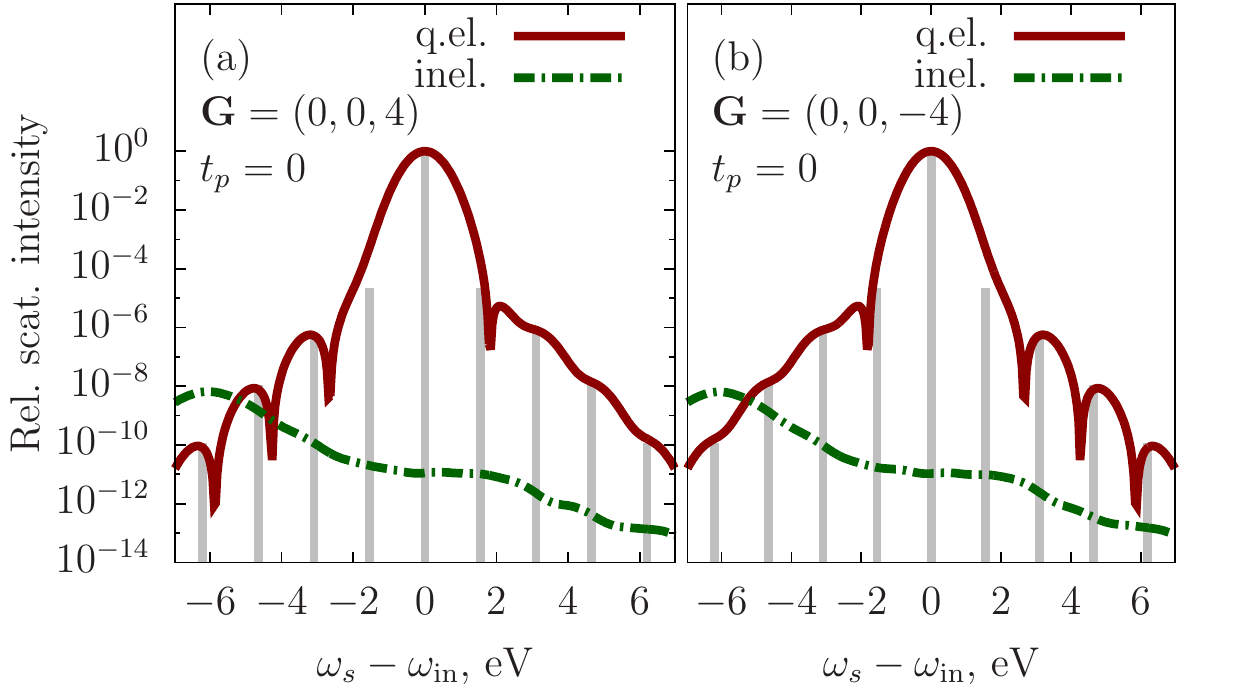}
\caption{Intensities of quasielastic and inelastic x-ray scattering signals at $\mathbf G = (0,0,4)$ and $\mathbf G = (0,0,-4)$ from a laser-dressed MgO crystal at different probe-pulse arrival times as a function of $\omega_\s-\omega_\i$. The intensities are normalized to the intensity of the main Bragg peak of MgO at $\mathbf G = (0,0,4)$. The gray vertical lines are situated at the positions of the side peaks, $\mu\omega$, and their heights correspond to their relative intensities.}
\label{Fig_Braggs_004}
\end{figure}

It may be surprising that $|\mathcal P^g_{2}(\bG)|>|\mathcal P^u_{1}(\bG)|$ at $\bG=(0,0,2)$, although the maximum amplitude of the first-order electron current density is higher than that of the second-order electron current density in Ref.~\onlinecite{CitepaperShort}.  As we discussed, $\mathcal P^{u(g)}_{\mu}(\bG)$ is proportional to $\bG\cdot \int d^3 r e^{i\mathbf G\cdot\mathbf r} \bmj_{\mu}(\br)$ [cf.~\eq{Eq_Four_curr}]. The first-order electron current density of MgO is localized around oxygen atoms. This means that its Fourier transform should be a delocalized function of $\bG$. Thus, $\mathcal P^u_{1}(\bG)$ should remain considerable at increasing $\bG$ parallel to $(0,0,1)$. The second-order electron current density is less localized in comparison to the first-order electron current density. Thus, its Fourier transform should be more localized in comparison to the Fourier transform of the first-order electron current density. $|\mathcal P^g_{2}(\bG)|$ should decrease faster with increasing $\bG$ than $|\mathcal P^u_{1}(\bG)|$ does.

Figure \ref{Fig_Braggs_004} shows the energy-resolved quasielastic and inelastic scattering signals at $\bG=(0,0,4)$ and $\bG=(0,0,-4)$ at $t_p=0$ normalized to the intensity of the main Bragg peak at $\bG=(0,0,4)$. The gray lines on the plot show the relative intensities of the $\mu$th-order side peaks that are proportional to $|\mathcal P^{u(g)}_{\mu}(\bG)|^2$. In agreement with the above considerations, the intensity of the first-order side peak remains considerable at $\bG=(0,0,4)$, whereas the intensity of the second-order side peak is strongly reduced. 

As mentioned above, when calculating the time-independent side peaks to the main Bragg peak of the laser-driven MgO in Ref.~\onlinecite{Popova-GorelovaPRB18}, we found several phenomena that we could not explain. We can now understand the behavior of the side peaks using results of Sec.~\ref{Sec_microscopic_within_FloqBloch}, and the microscopic optical response of MgO shown in Fig.~\ref{MgO_Den_Curr} and in Ref.~\onlinecite{CitepaperShort}. The first unexpected result was that the intensities of even-order side peaks  were nonzero, although even-order harmonics of MgO are zero. As we found in Sec.~\ref{Sec_MgO_GaAs}, the microscopic even-order optical response of MgO is nonzero although it results in zero macroscopic optical response. In the x-ray-optical wave-mixing experiment, x rays give access to the atomic scale and reveal the microscopic optical response.

The second surprising observation was that the intensities of the odd-order side peaks were zero at $\bG$ perpendicular to the driving-field polarization $\boldsymbol\epsilon$, whereas the intensities of the even-order side peaks did not change considerably at $\bG\perp\boldsymbol\epsilon$. Comparing the odd-order and even-order amplitudes of the electron density and of the electron current density shown in Fig.~\ref{MgO_Den_Curr} and in Ref.~\onlinecite{CitepaperShort}, this behavior becomes clear. The odd-order electron current densities are aligned along the driving-field polarization direction, such that $\bG\cdot \int d^3 r e^{i\mathbf G\cdot\mathbf r} \bmj_{\mu}(\br)$ is zero at $\bG\perp\boldsymbol\epsilon$. The even-order density amplitudes are close to a spherically symmetric distribution and their Fourier components do not strongly depend on an angle between $\bG$ and $\boldsymbol\epsilon$. Thus, just the dependence of x-ray-optical wave-mixing on the angle between the scattering direction and the driving-field polarization can reveal valuable information about microscopic optical response.


\subsubsection{Discussion}

To sum up, we have considered subcycle-resolved x-ray scattering from laser-driven crystals with a temporal resolution that is sufficient to resolve oscillations at the driving frequency $\omega$. The total scattering signal is the sum of the inelastic scattering and the quasielastic scattering signal $P_{\qe}(\bG)$. The quasielastic scattering signal is the x-ray-optical wave-mixing signal and contains information about optically-induced charge distributions and microscopic electron currents. It dominates the signal at scattered energies larger than the incoming x-ray photon energy. We found that the quasielastic scattering signal is notably noncentrosymmetric with respect to the scattering vector $\bG$ at a fixed scattered energy. It contains the antisymmetric part $(P_{\qe}(\bG)-P_{\qe}(-\bG))/2$, and the centrosymmetric part $(P_{\qe}(\bG)+P_{\qe}(-\bG))/2$. The temporal evolution of the antisymmetric part follows the temporal evolution of the first-order oscillations of the electron current density $-\bmj_{1}(\br)\cos(\omega t)$ [cf.~\eq{Eq_osc_current}]. It is directly connected to the Fourier transform of the electron current density if the crystal has inversion symmetry.  For crystals with broken inversion symmetry, the temporal dependence of the centrosymmetric part follows the first-order electron density $-\varrho_{1}(\br)\sin(\omega t)$ [cf.~\eq{Eq_osc_den}], whereas for crystals with inversion symmetry, this part is constant.


When the temporal resolution of the measurement is higher, the temporal dependence of the x-ray-optical wave mixing signal involves higher-order oscillations. The connection of oscillations of the antisymmetric and centrosymmetric part to the oscillations of the electron current density and of the electron density, respectively, also holds in this case. Higher-order oscillations of the centrosymmetric part involve only periodic functions that enter the expression for the oscillations of the electron density $\rho(\br,t_p)$, namely,  $\sin(\muodd\omega t_p)$ and $\cos(\mueven\omega t_p)$. The higher-order oscillations of  the antisymmetric part involve only periodic functions $\cos(\muodd\omega t_p)$ and $\sin(\mueven\omega t_p)$ that enter the expression for the oscillations of the electron current density $\mathbf j(\br,t_p)$. Other time- and momentum-resolved techniques for measuring freely evolving electron dynamics have a similar connection to the temporal evolution of electron density and electron current density \cite{Popova-GorelovaAppSci18}. 



The other finding of this Section is a method to reconstruct the Fourier transform of optically-induced charge distributions including its phase. We found that the x-ray-optical wave mixing signal oscillates out of phase with the electric field of the optical pulse and the phase shift depends on the spectral range and scattering vector $\bG$. The phase shift in  the spectral range $[\omega_\i+\mu\omega: \omega_\i+(\mu+1)\omega]$ is the phase difference between $\bG$ components of the Fourier transform of the $(\mu+1)$th and $\mu$th-order charge distributions. Thus, if the Fourier transform of the unperturbed density is known, phases of $\int d^3 r e^{i \bG\cdot\br}\varrho_\mu(\br)$ can be reconstructed. The amplitudes of the $\bG$ components of the Fourier transform can be reconstructed either from the time-independent part of the x-ray-optical wave mixing signal or from a subcycle-unresolved measurement.


\section{Discussion and Conclusions}

The optical response of crystals has been extensively investigated for more than a hundred years. Such studies have predominantly concentrated on the macroscopic optical response of a crystal, since it determines typical experimentally-detectable observables, such as harmonic generation. The macroscopic optical response of a crystal in most cases results from an induced dipole moment. Typically, the radiation power produced by the oscillating dipole moment dominates over the radiation power produced by higher-order moments in the optical regime \cite{Jackson}, and the macroscopic polarization in a dielectric material results from the induced dipole moment \cite{LandauElectrodynamics}. For this reason, it is customary to relate linear and nonlinear optical response to the induction of dipole moments. 

In this study, we reconsidered nonlinear optical response of band-gap crystals by focusing on its properties on the atomic scale. Our study applies to the regime of light-matter interaction that is either perturbative and describes conventional nonlinear optics experiments, or is non-perturbative, but still not sufficiently strong to considerably affect the band structure of the crystal considered. We developed a method to measure the microscopic optical response by means of ultrafast nonresonant x-ray scattering.

We found that, on the atomic scale, optically-induced charge distributions go far beyond the concept of a dipole and have a complex spatial structure. This structure has several interesting properties determined by the symmetry of the crystal. Time-reversal symmetry determines the phase of $\mu$th-order oscillations of the optically-induced charge distribution. Even-order charge distributions evolve as harmonics in phase with the vector potential of the optical field, whereas odd-order charge distributions evolve as harmonics in phase with the electric field. Spatial inversion symmetry of the crystal leads to the spatial inversion symmetry of $\mu$th-order optically-induced charge distributions. Thereby, even-order distributions are symmetric with respect to the transformation $\br\rightarrow-\br$, and odd-order distributions are antisymmetric. As a result, odd-order optically-induced charge distributions are aligned in such a way that macroscopic polarization is induced, and even-order distributions lead to a vanishing macroscopic polarization. Thus, even when macroscopic optical response is forbidden, charges still rearrange within the unit cell of the crystal. 

The microscopic optical response can be accessed by x rays with a wave length comparable to interatomic distances. Here, we developed a method to measure laser-driven electron dynamics on the atomic scale by means of ultrafast x-ray-optical wave mixing, {i.e.}~ultrafast x-ray scattering during the interaction of a crystal with an optical pulse. First, we have shown that charge flow manifests itself in a notable noncentrosymmetry of the subcycle-resolved x-ray-optical wave mixing signal with respect to the scattering vector. $\mu$th-order temporal oscillations of the anticentrosymmetric part of the signal are in phase with the $\mu$th-order oscillations of the electron current density. In the case of  a crystal with inversion symmetry, the anticentrosymmetric part of the x-ray-optical wave mixing signal at scattering vector $\bG$ is directly connected to the $\bG$ component of the Fourier transform of the electron current density.

We developed a procedure to reconstruct $\mu$th-order optically-induced charge distributions $\varrho_\mu(\br)$ from the subcycle-resolved x-ray-optical wave mixing signal. To this end, we propose to study scattering signals obtained with a temporal resolution that resolves signal oscillations with the frequency $\omega$. Such a signal comprises the Bragg peaks of the crystal, their side peaks centered at scattered energies $\omega_\i+\mu\omega$ and the interference terms between nearest-neighbor peaks. The amplitudes of the Fourier transfrom $\int d^3 r e^{i \bG\cdot\br}\varrho_\mu(\br)$ are obtained from the time-independent part of the x-ray-optical wave mixing signal at scattering vector $\bG$. The phases $\alpha_\mu(\bG)$ of the Fourier transform $\int d^3 r e^{i \bG\cdot\br}\varrho_\mu(\br) = |\int d^3 r e^{i \bG\cdot\br}\varrho_\mu(\br)|e^{i\alpha_\mu(\bG)}$ are retrieved from the phases of the temporal oscillations of the interference terms. X-ray-optical wave-mixing signals reveal even those optically-induced charge distributions that do not result in a macroscopic optical response, such as the even-order microscopic optical response of crystals with inversion symmetry.


Even though we focused our considerations to simple band-gap crystals exposed to a periodic optical excitation, we found many nontrivial properties of microscopic optical response that can be revealed with x-ray scattering. This demonstrates that x-ray-optical wave mixing techniques are powerful tools for obtaining exclusive insights into laser-driven dynamics in periodic materials. 
\section{Acknowledgment}
We acknowledge valuable discussions with David A.~Reis and Matthias Fuchs. Daria Popova-Gorelova acknowledges the funding from the Volkswagen Foundation through a Freigeist Fellowship.


\begin{thebibliography}{61}%
\makeatletter
\providecommand \@ifxundefined [1]{%
 \@ifx{#1\undefined}
}%
\providecommand \@ifnum [1]{%
 \ifnum #1\expandafter \@firstoftwo
 \else \expandafter \@secondoftwo
 \fi
}%
\providecommand \@ifx [1]{%
 \ifx #1\expandafter \@firstoftwo
 \else \expandafter \@secondoftwo
 \fi
}%
\providecommand \natexlab [1]{#1}%
\providecommand \enquote  [1]{``#1''}%
\providecommand \bibnamefont  [1]{#1}%
\providecommand \bibfnamefont [1]{#1}%
\providecommand \citenamefont [1]{#1}%
\providecommand \href@noop [0]{\@secondoftwo}%
\providecommand \href [0]{\begingroup \@sanitize@url \@href}%
\providecommand \@href[1]{\@@startlink{#1}\@@href}%
\providecommand \@@href[1]{\endgroup#1\@@endlink}%
\providecommand \@sanitize@url [0]{\catcode `\\12\catcode `\$12\catcode
  `\&12\catcode `\#12\catcode `\^12\catcode `\_12\catcode `\%12\relax}%
\providecommand \@@startlink[1]{}%
\providecommand \@@endlink[0]{}%
\providecommand \url  [0]{\begingroup\@sanitize@url \@url }%
\providecommand \@url [1]{\endgroup\@href {#1}{\urlprefix }}%
\providecommand \urlprefix  [0]{URL }%
\providecommand \Eprint [0]{\href }%
\providecommand \doibase [0]{https://doi.org/}%
\providecommand \selectlanguage [0]{\@gobble}%
\providecommand \bibinfo  [0]{\@secondoftwo}%
\providecommand \bibfield  [0]{\@secondoftwo}%
\providecommand \translation [1]{[#1]}%
\providecommand \BibitemOpen [0]{}%
\providecommand \bibitemStop [0]{}%
\providecommand \bibitemNoStop [0]{.\EOS\space}%
\providecommand \EOS [0]{\spacefactor3000\relax}%
\providecommand \BibitemShut  [1]{\csname bibitem#1\endcsname}%
\let\auto@bib@innerbib\@empty
\bibitem [{\citenamefont {Schiffrin}\ \emph {et~al.}(2012)\citenamefont
  {Schiffrin}, \citenamefont {Paasch-Colberg}, \citenamefont {Karpowicz},
  \citenamefont {Apalkov}, \citenamefont {Gerster}, \citenamefont
  {M\"{u}hlbrandt}, \citenamefont {Korbman}, \citenamefont {Reichert},
  \citenamefont {Schultze}, \citenamefont {Holzner}, \citenamefont {Barth},
  \citenamefont {Kienberger}, \citenamefont {Ernstorfer}, \citenamefont
  {Yakovlev}, \citenamefont {Stockman},\ and\ \citenamefont
  {Krausz}}]{SchiffrinNature12}%
  \BibitemOpen
  \bibfield  {author} {\bibinfo {author} {\bibfnamefont {A.}~\bibnamefont
  {Schiffrin}}, \bibinfo {author} {\bibfnamefont {T.}~\bibnamefont
  {Paasch-Colberg}}, \bibinfo {author} {\bibfnamefont {N.}~\bibnamefont
  {Karpowicz}}, \bibinfo {author} {\bibfnamefont {V.}~\bibnamefont {Apalkov}},
  \bibinfo {author} {\bibfnamefont {D.}~\bibnamefont {Gerster}}, \bibinfo
  {author} {\bibfnamefont {S.}~\bibnamefont {M\"{u}hlbrandt}}, \bibinfo
  {author} {\bibfnamefont {M.}~\bibnamefont {Korbman}}, \bibinfo {author}
  {\bibfnamefont {J.}~\bibnamefont {Reichert}}, \bibinfo {author}
  {\bibfnamefont {M.}~\bibnamefont {Schultze}}, \bibinfo {author}
  {\bibfnamefont {S.}~\bibnamefont {Holzner}}, \bibinfo {author} {\bibfnamefont
  {J.~V.}\ \bibnamefont {Barth}}, \bibinfo {author} {\bibfnamefont
  {R.}~\bibnamefont {Kienberger}}, \bibinfo {author} {\bibfnamefont
  {R.}~\bibnamefont {Ernstorfer}}, \bibinfo {author} {\bibfnamefont {V.~S.}\
  \bibnamefont {Yakovlev}}, \bibinfo {author} {\bibfnamefont {M.~I.}\
  \bibnamefont {Stockman}},\ and\ \bibinfo {author} {\bibfnamefont
  {F.}~\bibnamefont {Krausz}},\ }\bibfield  {title} {\bibinfo {title} {Optical
  field-induced current in dielectrics},\ }\href
  {https://doi.org/10.1038/nature11567} {\bibfield  {journal} {\bibinfo
  {journal} {Nature}\ }\textbf {\bibinfo {volume} {493}},\ \bibinfo {pages}
  {70} (\bibinfo {year} {2012})}\BibitemShut {NoStop}%
\bibitem [{\citenamefont {Schultze}\ \emph {et~al.}(2012)\citenamefont
  {Schultze}, \citenamefont {Bothschafter}, \citenamefont {Sommer},
  \citenamefont {Holzner}, \citenamefont {Schweinberger}, \citenamefont
  {Fiess}, \citenamefont {Hofstetter}, \citenamefont {Kienberger},
  \citenamefont {Apalkov}, \citenamefont {Yakovlev}, \citenamefont {Stockman},\
  and\ \citenamefont {Krausz}}]{SchultzeNature12}%
  \BibitemOpen
  \bibfield  {author} {\bibinfo {author} {\bibfnamefont {M.}~\bibnamefont
  {Schultze}}, \bibinfo {author} {\bibfnamefont {E.~M.}\ \bibnamefont
  {Bothschafter}}, \bibinfo {author} {\bibfnamefont {A.}~\bibnamefont
  {Sommer}}, \bibinfo {author} {\bibfnamefont {S.}~\bibnamefont {Holzner}},
  \bibinfo {author} {\bibfnamefont {W.}~\bibnamefont {Schweinberger}}, \bibinfo
  {author} {\bibfnamefont {M.}~\bibnamefont {Fiess}}, \bibinfo {author}
  {\bibfnamefont {M.}~\bibnamefont {Hofstetter}}, \bibinfo {author}
  {\bibfnamefont {R.}~\bibnamefont {Kienberger}}, \bibinfo {author}
  {\bibfnamefont {V.}~\bibnamefont {Apalkov}}, \bibinfo {author} {\bibfnamefont
  {V.~S.}\ \bibnamefont {Yakovlev}}, \bibinfo {author} {\bibfnamefont {M.~I.}\
  \bibnamefont {Stockman}},\ and\ \bibinfo {author} {\bibfnamefont
  {F.}~\bibnamefont {Krausz}},\ }\bibfield  {title} {\bibinfo {title}
  {Controlling dielectrics with the electric field of light},\ }\href
  {http://dx.doi.org/10.1038/nature11720} {\bibfield  {journal} {\bibinfo
  {journal} {Nature}\ }\textbf {\bibinfo {volume} {493}},\ \bibinfo {pages}
  {75} (\bibinfo {year} {2012})}\BibitemShut {NoStop}%
\bibitem [{\citenamefont {Chai}\ \emph {et~al.}(2018)\citenamefont {Chai},
  \citenamefont {Ropagnol}, \citenamefont {Raeis-Zadeh}, \citenamefont {Reid},
  \citenamefont {Safavi-Naeini},\ and\ \citenamefont {Ozaki}}]{ChaiPRL18}%
  \BibitemOpen
  \bibfield  {author} {\bibinfo {author} {\bibfnamefont {X.}~\bibnamefont
  {Chai}}, \bibinfo {author} {\bibfnamefont {X.}~\bibnamefont {Ropagnol}},
  \bibinfo {author} {\bibfnamefont {S.~M.}\ \bibnamefont {Raeis-Zadeh}},
  \bibinfo {author} {\bibfnamefont {M.}~\bibnamefont {Reid}}, \bibinfo {author}
  {\bibfnamefont {S.}~\bibnamefont {Safavi-Naeini}},\ and\ \bibinfo {author}
  {\bibfnamefont {T.}~\bibnamefont {Ozaki}},\ }\bibfield  {title} {\bibinfo
  {title} {Subcycle terahertz nonlinear optics},\ }\href
  {https://doi.org/10.1103/PhysRevLett.121.143901} {\bibfield  {journal}
  {\bibinfo  {journal} {Phys. Rev. Lett.}\ }\textbf {\bibinfo {volume} {121}},\
  \bibinfo {pages} {143901} (\bibinfo {year} {2018})}\BibitemShut {NoStop}%
\bibitem [{\citenamefont {Sederberg}\ \emph {et~al.}(2020)\citenamefont
  {Sederberg}, \citenamefont {Kong}, \citenamefont {Hufnagel}, \citenamefont
  {Zhang}, \citenamefont {Karimi},\ and\ \citenamefont
  {Corkum}}]{SederbergNatPhot20}%
  \BibitemOpen
  \bibfield  {author} {\bibinfo {author} {\bibfnamefont {S.}~\bibnamefont
  {Sederberg}}, \bibinfo {author} {\bibfnamefont {F.}~\bibnamefont {Kong}},
  \bibinfo {author} {\bibfnamefont {F.}~\bibnamefont {Hufnagel}}, \bibinfo
  {author} {\bibfnamefont {C.}~\bibnamefont {Zhang}}, \bibinfo {author}
  {\bibfnamefont {E.}~\bibnamefont {Karimi}},\ and\ \bibinfo {author}
  {\bibfnamefont {P.~B.}\ \bibnamefont {Corkum}},\ }\bibfield  {title}
  {\bibinfo {title} {Vectorized optoelectronic control and metrology in a
  semiconductor},\ }\bibfield  {journal} {\bibinfo  {journal} {Nature
  Photonics}\ }\href {https://doi.org/10.1038/s41566-020-0690-1}
  {10.1038/s41566-020-0690-1} (\bibinfo {year} {2020})\BibitemShut {NoStop}%
\bibitem [{\citenamefont {Kuehn}\ \emph {et~al.}(2010)\citenamefont {Kuehn},
  \citenamefont {Gaal}, \citenamefont {Reimann}, \citenamefont {Woerner},
  \citenamefont {Elsaesser},\ and\ \citenamefont {Hey}}]{KuehnPRL10}%
  \BibitemOpen
  \bibfield  {author} {\bibinfo {author} {\bibfnamefont {W.}~\bibnamefont
  {Kuehn}}, \bibinfo {author} {\bibfnamefont {P.}~\bibnamefont {Gaal}},
  \bibinfo {author} {\bibfnamefont {K.}~\bibnamefont {Reimann}}, \bibinfo
  {author} {\bibfnamefont {M.}~\bibnamefont {Woerner}}, \bibinfo {author}
  {\bibfnamefont {T.}~\bibnamefont {Elsaesser}},\ and\ \bibinfo {author}
  {\bibfnamefont {R.}~\bibnamefont {Hey}},\ }\bibfield  {title} {\bibinfo
  {title} {Coherent ballistic motion of electrons in a periodic potential},\
  }\href {https://doi.org/10.1103/PhysRevLett.104.146602} {\bibfield  {journal}
  {\bibinfo  {journal} {Phys. Rev. Lett.}\ }\textbf {\bibinfo {volume} {104}},\
  \bibinfo {pages} {146602} (\bibinfo {year} {2010})}\BibitemShut {NoStop}%
\bibitem [{\citenamefont {Schubert}\ \emph {et~al.}(2014)\citenamefont
  {Schubert}, \citenamefont {Hohenleutner}, \citenamefont {Langer},
  \citenamefont {Urbanek}, \citenamefont {Lange}, \citenamefont {Huttner},
  \citenamefont {Golde}, \citenamefont {Meier}, \citenamefont {Kira},
  \citenamefont {Koch},\ and\ \citenamefont {Huber}}]{SchubertNature14}%
  \BibitemOpen
  \bibfield  {author} {\bibinfo {author} {\bibfnamefont {O.}~\bibnamefont
  {Schubert}}, \bibinfo {author} {\bibfnamefont {M.}~\bibnamefont
  {Hohenleutner}}, \bibinfo {author} {\bibfnamefont {F.}~\bibnamefont
  {Langer}}, \bibinfo {author} {\bibfnamefont {B.}~\bibnamefont {Urbanek}},
  \bibinfo {author} {\bibfnamefont {C.}~\bibnamefont {Lange}}, \bibinfo
  {author} {\bibfnamefont {U.}~\bibnamefont {Huttner}}, \bibinfo {author}
  {\bibfnamefont {D.}~\bibnamefont {Golde}}, \bibinfo {author} {\bibfnamefont
  {T.}~\bibnamefont {Meier}}, \bibinfo {author} {\bibfnamefont
  {M.}~\bibnamefont {Kira}}, \bibinfo {author} {\bibfnamefont {S.~W.}\
  \bibnamefont {Koch}},\ and\ \bibinfo {author} {\bibfnamefont
  {R.}~\bibnamefont {Huber}},\ }\bibfield  {title} {\bibinfo {title} {Sub-cycle
  control of terahertz high-harmonic generation by dynamical bloch
  oscillations},\ }\href {https://doi.org/10.1038/nphoton.2013.349} {\bibfield
  {journal} {\bibinfo  {journal} {Nature Photonics}\ }\textbf {\bibinfo
  {volume} {8}},\ \bibinfo {pages} {119} (\bibinfo {year} {2014})}\BibitemShut
  {NoStop}%
\bibitem [{\citenamefont {Sommer}\ \emph {et~al.}(2016)\citenamefont {Sommer},
  \citenamefont {Bothschafter}, \citenamefont {Sato}, \citenamefont {Jakubeit},
  \citenamefont {Latka}, \citenamefont {Razskazovskaya}, \citenamefont
  {Fattahi}, \citenamefont {Jobst}, \citenamefont {Schweinberger},
  \citenamefont {Shirvanyan}, \citenamefont {Yakovlev}, \citenamefont
  {Kienberger}, \citenamefont {Yabana}, \citenamefont {Karpowicz},
  \citenamefont {Schultze},\ and\ \citenamefont {Krausz}}]{SommerNature16}%
  \BibitemOpen
  \bibfield  {author} {\bibinfo {author} {\bibfnamefont {A.}~\bibnamefont
  {Sommer}}, \bibinfo {author} {\bibfnamefont {E.~M.}\ \bibnamefont
  {Bothschafter}}, \bibinfo {author} {\bibfnamefont {S.~A.}\ \bibnamefont
  {Sato}}, \bibinfo {author} {\bibfnamefont {C.}~\bibnamefont {Jakubeit}},
  \bibinfo {author} {\bibfnamefont {T.}~\bibnamefont {Latka}}, \bibinfo
  {author} {\bibfnamefont {O.}~\bibnamefont {Razskazovskaya}}, \bibinfo
  {author} {\bibfnamefont {H.}~\bibnamefont {Fattahi}}, \bibinfo {author}
  {\bibfnamefont {M.}~\bibnamefont {Jobst}}, \bibinfo {author} {\bibfnamefont
  {W.}~\bibnamefont {Schweinberger}}, \bibinfo {author} {\bibfnamefont
  {V.}~\bibnamefont {Shirvanyan}}, \bibinfo {author} {\bibfnamefont {V.~S.}\
  \bibnamefont {Yakovlev}}, \bibinfo {author} {\bibfnamefont {R.}~\bibnamefont
  {Kienberger}}, \bibinfo {author} {\bibfnamefont {K.}~\bibnamefont {Yabana}},
  \bibinfo {author} {\bibfnamefont {N.}~\bibnamefont {Karpowicz}}, \bibinfo
  {author} {\bibfnamefont {M.}~\bibnamefont {Schultze}},\ and\ \bibinfo
  {author} {\bibfnamefont {F.}~\bibnamefont {Krausz}},\ }\bibfield  {title}
  {\bibinfo {title} {Attosecond nonlinear polarization and light--matter energy
  transfer in solids},\ }\href {https://doi.org/10.1038/nature17650} {\bibfield
   {journal} {\bibinfo  {journal} {Nature}\ }\textbf {\bibinfo {volume}
  {534}},\ \bibinfo {pages} {86} (\bibinfo {year} {2016})}\BibitemShut
  {NoStop}%
\bibitem [{\citenamefont {Schlaepfer}\ \emph {et~al.}(2018)\citenamefont
  {Schlaepfer}, \citenamefont {Lucchini}, \citenamefont {Sato}, \citenamefont
  {Volkov}, \citenamefont {Kasmi}, \citenamefont {Hartmann}, \citenamefont
  {Rubio}, \citenamefont {Gallmann},\ and\ \citenamefont
  {Keller}}]{SchlaepferNature18}%
  \BibitemOpen
  \bibfield  {author} {\bibinfo {author} {\bibfnamefont {F.}~\bibnamefont
  {Schlaepfer}}, \bibinfo {author} {\bibfnamefont {M.}~\bibnamefont
  {Lucchini}}, \bibinfo {author} {\bibfnamefont {S.~A.}\ \bibnamefont {Sato}},
  \bibinfo {author} {\bibfnamefont {M.}~\bibnamefont {Volkov}}, \bibinfo
  {author} {\bibfnamefont {L.}~\bibnamefont {Kasmi}}, \bibinfo {author}
  {\bibfnamefont {N.}~\bibnamefont {Hartmann}}, \bibinfo {author}
  {\bibfnamefont {A.}~\bibnamefont {Rubio}}, \bibinfo {author} {\bibfnamefont
  {L.}~\bibnamefont {Gallmann}},\ and\ \bibinfo {author} {\bibfnamefont
  {U.}~\bibnamefont {Keller}},\ }\bibfield  {title} {\bibinfo {title}
  {Attosecond optical-field-enhanced carrier injection into the gaas conduction
  band},\ }\href {https://doi.org/10.1038/s41567-018-0069-0} {\bibfield
  {journal} {\bibinfo  {journal} {Nature Physics}\ }\textbf {\bibinfo {volume}
  {14}},\ \bibinfo {pages} {560} (\bibinfo {year} {2018})}\BibitemShut
  {NoStop}%
\bibitem [{\citenamefont {Oka}\ and\ \citenamefont
  {Kitamura}(2019)}]{OkaARCMP19}%
  \BibitemOpen
  \bibfield  {author} {\bibinfo {author} {\bibfnamefont {T.}~\bibnamefont
  {Oka}}\ and\ \bibinfo {author} {\bibfnamefont {S.}~\bibnamefont {Kitamura}},\
  }\bibfield  {title} {\bibinfo {title} {Floquet engineering of quantum
  materials},\ }\href
  {https://doi.org/10.1146/annurev-conmatphys-031218-013423} {\bibfield
  {journal} {\bibinfo  {journal} {Annual Review of Condensed Matter Physics}\
  }\textbf {\bibinfo {volume} {10}},\ \bibinfo {pages} {387} (\bibinfo {year}
  {2019})}\BibitemShut {NoStop}%
\bibitem [{\citenamefont {Uzan}\ \emph {et~al.}(2020)\citenamefont {Uzan},
  \citenamefont {Orenstein}, \citenamefont {Jim{\'e}nez-Gal{\'a}n},
  \citenamefont {McDonald}, \citenamefont {Silva}, \citenamefont {Bruner},
  \citenamefont {Klimkin}, \citenamefont {Blanchet}, \citenamefont
  {Arusi-Parpar}, \citenamefont {Kr{\"u}ger}, \citenamefont {Rubtsov},
  \citenamefont {Smirnova}, \citenamefont {Ivanov}, \citenamefont {Yan},
  \citenamefont {Brabec},\ and\ \citenamefont {Dudovich}}]{UzanNatPhot20}%
  \BibitemOpen
  \bibfield  {author} {\bibinfo {author} {\bibfnamefont {A.~J.}\ \bibnamefont
  {Uzan}}, \bibinfo {author} {\bibfnamefont {G.}~\bibnamefont {Orenstein}},
  \bibinfo {author} {\bibfnamefont {{\'A}.}~\bibnamefont
  {Jim{\'e}nez-Gal{\'a}n}}, \bibinfo {author} {\bibfnamefont {C.}~\bibnamefont
  {McDonald}}, \bibinfo {author} {\bibfnamefont {R.~E.~F.}\ \bibnamefont
  {Silva}}, \bibinfo {author} {\bibfnamefont {B.~D.}\ \bibnamefont {Bruner}},
  \bibinfo {author} {\bibfnamefont {N.~D.}\ \bibnamefont {Klimkin}}, \bibinfo
  {author} {\bibfnamefont {V.}~\bibnamefont {Blanchet}}, \bibinfo {author}
  {\bibfnamefont {T.}~\bibnamefont {Arusi-Parpar}}, \bibinfo {author}
  {\bibfnamefont {M.}~\bibnamefont {Kr{\"u}ger}}, \bibinfo {author}
  {\bibfnamefont {A.~N.}\ \bibnamefont {Rubtsov}}, \bibinfo {author}
  {\bibfnamefont {O.}~\bibnamefont {Smirnova}}, \bibinfo {author}
  {\bibfnamefont {M.}~\bibnamefont {Ivanov}}, \bibinfo {author} {\bibfnamefont
  {B.}~\bibnamefont {Yan}}, \bibinfo {author} {\bibfnamefont {T.}~\bibnamefont
  {Brabec}},\ and\ \bibinfo {author} {\bibfnamefont {N.}~\bibnamefont
  {Dudovich}},\ }\bibfield  {title} {\bibinfo {title} {Attosecond spectral
  singularities in solid-state high-harmonic generation},\ }\href
  {https://doi.org/10.1038/s41566-019-0574-4} {\bibfield  {journal} {\bibinfo
  {journal} {Nature Photonics}\ }\textbf {\bibinfo {volume} {14}},\ \bibinfo
  {pages} {183} (\bibinfo {year} {2020})}\BibitemShut {NoStop}%
\bibitem [{\citenamefont {Ghimire}\ \emph {et~al.}(2010)\citenamefont
  {Ghimire}, \citenamefont {DiChiara}, \citenamefont {Sistrunk}, \citenamefont
  {Agostini}, \citenamefont {DiMauro},\ and\ \citenamefont
  {Reis}}]{GhimireNature11}%
  \BibitemOpen
  \bibfield  {author} {\bibinfo {author} {\bibfnamefont {S.}~\bibnamefont
  {Ghimire}}, \bibinfo {author} {\bibfnamefont {A.~D.}\ \bibnamefont
  {DiChiara}}, \bibinfo {author} {\bibfnamefont {E.}~\bibnamefont {Sistrunk}},
  \bibinfo {author} {\bibfnamefont {P.}~\bibnamefont {Agostini}}, \bibinfo
  {author} {\bibfnamefont {L.~F.}\ \bibnamefont {DiMauro}},\ and\ \bibinfo
  {author} {\bibfnamefont {D.~A.}\ \bibnamefont {Reis}},\ }\bibfield  {title}
  {\bibinfo {title} {Observation of high-order harmonic generation in a bulk
  crystal},\ }\href {https://doi.org/10.1038/nphys1847} {\bibfield  {journal}
  {\bibinfo  {journal} {Nature Physics}\ }\textbf {\bibinfo {volume} {7}},\
  \bibinfo {pages} {138} (\bibinfo {year} {2010})}\BibitemShut {NoStop}%
\bibitem [{\citenamefont {Schoetz}\ \emph {et~al.}(2019)\citenamefont
  {Schoetz}, \citenamefont {Wang}, \citenamefont {Pisanty}, \citenamefont
  {Lewenstein}, \citenamefont {Kling},\ and\ \citenamefont
  {Ciappina}}]{SchoetzACSPhotonics19}%
  \BibitemOpen
  \bibfield  {author} {\bibinfo {author} {\bibfnamefont {J.}~\bibnamefont
  {Schoetz}}, \bibinfo {author} {\bibfnamefont {Z.}~\bibnamefont {Wang}},
  \bibinfo {author} {\bibfnamefont {E.}~\bibnamefont {Pisanty}}, \bibinfo
  {author} {\bibfnamefont {M.}~\bibnamefont {Lewenstein}}, \bibinfo {author}
  {\bibfnamefont {M.~F.}\ \bibnamefont {Kling}},\ and\ \bibinfo {author}
  {\bibfnamefont {M.~F.}\ \bibnamefont {Ciappina}},\ }\bibfield  {title}
  {\bibinfo {title} {Perspective on petahertz electronics and attosecond
  nanoscopy},\ }\bibfield  {booktitle} {\emph {\bibinfo {booktitle} {ACS
  Photonics}},\ }\href {https://doi.org/10.1021/acsphotonics.9b01188}
  {\bibfield  {journal} {\bibinfo  {journal} {ACS Photonics}\ }\textbf
  {\bibinfo {volume} {6}},\ \bibinfo {pages} {3057} (\bibinfo {year}
  {2019})}\BibitemShut {NoStop}%
\bibitem [{\citenamefont {Kruchinin}\ \emph {et~al.}(2018)\citenamefont
  {Kruchinin}, \citenamefont {Krausz},\ and\ \citenamefont
  {Yakovlev}}]{KruchininRMP18}%
  \BibitemOpen
  \bibfield  {author} {\bibinfo {author} {\bibfnamefont {S.~Y.}\ \bibnamefont
  {Kruchinin}}, \bibinfo {author} {\bibfnamefont {F.}~\bibnamefont {Krausz}},\
  and\ \bibinfo {author} {\bibfnamefont {V.~S.}\ \bibnamefont {Yakovlev}},\
  }\bibfield  {title} {\bibinfo {title} {Colloquium: Strong-field phenomena in
  periodic systems},\ }\href {https://doi.org/10.1103/RevModPhys.90.021002}
  {\bibfield  {journal} {\bibinfo  {journal} {Rev. Mod. Phys.}\ }\textbf
  {\bibinfo {volume} {90}},\ \bibinfo {pages} {021002} (\bibinfo {year}
  {2018})}\BibitemShut {NoStop}%
\bibitem [{\citenamefont {You}\ \emph {et~al.}(2016)\citenamefont {You},
  \citenamefont {Reis},\ and\ \citenamefont {Ghimire}}]{YouNature16}%
  \BibitemOpen
  \bibfield  {author} {\bibinfo {author} {\bibfnamefont {Y.~S.}\ \bibnamefont
  {You}}, \bibinfo {author} {\bibfnamefont {D.~A.}\ \bibnamefont {Reis}},\ and\
  \bibinfo {author} {\bibfnamefont {S.}~\bibnamefont {Ghimire}},\ }\bibfield
  {title} {\bibinfo {title} {Anisotropic high-harmonic generation in bulk
  crystals},\ }\href {https://doi.org/10.1038/nphys3955} {\bibfield  {journal}
  {\bibinfo  {journal} {Nature Physics}\ }\textbf {\bibinfo {volume} {13}},\
  \bibinfo {pages} {345} (\bibinfo {year} {2016})}\BibitemShut {NoStop}%
\bibitem [{\citenamefont {Ndabashimiye}\ \emph {et~al.}(2016)\citenamefont
  {Ndabashimiye}, \citenamefont {Ghimire}, \citenamefont {Wu}, \citenamefont
  {Browne}, \citenamefont {Schafer}, \citenamefont {Gaarde},\ and\
  \citenamefont {Reis}}]{NdabashimiyeNature16}%
  \BibitemOpen
  \bibfield  {author} {\bibinfo {author} {\bibfnamefont {G.}~\bibnamefont
  {Ndabashimiye}}, \bibinfo {author} {\bibfnamefont {S.}~\bibnamefont
  {Ghimire}}, \bibinfo {author} {\bibfnamefont {M.}~\bibnamefont {Wu}},
  \bibinfo {author} {\bibfnamefont {D.~A.}\ \bibnamefont {Browne}}, \bibinfo
  {author} {\bibfnamefont {K.~J.}\ \bibnamefont {Schafer}}, \bibinfo {author}
  {\bibfnamefont {M.~B.}\ \bibnamefont {Gaarde}},\ and\ \bibinfo {author}
  {\bibfnamefont {D.~A.}\ \bibnamefont {Reis}},\ }\bibfield  {title} {\bibinfo
  {title} {Solid-state harmonics beyond the atomic limit},\ }\href
  {https://doi.org/10.1038/nature17660} {\bibfield  {journal} {\bibinfo
  {journal} {Nature}\ }\textbf {\bibinfo {volume} {534}},\ \bibinfo {pages}
  {520} (\bibinfo {year} {2016})}\BibitemShut {NoStop}%
\bibitem [{\citenamefont {Lakhotia}\ \emph {et~al.}(2020)\citenamefont
  {Lakhotia}, \citenamefont {Kim}, \citenamefont {Zhan}, \citenamefont {Hu},
  \citenamefont {Meng},\ and\ \citenamefont {Goulielmakis}}]{LakhotiaNature20}%
  \BibitemOpen
  \bibfield  {author} {\bibinfo {author} {\bibfnamefont {H.}~\bibnamefont
  {Lakhotia}}, \bibinfo {author} {\bibfnamefont {H.~Y.}\ \bibnamefont {Kim}},
  \bibinfo {author} {\bibfnamefont {M.}~\bibnamefont {Zhan}}, \bibinfo {author}
  {\bibfnamefont {S.}~\bibnamefont {Hu}}, \bibinfo {author} {\bibfnamefont
  {S.}~\bibnamefont {Meng}},\ and\ \bibinfo {author} {\bibfnamefont
  {E.}~\bibnamefont {Goulielmakis}},\ }\bibfield  {title} {\bibinfo {title}
  {Laser picoscopy of valence electrons in solids},\ }\href
  {https://doi.org/10.1038/s41586-020-2429-z} {\bibfield  {journal} {\bibinfo
  {journal} {Nature}\ }\textbf {\bibinfo {volume} {583}},\ \bibinfo {pages}
  {55} (\bibinfo {year} {2020})}\BibitemShut {NoStop}%
\bibitem [{\citenamefont {Freund}\ and\ \citenamefont
  {Levine}(1970)}]{FreundPRL70}%
  \BibitemOpen
  \bibfield  {author} {\bibinfo {author} {\bibfnamefont {I.}~\bibnamefont
  {Freund}}\ and\ \bibinfo {author} {\bibfnamefont {B.~F.}\ \bibnamefont
  {Levine}},\ }\bibfield  {title} {\bibinfo {title} {Optically modulated x-ray
  diffraction},\ }\href {https://doi.org/10.1103/PhysRevLett.25.1241}
  {\bibfield  {journal} {\bibinfo  {journal} {Phys. Rev. Lett.}\ }\textbf
  {\bibinfo {volume} {25}},\ \bibinfo {pages} {1241} (\bibinfo {year}
  {1970})}\BibitemShut {NoStop}%
\bibitem [{\citenamefont {Eisenberger}\ and\ \citenamefont
  {McCall}(1971)}]{EisenbergerPRA71}%
  \BibitemOpen
  \bibfield  {author} {\bibinfo {author} {\bibfnamefont {P.~M.}\ \bibnamefont
  {Eisenberger}}\ and\ \bibinfo {author} {\bibfnamefont {S.~L.}\ \bibnamefont
  {McCall}},\ }\bibfield  {title} {\bibinfo {title} {Mixing of x-ray and
  optical photons},\ }\href {https://doi.org/10.1103/PhysRevA.3.1145}
  {\bibfield  {journal} {\bibinfo  {journal} {Phys. Rev. A}\ }\textbf {\bibinfo
  {volume} {3}},\ \bibinfo {pages} {1145} (\bibinfo {year} {1971})}\BibitemShut
  {NoStop}%
\bibitem [{\citenamefont {Glover}\ \emph {et~al.}(2012)\citenamefont {Glover},
  \citenamefont {Fritz}, \citenamefont {Cammarata}, \citenamefont {Allison},
  \citenamefont {Coh}, \citenamefont {Feldkamp}, \citenamefont {Lemke},
  \citenamefont {Zhu}, \citenamefont {Feng}, \citenamefont {Coffee},
  \citenamefont {Fuchs}, \citenamefont {Ghimire}, \citenamefont {Chen},
  \citenamefont {Shwartz}, \citenamefont {Reis}, \citenamefont {Harris},\ and\
  \citenamefont {Hastings}}]{GloverNature12}%
  \BibitemOpen
  \bibfield  {author} {\bibinfo {author} {\bibfnamefont {T.~E.}\ \bibnamefont
  {Glover}}, \bibinfo {author} {\bibfnamefont {D.~M.}\ \bibnamefont {Fritz}},
  \bibinfo {author} {\bibfnamefont {M.}~\bibnamefont {Cammarata}}, \bibinfo
  {author} {\bibfnamefont {T.~K.}\ \bibnamefont {Allison}}, \bibinfo {author}
  {\bibfnamefont {S.}~\bibnamefont {Coh}}, \bibinfo {author} {\bibfnamefont
  {J.~M.}\ \bibnamefont {Feldkamp}}, \bibinfo {author} {\bibfnamefont
  {H.}~\bibnamefont {Lemke}}, \bibinfo {author} {\bibfnamefont
  {D.}~\bibnamefont {Zhu}}, \bibinfo {author} {\bibfnamefont {Y.}~\bibnamefont
  {Feng}}, \bibinfo {author} {\bibfnamefont {R.~N.}\ \bibnamefont {Coffee}},
  \bibinfo {author} {\bibfnamefont {M.}~\bibnamefont {Fuchs}}, \bibinfo
  {author} {\bibfnamefont {S.}~\bibnamefont {Ghimire}}, \bibinfo {author}
  {\bibfnamefont {J.}~\bibnamefont {Chen}}, \bibinfo {author} {\bibfnamefont
  {S.}~\bibnamefont {Shwartz}}, \bibinfo {author} {\bibfnamefont {D.~A.}\
  \bibnamefont {Reis}}, \bibinfo {author} {\bibfnamefont {S.~E.}\ \bibnamefont
  {Harris}},\ and\ \bibinfo {author} {\bibfnamefont {J.~B.}\ \bibnamefont
  {Hastings}},\ }\bibfield  {title} {\bibinfo {title} {X-ray and optical wave
  mixing},\ }\href {https://doi.org/10.1038/nature11340} {\bibfield  {journal}
  {\bibinfo  {journal} {Nature}\ }\textbf {\bibinfo {volume} {488}},\ \bibinfo
  {pages} {603} (\bibinfo {year} {2012})}\BibitemShut {NoStop}%
\bibitem [{\citenamefont {Schori}\ \emph {et~al.}(2017)\citenamefont {Schori},
  \citenamefont {B\"omer}, \citenamefont {Borodin}, \citenamefont {Collins},
  \citenamefont {Detlefs}, \citenamefont {Moretti~Sala}, \citenamefont
  {Yudovich},\ and\ \citenamefont {Shwartz}}]{SchoriPRL17}%
  \BibitemOpen
  \bibfield  {author} {\bibinfo {author} {\bibfnamefont {A.}~\bibnamefont
  {Schori}}, \bibinfo {author} {\bibfnamefont {C.}~\bibnamefont {B\"omer}},
  \bibinfo {author} {\bibfnamefont {D.}~\bibnamefont {Borodin}}, \bibinfo
  {author} {\bibfnamefont {S.~P.}\ \bibnamefont {Collins}}, \bibinfo {author}
  {\bibfnamefont {B.}~\bibnamefont {Detlefs}}, \bibinfo {author} {\bibfnamefont
  {M.}~\bibnamefont {Moretti~Sala}}, \bibinfo {author} {\bibfnamefont
  {S.}~\bibnamefont {Yudovich}},\ and\ \bibinfo {author} {\bibfnamefont
  {S.}~\bibnamefont {Shwartz}},\ }\bibfield  {title} {\bibinfo {title}
  {Parametric down-conversion of x rays into the optical regime},\ }\href
  {https://doi.org/10.1103/PhysRevLett.119.253902} {\bibfield  {journal}
  {\bibinfo  {journal} {Phys. Rev. Lett.}\ }\textbf {\bibinfo {volume} {119}},\
  \bibinfo {pages} {253902} (\bibinfo {year} {2017})}\BibitemShut {NoStop}%
\bibitem [{\citenamefont {Rouxel}\ \emph {et~al.}(2018)\citenamefont {Rouxel},
  \citenamefont {Kowalewski}, \citenamefont {Bennett},\ and\ \citenamefont
  {Mukamel}}]{RouxelPRL18}%
  \BibitemOpen
  \bibfield  {author} {\bibinfo {author} {\bibfnamefont {J.~R.}\ \bibnamefont
  {Rouxel}}, \bibinfo {author} {\bibfnamefont {M.}~\bibnamefont {Kowalewski}},
  \bibinfo {author} {\bibfnamefont {K.}~\bibnamefont {Bennett}},\ and\ \bibinfo
  {author} {\bibfnamefont {S.}~\bibnamefont {Mukamel}},\ }\bibfield  {title}
  {\bibinfo {title} {X-ray sum frequency diffraction for direct imaging of
  ultrafast electron dynamics},\ }\href
  {https://doi.org/10.1103/PhysRevLett.120.243902} {\bibfield  {journal}
  {\bibinfo  {journal} {Phys. Rev. Lett.}\ }\textbf {\bibinfo {volume} {120}},\
  \bibinfo {pages} {243902} (\bibinfo {year} {2018})}\BibitemShut {NoStop}%
\bibitem [{\citenamefont {Cohen}\ and\ \citenamefont
  {Shwartz}(2019)}]{CohenPRR19}%
  \BibitemOpen
  \bibfield  {author} {\bibinfo {author} {\bibfnamefont {R.}~\bibnamefont
  {Cohen}}\ and\ \bibinfo {author} {\bibfnamefont {S.}~\bibnamefont
  {Shwartz}},\ }\bibfield  {title} {\bibinfo {title} {Theory of nonlinear
  interactions between x rays and optical radiation in crystals},\ }\href
  {https://doi.org/10.1103/PhysRevResearch.1.033133} {\bibfield  {journal}
  {\bibinfo  {journal} {Phys. Rev. Research}\ }\textbf {\bibinfo {volume}
  {1}},\ \bibinfo {pages} {033133} (\bibinfo {year} {2019})}\BibitemShut
  {NoStop}%
\bibitem [{\citenamefont {Popova-Gorelova}\ \emph {et~al.}(2018)\citenamefont
  {Popova-Gorelova}, \citenamefont {Reis},\ and\ \citenamefont
  {Santra}}]{Popova-GorelovaPRB18}%
  \BibitemOpen
  \bibfield  {author} {\bibinfo {author} {\bibfnamefont {D.}~\bibnamefont
  {Popova-Gorelova}}, \bibinfo {author} {\bibfnamefont {D.~A.}\ \bibnamefont
  {Reis}},\ and\ \bibinfo {author} {\bibfnamefont {R.}~\bibnamefont {Santra}},\
  }\bibfield  {title} {\bibinfo {title} {Theory of x-ray scattering from
  laser-driven electronic systems},\ }\href
  {https://doi.org/10.1103/PhysRevB.98.224302} {\bibfield  {journal} {\bibinfo
  {journal} {Phys. Rev. B}\ }\textbf {\bibinfo {volume} {98}},\ \bibinfo
  {pages} {224302} (\bibinfo {year} {2018})}\BibitemShut {NoStop}%
\bibitem [{\citenamefont {Teichmann}\ \emph {et~al.}(2016)\citenamefont
  {Teichmann}, \citenamefont {Silva}, \citenamefont {Cousin}, \citenamefont
  {Hemmer},\ and\ \citenamefont {Biegert}}]{TeichmannNatComm16}%
  \BibitemOpen
  \bibfield  {author} {\bibinfo {author} {\bibfnamefont {S.~M.}\ \bibnamefont
  {Teichmann}}, \bibinfo {author} {\bibfnamefont {F.}~\bibnamefont {Silva}},
  \bibinfo {author} {\bibfnamefont {S.~L.}\ \bibnamefont {Cousin}}, \bibinfo
  {author} {\bibfnamefont {M.}~\bibnamefont {Hemmer}},\ and\ \bibinfo {author}
  {\bibfnamefont {J.}~\bibnamefont {Biegert}},\ }\bibfield  {title} {\bibinfo
  {title} {0.5-kev soft x-ray attosecond continua},\ }\href
  {https://doi.org/10.1038/ncomms11493} {\bibfield  {journal} {\bibinfo
  {journal} {Nature Communications}\ }\textbf {\bibinfo {volume} {7}},\
  \bibinfo {pages} {11493} (\bibinfo {year} {2016})}\BibitemShut {NoStop}%
\bibitem [{\citenamefont {Huang}\ \emph {et~al.}(2017)\citenamefont {Huang},
  \citenamefont {Ding}, \citenamefont {Feng}, \citenamefont {Hemsing},
  \citenamefont {Huang}, \citenamefont {Krzywinski}, \citenamefont {Lutman},
  \citenamefont {Marinelli}, \citenamefont {Maxwell},\ and\ \citenamefont
  {Zhu}}]{HuangPRL17}%
  \BibitemOpen
  \bibfield  {author} {\bibinfo {author} {\bibfnamefont {S.}~\bibnamefont
  {Huang}}, \bibinfo {author} {\bibfnamefont {Y.}~\bibnamefont {Ding}},
  \bibinfo {author} {\bibfnamefont {Y.}~\bibnamefont {Feng}}, \bibinfo {author}
  {\bibfnamefont {E.}~\bibnamefont {Hemsing}}, \bibinfo {author} {\bibfnamefont
  {Z.}~\bibnamefont {Huang}}, \bibinfo {author} {\bibfnamefont
  {J.}~\bibnamefont {Krzywinski}}, \bibinfo {author} {\bibfnamefont {A.~A.}\
  \bibnamefont {Lutman}}, \bibinfo {author} {\bibfnamefont {A.}~\bibnamefont
  {Marinelli}}, \bibinfo {author} {\bibfnamefont {T.~J.}\ \bibnamefont
  {Maxwell}},\ and\ \bibinfo {author} {\bibfnamefont {D.}~\bibnamefont {Zhu}},\
  }\bibfield  {title} {\bibinfo {title} {Generating single-spike hard x-ray
  pulses with nonlinear bunch compression in free-electron lasers},\ }\href
  {https://doi.org/10.1103/PhysRevLett.119.154801} {\bibfield  {journal}
  {\bibinfo  {journal} {Phys. Rev. Lett.}\ }\textbf {\bibinfo {volume} {119}},\
  \bibinfo {pages} {154801} (\bibinfo {year} {2017})}\BibitemShut {NoStop}%
\bibitem [{\citenamefont {Parc}\ \emph {et~al.}(2018)\citenamefont {Parc},
  \citenamefont {Shim},\ and\ \citenamefont {Kim}}]{ParcApplSci18}%
  \BibitemOpen
  \bibfield  {author} {\bibinfo {author} {\bibfnamefont {Y.~W.}\ \bibnamefont
  {Parc}}, \bibinfo {author} {\bibfnamefont {C.~H.}\ \bibnamefont {Shim}},\
  and\ \bibinfo {author} {\bibfnamefont {D.~E.}\ \bibnamefont {Kim}},\
  }\bibfield  {title} {\bibinfo {title} {Toward the generation of an isolated
  tw-attosecond x-ray pulse in xfel},\ }\bibfield  {journal} {\bibinfo
  {journal} {Applied Sciences}\ }\textbf {\bibinfo {volume} {8}},\ \href
  {https://doi.org/10.3390/app8091588} {10.3390/app8091588} (\bibinfo {year}
  {2018})\BibitemShut {NoStop}%
\bibitem [{\citenamefont {Li}\ \emph {et~al.}(2017)\citenamefont {Li},
  \citenamefont {Ren}, \citenamefont {Yin}, \citenamefont {Zhao}, \citenamefont
  {Chew}, \citenamefont {Cheng}, \citenamefont {Cunningham}, \citenamefont
  {Wang}, \citenamefont {Hu}, \citenamefont {Wu}, \citenamefont {Chini},\ and\
  \citenamefont {Chang}}]{LiNatComm17}%
  \BibitemOpen
  \bibfield  {author} {\bibinfo {author} {\bibfnamefont {J.}~\bibnamefont
  {Li}}, \bibinfo {author} {\bibfnamefont {X.}~\bibnamefont {Ren}}, \bibinfo
  {author} {\bibfnamefont {Y.}~\bibnamefont {Yin}}, \bibinfo {author}
  {\bibfnamefont {K.}~\bibnamefont {Zhao}}, \bibinfo {author} {\bibfnamefont
  {A.}~\bibnamefont {Chew}}, \bibinfo {author} {\bibfnamefont {Y.}~\bibnamefont
  {Cheng}}, \bibinfo {author} {\bibfnamefont {E.}~\bibnamefont {Cunningham}},
  \bibinfo {author} {\bibfnamefont {Y.}~\bibnamefont {Wang}}, \bibinfo {author}
  {\bibfnamefont {S.}~\bibnamefont {Hu}}, \bibinfo {author} {\bibfnamefont
  {Y.}~\bibnamefont {Wu}}, \bibinfo {author} {\bibfnamefont {M.}~\bibnamefont
  {Chini}},\ and\ \bibinfo {author} {\bibfnamefont {Z.}~\bibnamefont {Chang}},\
  }\bibfield  {title} {\bibinfo {title} {53-attosecond x-ray pulses reach the
  carbon k-edge},\ }\href {https://doi.org/10.1038/s41467-017-00321-0}
  {\bibfield  {journal} {\bibinfo  {journal} {Nature communications}\ }\textbf
  {\bibinfo {volume} {8}},\ \bibinfo {pages} {186} (\bibinfo {year}
  {2017})}\BibitemShut {NoStop}%
\bibitem [{\citenamefont {Kaertner}\ \emph {et~al.}(2016)\citenamefont
  {Kaertner}, \citenamefont {Ahr}, \citenamefont {Calendron}, \citenamefont
  {\c{C}ankaya}, \citenamefont {Carbajo}, \citenamefont {Chang}, \citenamefont
  {Cirmi}, \citenamefont {Doerner}, \citenamefont {Dorda}, \citenamefont
  {Fallahi}, \citenamefont {Hartin}, \citenamefont {Hemmer}, \citenamefont
  {Hobbs}, \citenamefont {Hua}, \citenamefont {Huang}, \citenamefont {Letrun},
  \citenamefont {Matlis}, \citenamefont {Mazalova}, \citenamefont {Muecke},
  \citenamefont {Nanni}, \citenamefont {Putnam}, \citenamefont {Ravi},
  \citenamefont {Reichert}, \citenamefont {Sarrou}, \citenamefont {Wu},
  \citenamefont {Yahaghi}, \citenamefont {Ye}, \citenamefont {Zapata},
  \citenamefont {Zhang}, \citenamefont {Zhou}, \citenamefont {Miller},
  \citenamefont {Berggren}, \citenamefont {Graafsma}, \citenamefont {Meents},
  \citenamefont {Assmann}, \citenamefont {Chapman},\ and\ \citenamefont
  {Fromme}}]{KaertnerNIMPRSA16}%
  \BibitemOpen
  \bibfield  {author} {\bibinfo {author} {\bibfnamefont {F.}~\bibnamefont
  {Kaertner}}, \bibinfo {author} {\bibfnamefont {F.}~\bibnamefont {Ahr}},
  \bibinfo {author} {\bibfnamefont {A.-L.}\ \bibnamefont {Calendron}}, \bibinfo
  {author} {\bibfnamefont {H.}~\bibnamefont {\c{C}ankaya}}, \bibinfo {author}
  {\bibfnamefont {S.}~\bibnamefont {Carbajo}}, \bibinfo {author} {\bibfnamefont
  {G.}~\bibnamefont {Chang}}, \bibinfo {author} {\bibfnamefont
  {G.}~\bibnamefont {Cirmi}}, \bibinfo {author} {\bibfnamefont
  {K.}~\bibnamefont {Doerner}}, \bibinfo {author} {\bibfnamefont
  {U.}~\bibnamefont {Dorda}}, \bibinfo {author} {\bibfnamefont
  {A.}~\bibnamefont {Fallahi}}, \bibinfo {author} {\bibfnamefont
  {A.}~\bibnamefont {Hartin}}, \bibinfo {author} {\bibfnamefont
  {M.}~\bibnamefont {Hemmer}}, \bibinfo {author} {\bibfnamefont
  {R.}~\bibnamefont {Hobbs}}, \bibinfo {author} {\bibfnamefont
  {Y.}~\bibnamefont {Hua}}, \bibinfo {author} {\bibfnamefont {W.~R.}\
  \bibnamefont {Huang}}, \bibinfo {author} {\bibfnamefont {R.}~\bibnamefont
  {Letrun}}, \bibinfo {author} {\bibfnamefont {N.}~\bibnamefont {Matlis}},
  \bibinfo {author} {\bibfnamefont {V.}~\bibnamefont {Mazalova}}, \bibinfo
  {author} {\bibfnamefont {O.}~\bibnamefont {Muecke}}, \bibinfo {author}
  {\bibfnamefont {E.}~\bibnamefont {Nanni}}, \bibinfo {author} {\bibfnamefont
  {W.}~\bibnamefont {Putnam}}, \bibinfo {author} {\bibfnamefont
  {K.}~\bibnamefont {Ravi}}, \bibinfo {author} {\bibfnamefont {F.}~\bibnamefont
  {Reichert}}, \bibinfo {author} {\bibfnamefont {I.}~\bibnamefont {Sarrou}},
  \bibinfo {author} {\bibfnamefont {X.}~\bibnamefont {Wu}}, \bibinfo {author}
  {\bibfnamefont {A.}~\bibnamefont {Yahaghi}}, \bibinfo {author} {\bibfnamefont
  {H.}~\bibnamefont {Ye}}, \bibinfo {author} {\bibfnamefont {L.}~\bibnamefont
  {Zapata}}, \bibinfo {author} {\bibfnamefont {D.}~\bibnamefont {Zhang}},
  \bibinfo {author} {\bibfnamefont {C.}~\bibnamefont {Zhou}}, \bibinfo {author}
  {\bibfnamefont {R.~J.~D.}\ \bibnamefont {Miller}}, \bibinfo {author}
  {\bibfnamefont {K.~K.}\ \bibnamefont {Berggren}}, \bibinfo {author}
  {\bibfnamefont {H.}~\bibnamefont {Graafsma}}, \bibinfo {author}
  {\bibfnamefont {A.}~\bibnamefont {Meents}}, \bibinfo {author} {\bibfnamefont
  {R.}~\bibnamefont {Assmann}}, \bibinfo {author} {\bibfnamefont {H.~N.}\
  \bibnamefont {Chapman}},\ and\ \bibinfo {author} {\bibfnamefont
  {P.}~\bibnamefont {Fromme}},\ }\bibfield  {title} {\bibinfo {title} {{AXSIS}:
  {E}xploring the frontiers in attosecond {X}-ray science, imaging and
  spectroscopy},\ }\href
  {https://doi.org/https://doi.org/10.1016/j.nima.2016.02.080} {\bibfield
  {journal} {\bibinfo  {journal} {Nuclear Instruments and Methods in Physics
  Research Section A: Accelerators, Spectrometers, Detectors and Associated
  Equipment}\ }\textbf {\bibinfo {volume} {829}},\ \bibinfo {pages} {24 }
  (\bibinfo {year} {2016})}\BibitemShut {NoStop}%
\bibitem [{\citenamefont {Duris}\ \emph {et~al.}(2020)\citenamefont {Duris},
  \citenamefont {Li}, \citenamefont {Driver}, \citenamefont {Champenois},
  \citenamefont {MacArthur}, \citenamefont {Lutman}, \citenamefont {Zhang},
  \citenamefont {Rosenberger}, \citenamefont {Aldrich}, \citenamefont {Coffee},
  \citenamefont {Coslovich}, \citenamefont {Decker}, \citenamefont {Glownia},
  \citenamefont {Hartmann}, \citenamefont {Helml}, \citenamefont {Kamalov},
  \citenamefont {Knurr}, \citenamefont {Krzywinski}, \citenamefont {Lin},
  \citenamefont {Marangos}, \citenamefont {Nantel}, \citenamefont {Natan},
  \citenamefont {O'Neal}, \citenamefont {Shivaram}, \citenamefont {Walter},
  \citenamefont {Wang}, \citenamefont {Welch}, \citenamefont {Wolf},
  \citenamefont {Xu}, \citenamefont {Kling}, \citenamefont {Bucksbaum},
  \citenamefont {Zholents}, \citenamefont {Huang}, \citenamefont {Cryan},\ and\
  \citenamefont {Marinelli}}]{DurisNatPhot20}%
  \BibitemOpen
  \bibfield  {author} {\bibinfo {author} {\bibfnamefont {J.}~\bibnamefont
  {Duris}}, \bibinfo {author} {\bibfnamefont {S.}~\bibnamefont {Li}}, \bibinfo
  {author} {\bibfnamefont {T.}~\bibnamefont {Driver}}, \bibinfo {author}
  {\bibfnamefont {E.~G.}\ \bibnamefont {Champenois}}, \bibinfo {author}
  {\bibfnamefont {J.~P.}\ \bibnamefont {MacArthur}}, \bibinfo {author}
  {\bibfnamefont {A.~A.}\ \bibnamefont {Lutman}}, \bibinfo {author}
  {\bibfnamefont {Z.}~\bibnamefont {Zhang}}, \bibinfo {author} {\bibfnamefont
  {P.}~\bibnamefont {Rosenberger}}, \bibinfo {author} {\bibfnamefont {J.~W.}\
  \bibnamefont {Aldrich}}, \bibinfo {author} {\bibfnamefont {R.}~\bibnamefont
  {Coffee}}, \bibinfo {author} {\bibfnamefont {G.}~\bibnamefont {Coslovich}},
  \bibinfo {author} {\bibfnamefont {F.-J.}\ \bibnamefont {Decker}}, \bibinfo
  {author} {\bibfnamefont {J.~M.}\ \bibnamefont {Glownia}}, \bibinfo {author}
  {\bibfnamefont {G.}~\bibnamefont {Hartmann}}, \bibinfo {author}
  {\bibfnamefont {W.}~\bibnamefont {Helml}}, \bibinfo {author} {\bibfnamefont
  {A.}~\bibnamefont {Kamalov}}, \bibinfo {author} {\bibfnamefont
  {J.}~\bibnamefont {Knurr}}, \bibinfo {author} {\bibfnamefont
  {J.}~\bibnamefont {Krzywinski}}, \bibinfo {author} {\bibfnamefont {M.-F.}\
  \bibnamefont {Lin}}, \bibinfo {author} {\bibfnamefont {J.~P.}\ \bibnamefont
  {Marangos}}, \bibinfo {author} {\bibfnamefont {M.}~\bibnamefont {Nantel}},
  \bibinfo {author} {\bibfnamefont {A.}~\bibnamefont {Natan}}, \bibinfo
  {author} {\bibfnamefont {J.~T.}\ \bibnamefont {O'Neal}}, \bibinfo {author}
  {\bibfnamefont {N.}~\bibnamefont {Shivaram}}, \bibinfo {author}
  {\bibfnamefont {P.}~\bibnamefont {Walter}}, \bibinfo {author} {\bibfnamefont
  {A.~L.}\ \bibnamefont {Wang}}, \bibinfo {author} {\bibfnamefont {J.~J.}\
  \bibnamefont {Welch}}, \bibinfo {author} {\bibfnamefont {T.~J.~A.}\
  \bibnamefont {Wolf}}, \bibinfo {author} {\bibfnamefont {J.~Z.}\ \bibnamefont
  {Xu}}, \bibinfo {author} {\bibfnamefont {M.~F.}\ \bibnamefont {Kling}},
  \bibinfo {author} {\bibfnamefont {P.~H.}\ \bibnamefont {Bucksbaum}}, \bibinfo
  {author} {\bibfnamefont {A.}~\bibnamefont {Zholents}}, \bibinfo {author}
  {\bibfnamefont {Z.}~\bibnamefont {Huang}}, \bibinfo {author} {\bibfnamefont
  {J.~P.}\ \bibnamefont {Cryan}},\ and\ \bibinfo {author} {\bibfnamefont
  {A.}~\bibnamefont {Marinelli}},\ }\bibfield  {title} {\bibinfo {title}
  {Tunable isolated attosecond x-ray pulses with gigawatt peak power from a
  free-electron laser},\ }\href {https://doi.org/10.1038/s41566-019-0549-5}
  {\bibfield  {journal} {\bibinfo  {journal} {Nature Photonics}\ }\textbf
  {\bibinfo {volume} {14}},\ \bibinfo {pages} {30} (\bibinfo {year}
  {2020})}\BibitemShut {NoStop}%
\bibitem [{\citenamefont {Popova-Gorelova}\ \emph {et~al.}()\citenamefont
  {Popova-Gorelova}, \citenamefont {Guskov},\ and\ \citenamefont
  {Santra}}]{CitepaperShort}%
  \BibitemOpen
  \bibfield  {author} {\bibinfo {author} {\bibfnamefont {D.}~\bibnamefont
  {Popova-Gorelova}}, \bibinfo {author} {\bibfnamefont {V.~A.}\ \bibnamefont
  {Guskov}},\ and\ \bibinfo {author} {\bibfnamefont {R.}~\bibnamefont
  {Santra}},\ }\href@noop {} {\bibinfo {title} {Microscopic electron dynamics
  in nonlinear optical response of solids}},\ \bibinfo {note} {jointly
  submitted with this manuscript}\BibitemShut {NoStop}%
\bibitem [{\citenamefont {Hsu}\ and\ \citenamefont {Reichl}(2006)}]{HsuPRB06}%
  \BibitemOpen
  \bibfield  {author} {\bibinfo {author} {\bibfnamefont {H.}~\bibnamefont
  {Hsu}}\ and\ \bibinfo {author} {\bibfnamefont {L.~E.}\ \bibnamefont
  {Reichl}},\ }\bibfield  {title} {\bibinfo {title} {Floquet-bloch states,
  quasienergy bands, and high-order harmonic generation for single-walled
  carbon nanotubes under intense laser fields},\ }\href
  {https://doi.org/10.1103/PhysRevB.74.115406} {\bibfield  {journal} {\bibinfo
  {journal} {Phys. Rev. B}\ }\textbf {\bibinfo {volume} {74}},\ \bibinfo
  {pages} {115406} (\bibinfo {year} {2006})}\BibitemShut {NoStop}%
\bibitem [{\citenamefont {Faisal}\ and\ \citenamefont
  {Kami\ifmmode~\acute{n}\else \'{n}\fi{}ski}(1997)}]{FaisalPRA97}%
  \BibitemOpen
  \bibfield  {author} {\bibinfo {author} {\bibfnamefont {F.~H.~M.}\
  \bibnamefont {Faisal}}\ and\ \bibinfo {author} {\bibfnamefont {J.~Z.}\
  \bibnamefont {Kami\ifmmode~\acute{n}\else \'{n}\fi{}ski}},\ }\bibfield
  {title} {\bibinfo {title} {Floquet-bloch theory of high-harmonic generation
  in periodic structures},\ }\href {https://doi.org/10.1103/PhysRevA.56.748}
  {\bibfield  {journal} {\bibinfo  {journal} {Phys. Rev. A}\ }\textbf {\bibinfo
  {volume} {56}},\ \bibinfo {pages} {748} (\bibinfo {year} {1997})}\BibitemShut
  {NoStop}%
\bibitem [{\citenamefont {Tzoar}\ and\ \citenamefont
  {Gersten}(1975)}]{TzoarPRB75}%
  \BibitemOpen
  \bibfield  {author} {\bibinfo {author} {\bibfnamefont {N.}~\bibnamefont
  {Tzoar}}\ and\ \bibinfo {author} {\bibfnamefont {J.~I.}\ \bibnamefont
  {Gersten}},\ }\bibfield  {title} {\bibinfo {title} {Theory of electronic band
  structure in intense laser fields},\ }\href
  {https://doi.org/10.1103/PhysRevB.12.1132} {\bibfield  {journal} {\bibinfo
  {journal} {Phys. Rev. B}\ }\textbf {\bibinfo {volume} {12}},\ \bibinfo
  {pages} {1132} (\bibinfo {year} {1975})}\BibitemShut {NoStop}%
\bibitem [{\citenamefont {Higuchi}\ \emph {et~al.}(2014)\citenamefont
  {Higuchi}, \citenamefont {Stockman},\ and\ \citenamefont
  {Hommelhoff}}]{HiguchiPRL14}%
  \BibitemOpen
  \bibfield  {author} {\bibinfo {author} {\bibfnamefont {T.}~\bibnamefont
  {Higuchi}}, \bibinfo {author} {\bibfnamefont {M.~I.}\ \bibnamefont
  {Stockman}},\ and\ \bibinfo {author} {\bibfnamefont {P.}~\bibnamefont
  {Hommelhoff}},\ }\bibfield  {title} {\bibinfo {title} {Strong-field
  perspective on high-harmonic radiation from bulk solids},\ }\href
  {https://doi.org/10.1103/PhysRevLett.113.213901} {\bibfield  {journal}
  {\bibinfo  {journal} {Phys. Rev. Lett.}\ }\textbf {\bibinfo {volume} {113}},\
  \bibinfo {pages} {213901} (\bibinfo {year} {2014})}\BibitemShut {NoStop}%
\bibitem [{\citenamefont {Moiseyev}(2015)}]{MoiseyevPRA15}%
  \BibitemOpen
  \bibfield  {author} {\bibinfo {author} {\bibfnamefont {N.}~\bibnamefont
  {Moiseyev}},\ }\bibfield  {title} {\bibinfo {title} {Selection rules for
  harmonic generation in solids},\ }\href
  {https://doi.org/10.1103/PhysRevA.91.053811} {\bibfield  {journal} {\bibinfo
  {journal} {Phys. Rev. A}\ }\textbf {\bibinfo {volume} {91}},\ \bibinfo
  {pages} {053811} (\bibinfo {year} {2015})}\BibitemShut {NoStop}%
\bibitem [{\citenamefont {Ikeda}\ \emph {et~al.}(2018)\citenamefont {Ikeda},
  \citenamefont {Chinzei},\ and\ \citenamefont {Tsunetsugu}}]{IkedaPRA18}%
  \BibitemOpen
  \bibfield  {author} {\bibinfo {author} {\bibfnamefont {T.~N.}\ \bibnamefont
  {Ikeda}}, \bibinfo {author} {\bibfnamefont {K.}~\bibnamefont {Chinzei}},\
  and\ \bibinfo {author} {\bibfnamefont {H.}~\bibnamefont {Tsunetsugu}},\
  }\bibfield  {title} {\bibinfo {title} {Floquet-theoretical formulation and
  analysis of high-order harmonic generation in solids},\ }\href
  {https://doi.org/10.1103/PhysRevA.98.063426} {\bibfield  {journal} {\bibinfo
  {journal} {Phys. Rev. A}\ }\textbf {\bibinfo {volume} {98}},\ \bibinfo
  {pages} {063426} (\bibinfo {year} {2018})}\BibitemShut {NoStop}%
\bibitem [{\citenamefont {Mandel}\ and\ \citenamefont {Wolf}(1995)}]{Mandel}%
  \BibitemOpen
  \bibfield  {author} {\bibinfo {author} {\bibfnamefont {L.}~\bibnamefont
  {Mandel}}\ and\ \bibinfo {author} {\bibfnamefont {E.}~\bibnamefont {Wolf}},\
  }\href@noop {} {\emph {\bibinfo {title} {Optical Coherence and Quantum
  Optics}}}\ (\bibinfo  {publisher} {Cambridge University Press},\ \bibinfo
  {address} {Cambridge},\ \bibinfo {year} {1995})\BibitemShut {NoStop}%
\bibitem [{\citenamefont {Dixit}\ \emph {et~al.}(2012)\citenamefont {Dixit},
  \citenamefont {Vendrell},\ and\ \citenamefont {Santra}}]{DixitPNAS12}%
  \BibitemOpen
  \bibfield  {author} {\bibinfo {author} {\bibfnamefont {G.}~\bibnamefont
  {Dixit}}, \bibinfo {author} {\bibfnamefont {O.}~\bibnamefont {Vendrell}},\
  and\ \bibinfo {author} {\bibfnamefont {R.}~\bibnamefont {Santra}},\
  }\bibfield  {title} {\bibinfo {title} {Imaging electronic quantum motion with
  light},\ }\href {https://doi.org/10.1073/pnas.1202226109} {\bibfield
  {journal} {\bibinfo  {journal} {Proceedings of the National Academy of
  Sciences}\ }\textbf {\bibinfo {volume} {109}},\ \bibinfo {pages} {11636}
  (\bibinfo {year} {2012})}\BibitemShut {NoStop}%
\bibitem [{\citenamefont {Popova-Gorelova}(2018)}]{Popova-GorelovaAppSci18}%
  \BibitemOpen
  \bibfield  {author} {\bibinfo {author} {\bibfnamefont {D.}~\bibnamefont
  {Popova-Gorelova}},\ }\bibfield  {title} {\bibinfo {title} {Imaging electron
  dynamics with ultrashort light pulses: A theory perspective},\ }\href
  {https://doi.org/10.3390/app8030318} {\bibfield  {journal} {\bibinfo
  {journal} {Applied Sciences}\ }\textbf {\bibinfo {volume} {8}},\ \bibinfo
  {pages} {318} (\bibinfo {year} {2018})}\BibitemShut {NoStop}%
\bibitem [{\citenamefont {Ben-Tal}\ \emph {et~al.}(1993)\citenamefont
  {Ben-Tal}, \citenamefont {Moiseyev},\ and\ \citenamefont
  {Beswick}}]{Ben-TalJPhB93}%
  \BibitemOpen
  \bibfield  {author} {\bibinfo {author} {\bibfnamefont {N.}~\bibnamefont
  {Ben-Tal}}, \bibinfo {author} {\bibfnamefont {N.}~\bibnamefont {Moiseyev}},\
  and\ \bibinfo {author} {\bibfnamefont {A.}~\bibnamefont {Beswick}},\
  }\bibfield  {title} {\bibinfo {title} {The effect of hamiltonian symmetry on
  generation of odd and even harmonics},\ }\href
  {https://doi.org/10.1088/0953-4075/26/18/012} {\bibfield  {journal} {\bibinfo
   {journal} {Journal of Physics B: Atomic, Molecular and Optical Physics}\
  }\textbf {\bibinfo {volume} {26}},\ \bibinfo {pages} {3017} (\bibinfo {year}
  {1993})}\BibitemShut {NoStop}%
\bibitem [{\citenamefont {Fleischer}\ and\ \citenamefont
  {Moiseyev}(2005)}]{FleischerPRA05}%
  \BibitemOpen
  \bibfield  {author} {\bibinfo {author} {\bibfnamefont {A.}~\bibnamefont
  {Fleischer}}\ and\ \bibinfo {author} {\bibfnamefont {N.}~\bibnamefont
  {Moiseyev}},\ }\bibfield  {title} {\bibinfo {title} {Adiabatic theorem for
  non-hermitian time-dependent open systems},\ }\href
  {https://doi.org/10.1103/PhysRevA.72.032103} {\bibfield  {journal} {\bibinfo
  {journal} {Phys. Rev. A}\ }\textbf {\bibinfo {volume} {72}},\ \bibinfo
  {pages} {032103} (\bibinfo {year} {2005})}\BibitemShut {NoStop}%
\bibitem [{\citenamefont {Breuer}\ and\ \citenamefont
  {Holthaus}(1989)}]{BreuerZphD89}%
  \BibitemOpen
  \bibfield  {author} {\bibinfo {author} {\bibfnamefont {H.~P.}\ \bibnamefont
  {Breuer}}\ and\ \bibinfo {author} {\bibfnamefont {M.}~\bibnamefont
  {Holthaus}},\ }\bibfield  {title} {\bibinfo {title} {Adiabatic processes in
  the ionization of highly excited hydrogen atoms},\ }\href
  {https://doi.org/10.1007/BF01436579} {\bibfield  {journal} {\bibinfo
  {journal} {Zeitschrift f{\"u}r Physik D Atoms, Molecules and Clusters}\
  }\textbf {\bibinfo {volume} {11}},\ \bibinfo {pages} {1} (\bibinfo {year}
  {1989})}\BibitemShut {NoStop}%
\bibitem [{\citenamefont {Luu}\ \emph {et~al.}(2015)\citenamefont {Luu},
  \citenamefont {Garg}, \citenamefont {Kruchinin}, \citenamefont {Moulet},
  \citenamefont {Hassan},\ and\ \citenamefont {Goulielmakis}}]{LuuNature15}%
  \BibitemOpen
  \bibfield  {author} {\bibinfo {author} {\bibfnamefont {T.~T.}\ \bibnamefont
  {Luu}}, \bibinfo {author} {\bibfnamefont {M.}~\bibnamefont {Garg}}, \bibinfo
  {author} {\bibfnamefont {S.~Y.}\ \bibnamefont {Kruchinin}}, \bibinfo {author}
  {\bibfnamefont {A.}~\bibnamefont {Moulet}}, \bibinfo {author} {\bibfnamefont
  {M.~T.}\ \bibnamefont {Hassan}},\ and\ \bibinfo {author} {\bibfnamefont
  {E.}~\bibnamefont {Goulielmakis}},\ }\bibfield  {title} {\bibinfo {title}
  {Extreme ultraviolet high-harmonic spectroscopy of solids},\ }\href
  {https://doi.org/10.1038/nature14456} {\bibfield  {journal} {\bibinfo
  {journal} {Nature}\ }\textbf {\bibinfo {volume} {521}},\ \bibinfo {pages}
  {498} (\bibinfo {year} {2015})}\BibitemShut {NoStop}%
\bibitem [{\citenamefont {Santra}\ and\ \citenamefont
  {Greene}(2004)}]{SantraPRA04}%
  \BibitemOpen
  \bibfield  {author} {\bibinfo {author} {\bibfnamefont {R.}~\bibnamefont
  {Santra}}\ and\ \bibinfo {author} {\bibfnamefont {C.~H.}\ \bibnamefont
  {Greene}},\ }\bibfield  {title} {\bibinfo {title} {Multiphoton ionization of
  xenon in the vuv regime},\ }\href
  {https://doi.org/10.1103/PhysRevA.70.053401} {\bibfield  {journal} {\bibinfo
  {journal} {Phys. Rev. A}\ }\textbf {\bibinfo {volume} {70}},\ \bibinfo
  {pages} {053401} (\bibinfo {year} {2004})}\BibitemShut {NoStop}%
\bibitem [{\citenamefont {Shirley}(1965)}]{ShirleyPR65}%
  \BibitemOpen
  \bibfield  {author} {\bibinfo {author} {\bibfnamefont {J.~H.}\ \bibnamefont
  {Shirley}},\ }\bibfield  {title} {\bibinfo {title} {Solution of the
  schr\"odinger equation with a hamiltonian periodic in time},\ }\href
  {https://doi.org/10.1103/PhysRev.138.B979} {\bibfield  {journal} {\bibinfo
  {journal} {Phys. Rev.}\ }\textbf {\bibinfo {volume} {138}},\ \bibinfo {pages}
  {B979} (\bibinfo {year} {1965})}\BibitemShut {NoStop}%
\bibitem [{\citenamefont {Kittel}(2004)}]{Kittel}%
  \BibitemOpen
  \bibfield  {author} {\bibinfo {author} {\bibfnamefont {C.}~\bibnamefont
  {Kittel}},\ }\href@noop {} {\emph {\bibinfo {title} {Introduction to Solid
  State Physics}}}\ (\bibinfo  {publisher} {Wiley},\ \bibinfo {address} {New
  York City},\ \bibinfo {year} {2004})\BibitemShut {NoStop}%
\bibitem [{\citenamefont {Shen}(2003)}]{ShenBook}%
  \BibitemOpen
  \bibfield  {author} {\bibinfo {author} {\bibfnamefont {Y.}~\bibnamefont
  {Shen}},\ }\href {http://books.google.de/books?id=xeg9AQAAIAAJ} {\emph
  {\bibinfo {title} {The principles of nonlinear optics}}},\ Wiley classics
  library\ (\bibinfo  {publisher} {Wiley-Interscience},\ \bibinfo {year}
  {2003})\BibitemShut {NoStop}%
\bibitem [{\citenamefont {Bloembergen}\ and\ \citenamefont
  {Shen}(1964)}]{BloembergenPhRev64}%
  \BibitemOpen
  \bibfield  {author} {\bibinfo {author} {\bibfnamefont {N.}~\bibnamefont
  {Bloembergen}}\ and\ \bibinfo {author} {\bibfnamefont {Y.~R.}\ \bibnamefont
  {Shen}},\ }\bibfield  {title} {\bibinfo {title} {Quantum-theoretical
  comparison of nonlinear susceptibilities in parametric media, lasers, and
  raman lasers},\ }\href {https://doi.org/10.1103/PhysRev.133.A37} {\bibfield
  {journal} {\bibinfo  {journal} {Phys. Rev.}\ }\textbf {\bibinfo {volume}
  {133}},\ \bibinfo {pages} {A37} (\bibinfo {year} {1964})}\BibitemShut
  {NoStop}%
\bibitem [{\citenamefont {Keller}(2012)}]{KellerBook}%
  \BibitemOpen
  \bibfield  {author} {\bibinfo {author} {\bibfnamefont {O.}~\bibnamefont
  {Keller}},\ }\href {https://books.google.de/books?id=v2ck\_\_wFOBEC} {\emph
  {\bibinfo {title} {Quantum Theory of Near-Field Electrodynamics}}},\
  Nano-Optics and Nanophotonics\ (\bibinfo  {publisher} {Springer Berlin
  Heidelberg},\ \bibinfo {address} {Berlin},\ \bibinfo {year}
  {2012})\BibitemShut {NoStop}%
\bibitem [{\citenamefont {Ernotte}\ \emph {et~al.}(2018)\citenamefont
  {Ernotte}, \citenamefont {Hammond},\ and\ \citenamefont
  {Taucer}}]{ErnottePRB18}%
  \BibitemOpen
  \bibfield  {author} {\bibinfo {author} {\bibfnamefont {G.}~\bibnamefont
  {Ernotte}}, \bibinfo {author} {\bibfnamefont {T.~J.}\ \bibnamefont
  {Hammond}},\ and\ \bibinfo {author} {\bibfnamefont {M.}~\bibnamefont
  {Taucer}},\ }\bibfield  {title} {\bibinfo {title} {A gauge-invariant
  formulation of interband and intraband currents in solids},\ }\href
  {https://doi.org/10.1103/PhysRevB.98.235202} {\bibfield  {journal} {\bibinfo
  {journal} {Phys. Rev. B}\ }\textbf {\bibinfo {volume} {98}},\ \bibinfo
  {pages} {235202} (\bibinfo {year} {2018})}\BibitemShut {NoStop}%
\bibitem [{\citenamefont {Yariv}\ and\ \citenamefont {Yeh}(1984)}]{YarivBook}%
  \BibitemOpen
  \bibfield  {author} {\bibinfo {author} {\bibfnamefont {A.}~\bibnamefont
  {Yariv}}\ and\ \bibinfo {author} {\bibfnamefont {P.}~\bibnamefont {Yeh}},\
  }\href {https://books.google.de/books?id=jjzxAAAAMAAJ} {\emph {\bibinfo
  {title} {Optical Waves in Crystals: Propagation and Control of Laser
  Radiation}}},\ A Wiley interscience publication\ (\bibinfo  {publisher}
  {Wiley},\ \bibinfo {address} {Hoboken},\ \bibinfo {year} {1984})\BibitemShut
  {NoStop}%
\bibitem [{\citenamefont {Gonze}\ \emph {et~al.}(2016)\citenamefont {Gonze},
  \citenamefont {Jollet}, \citenamefont {Abreu~Araujo}, \citenamefont {Adams},
  \citenamefont {Amadon}, \citenamefont {Applencourt}, \citenamefont {Audouze},
  \citenamefont {Beuken}, \citenamefont {Bieder}, \citenamefont {Bokhanchuk},
  \citenamefont {Bousquet}, \citenamefont {Bruneval}, \citenamefont {Caliste},
  \citenamefont {C{\^o}t{\'e}}, \citenamefont {Dahm}, \citenamefont {Da~Pieve},
  \citenamefont {Delaveau}, \citenamefont {Di~Gennaro}, \citenamefont {Dorado},
  \citenamefont {Espejo}, \citenamefont {Geneste}, \citenamefont {Genovese},
  \citenamefont {Gerossier}, \citenamefont {Giantomassi}, \citenamefont
  {Gillet}, \citenamefont {Hamann}, \citenamefont {He}, \citenamefont {Jomard},
  \citenamefont {Laflamme~Janssen}, \citenamefont {Le~Roux}, \citenamefont
  {Levitt}, \citenamefont {Lherbier}, \citenamefont {Liu}, \citenamefont
  {Luka{\v c}evi{\'c}}, \citenamefont {Martin}, \citenamefont {Martins},
  \citenamefont {Oliveira}, \citenamefont {Ponc{\'e}}, \citenamefont
  {Pouillon}, \citenamefont {Rangel}, \citenamefont {Rignanese}, \citenamefont
  {Romero}, \citenamefont {Rousseau}, \citenamefont {Rubel}, \citenamefont
  {Shukri}, \citenamefont {Stankovski}, \citenamefont {Torrent}, \citenamefont
  {Van~Setten}, \citenamefont {Van~Troeye}, \citenamefont {Verstraete},
  \citenamefont {Waroquiers}, \citenamefont {Wiktor}, \citenamefont {Xu},
  \citenamefont {Zhou},\ and\ \citenamefont {Zwanziger}}]{Gonze16}%
  \BibitemOpen
  \bibfield  {author} {\bibinfo {author} {\bibfnamefont {X.}~\bibnamefont
  {Gonze}}, \bibinfo {author} {\bibfnamefont {F.}~\bibnamefont {Jollet}},
  \bibinfo {author} {\bibfnamefont {F.}~\bibnamefont {Abreu~Araujo}}, \bibinfo
  {author} {\bibfnamefont {D.}~\bibnamefont {Adams}}, \bibinfo {author}
  {\bibfnamefont {B.}~\bibnamefont {Amadon}}, \bibinfo {author} {\bibfnamefont
  {T.}~\bibnamefont {Applencourt}}, \bibinfo {author} {\bibfnamefont
  {C.}~\bibnamefont {Audouze}}, \bibinfo {author} {\bibfnamefont {J.~M.}\
  \bibnamefont {Beuken}}, \bibinfo {author} {\bibfnamefont {J.}~\bibnamefont
  {Bieder}}, \bibinfo {author} {\bibfnamefont {A.}~\bibnamefont {Bokhanchuk}},
  \bibinfo {author} {\bibfnamefont {E.}~\bibnamefont {Bousquet}}, \bibinfo
  {author} {\bibfnamefont {F.}~\bibnamefont {Bruneval}}, \bibinfo {author}
  {\bibfnamefont {D.}~\bibnamefont {Caliste}}, \bibinfo {author} {\bibfnamefont
  {M.}~\bibnamefont {C{\^o}t{\'e}}}, \bibinfo {author} {\bibfnamefont
  {F.}~\bibnamefont {Dahm}}, \bibinfo {author} {\bibfnamefont {F.}~\bibnamefont
  {Da~Pieve}}, \bibinfo {author} {\bibfnamefont {M.}~\bibnamefont {Delaveau}},
  \bibinfo {author} {\bibfnamefont {M.}~\bibnamefont {Di~Gennaro}}, \bibinfo
  {author} {\bibfnamefont {B.}~\bibnamefont {Dorado}}, \bibinfo {author}
  {\bibfnamefont {C.}~\bibnamefont {Espejo}}, \bibinfo {author} {\bibfnamefont
  {G.}~\bibnamefont {Geneste}}, \bibinfo {author} {\bibfnamefont
  {L.}~\bibnamefont {Genovese}}, \bibinfo {author} {\bibfnamefont
  {A.}~\bibnamefont {Gerossier}}, \bibinfo {author} {\bibfnamefont
  {M.}~\bibnamefont {Giantomassi}}, \bibinfo {author} {\bibfnamefont
  {Y.}~\bibnamefont {Gillet}}, \bibinfo {author} {\bibfnamefont {D.~R.}\
  \bibnamefont {Hamann}}, \bibinfo {author} {\bibfnamefont {L.}~\bibnamefont
  {He}}, \bibinfo {author} {\bibfnamefont {G.}~\bibnamefont {Jomard}}, \bibinfo
  {author} {\bibfnamefont {J.}~\bibnamefont {Laflamme~Janssen}}, \bibinfo
  {author} {\bibfnamefont {S.}~\bibnamefont {Le~Roux}}, \bibinfo {author}
  {\bibfnamefont {A.}~\bibnamefont {Levitt}}, \bibinfo {author} {\bibfnamefont
  {A.}~\bibnamefont {Lherbier}}, \bibinfo {author} {\bibfnamefont
  {F.}~\bibnamefont {Liu}}, \bibinfo {author} {\bibfnamefont {I.}~\bibnamefont
  {Luka{\v c}evi{\'c}}}, \bibinfo {author} {\bibfnamefont {A.}~\bibnamefont
  {Martin}}, \bibinfo {author} {\bibfnamefont {C.}~\bibnamefont {Martins}},
  \bibinfo {author} {\bibfnamefont {M.~J.~T.}\ \bibnamefont {Oliveira}},
  \bibinfo {author} {\bibfnamefont {S.}~\bibnamefont {Ponc{\'e}}}, \bibinfo
  {author} {\bibfnamefont {Y.}~\bibnamefont {Pouillon}}, \bibinfo {author}
  {\bibfnamefont {T.}~\bibnamefont {Rangel}}, \bibinfo {author} {\bibfnamefont
  {G.~M.}\ \bibnamefont {Rignanese}}, \bibinfo {author} {\bibfnamefont {A.~H.}\
  \bibnamefont {Romero}}, \bibinfo {author} {\bibfnamefont {B.}~\bibnamefont
  {Rousseau}}, \bibinfo {author} {\bibfnamefont {O.}~\bibnamefont {Rubel}},
  \bibinfo {author} {\bibfnamefont {A.~A.}\ \bibnamefont {Shukri}}, \bibinfo
  {author} {\bibfnamefont {M.}~\bibnamefont {Stankovski}}, \bibinfo {author}
  {\bibfnamefont {M.}~\bibnamefont {Torrent}}, \bibinfo {author} {\bibfnamefont
  {M.~J.}\ \bibnamefont {Van~Setten}}, \bibinfo {author} {\bibfnamefont
  {B.}~\bibnamefont {Van~Troeye}}, \bibinfo {author} {\bibfnamefont {M.~J.}\
  \bibnamefont {Verstraete}}, \bibinfo {author} {\bibfnamefont
  {D.}~\bibnamefont {Waroquiers}}, \bibinfo {author} {\bibfnamefont
  {J.}~\bibnamefont {Wiktor}}, \bibinfo {author} {\bibfnamefont
  {B.}~\bibnamefont {Xu}}, \bibinfo {author} {\bibfnamefont {A.}~\bibnamefont
  {Zhou}},\ and\ \bibinfo {author} {\bibfnamefont {J.~W.}\ \bibnamefont
  {Zwanziger}},\ }\bibfield  {title} {\bibinfo {title} {Recent developments in
  the abinit software package},\ }\href
  {https://doi.org/http://dx.doi.org/10.1016/j.cpc.2016.04.003} {\bibfield
  {journal} {\bibinfo  {journal} {Computer Physics Communications}\ }\textbf
  {\bibinfo {volume} {205}},\ \bibinfo {pages} {106 } (\bibinfo {year}
  {2016})}\BibitemShut {NoStop}%
\bibitem [{\citenamefont {Gonze}\ \emph {et~al.}(2009)\citenamefont {Gonze},
  \citenamefont {Amadon}, \citenamefont {Anglade}, \citenamefont {Beuken},
  \citenamefont {Bottin}, \citenamefont {Boulanger}, \citenamefont {Bruneval},
  \citenamefont {Caliste}, \citenamefont {Caracas}, \citenamefont {C\'ot\'e},
  \citenamefont {Deutsch}, \citenamefont {Genovese}, \citenamefont {Ghosez},
  \citenamefont {Giantomassi}, \citenamefont {Goedecker}, \citenamefont
  {Hamann}, \citenamefont {Hermet}, \citenamefont {Jollet}, \citenamefont
  {Jomard}, \citenamefont {Leroux}, \citenamefont {Mancini}, \citenamefont
  {Mazevet}, \citenamefont {Oliveira}, \citenamefont {Onida}, \citenamefont
  {Pouillon}, \citenamefont {Rangel}, \citenamefont {Rignanese}, \citenamefont
  {Sangalli}, \citenamefont {Shaltaf}, \citenamefont {Torrent}, \citenamefont
  {Verstraete}, \citenamefont {Zerah},\ and\ \citenamefont
  {Zwanziger}}]{Gonze09}%
  \BibitemOpen
  \bibfield  {author} {\bibinfo {author} {\bibfnamefont {X.}~\bibnamefont
  {Gonze}}, \bibinfo {author} {\bibfnamefont {B.}~\bibnamefont {Amadon}},
  \bibinfo {author} {\bibfnamefont {P.-M.}\ \bibnamefont {Anglade}}, \bibinfo
  {author} {\bibfnamefont {J.-M.}\ \bibnamefont {Beuken}}, \bibinfo {author}
  {\bibfnamefont {F.}~\bibnamefont {Bottin}}, \bibinfo {author} {\bibfnamefont
  {P.}~\bibnamefont {Boulanger}}, \bibinfo {author} {\bibfnamefont
  {F.}~\bibnamefont {Bruneval}}, \bibinfo {author} {\bibfnamefont
  {D.}~\bibnamefont {Caliste}}, \bibinfo {author} {\bibfnamefont
  {R.}~\bibnamefont {Caracas}}, \bibinfo {author} {\bibfnamefont
  {M.}~\bibnamefont {C\'ot\'e}}, \bibinfo {author} {\bibfnamefont
  {T.}~\bibnamefont {Deutsch}}, \bibinfo {author} {\bibfnamefont
  {L.}~\bibnamefont {Genovese}}, \bibinfo {author} {\bibfnamefont
  {P.}~\bibnamefont {Ghosez}}, \bibinfo {author} {\bibfnamefont
  {M.}~\bibnamefont {Giantomassi}}, \bibinfo {author} {\bibfnamefont
  {S.}~\bibnamefont {Goedecker}}, \bibinfo {author} {\bibfnamefont
  {D.}~\bibnamefont {Hamann}}, \bibinfo {author} {\bibfnamefont
  {P.}~\bibnamefont {Hermet}}, \bibinfo {author} {\bibfnamefont
  {F.}~\bibnamefont {Jollet}}, \bibinfo {author} {\bibfnamefont
  {G.}~\bibnamefont {Jomard}}, \bibinfo {author} {\bibfnamefont
  {S.}~\bibnamefont {Leroux}}, \bibinfo {author} {\bibfnamefont
  {M.}~\bibnamefont {Mancini}}, \bibinfo {author} {\bibfnamefont
  {S.}~\bibnamefont {Mazevet}}, \bibinfo {author} {\bibfnamefont
  {M.}~\bibnamefont {Oliveira}}, \bibinfo {author} {\bibfnamefont
  {G.}~\bibnamefont {Onida}}, \bibinfo {author} {\bibfnamefont
  {Y.}~\bibnamefont {Pouillon}}, \bibinfo {author} {\bibfnamefont
  {T.}~\bibnamefont {Rangel}}, \bibinfo {author} {\bibfnamefont {G.-M.}\
  \bibnamefont {Rignanese}}, \bibinfo {author} {\bibfnamefont {D.}~\bibnamefont
  {Sangalli}}, \bibinfo {author} {\bibfnamefont {R.}~\bibnamefont {Shaltaf}},
  \bibinfo {author} {\bibfnamefont {M.}~\bibnamefont {Torrent}}, \bibinfo
  {author} {\bibfnamefont {M.}~\bibnamefont {Verstraete}}, \bibinfo {author}
  {\bibfnamefont {G.}~\bibnamefont {Zerah}},\ and\ \bibinfo {author}
  {\bibfnamefont {J.}~\bibnamefont {Zwanziger}},\ }\bibfield  {title} {\bibinfo
  {title} {Abinit: First-principles approach to material and nanosystem
  properties},\ }\href
  {https://doi.org/http://dx.doi.org/10.1016/j.cpc.2009.07.007} {\bibfield
  {journal} {\bibinfo  {journal} {Computer Physics Communications}\ }\textbf
  {\bibinfo {volume} {180}},\ \bibinfo {pages} {2582 } (\bibinfo {year}
  {2009})}\BibitemShut {NoStop}%
\bibitem [{\citenamefont {Gonze}\ \emph {et~al.}(2005)\citenamefont {Gonze},
  \citenamefont {Rignanese}, \citenamefont {Verstraete}, \citenamefont
  {Beuken}, \citenamefont {Pouillon}, \citenamefont {Caracas}, \citenamefont
  {Jollet}, \citenamefont {Torrent}, \citenamefont {Zerah}, \citenamefont
  {Mikami}, \citenamefont {Ghosez}, \citenamefont {Veithen}, \citenamefont
  {Raty}, \citenamefont {Olevano}, \citenamefont {Bruneval}, \citenamefont
  {Reining}, \citenamefont {Godby}, \citenamefont {Onida}, \citenamefont
  {Hamann},\ and\ \citenamefont {Allan}}]{Gonze05}%
  \BibitemOpen
  \bibfield  {author} {\bibinfo {author} {\bibfnamefont {X.}~\bibnamefont
  {Gonze}}, \bibinfo {author} {\bibfnamefont {G.}~\bibnamefont {Rignanese}},
  \bibinfo {author} {\bibfnamefont {M.}~\bibnamefont {Verstraete}}, \bibinfo
  {author} {\bibfnamefont {J.}~\bibnamefont {Beuken}}, \bibinfo {author}
  {\bibfnamefont {Y.}~\bibnamefont {Pouillon}}, \bibinfo {author}
  {\bibfnamefont {R.}~\bibnamefont {Caracas}}, \bibinfo {author} {\bibfnamefont
  {F.}~\bibnamefont {Jollet}}, \bibinfo {author} {\bibfnamefont
  {M.}~\bibnamefont {Torrent}}, \bibinfo {author} {\bibfnamefont
  {G.}~\bibnamefont {Zerah}}, \bibinfo {author} {\bibfnamefont
  {M.}~\bibnamefont {Mikami}}, \bibinfo {author} {\bibfnamefont
  {P.}~\bibnamefont {Ghosez}}, \bibinfo {author} {\bibfnamefont
  {M.}~\bibnamefont {Veithen}}, \bibinfo {author} {\bibfnamefont
  {J.}~\bibnamefont {Raty}}, \bibinfo {author} {\bibfnamefont {V.}~\bibnamefont
  {Olevano}}, \bibinfo {author} {\bibfnamefont {F.}~\bibnamefont {Bruneval}},
  \bibinfo {author} {\bibfnamefont {L.}~\bibnamefont {Reining}}, \bibinfo
  {author} {\bibfnamefont {R.}~\bibnamefont {Godby}}, \bibinfo {author}
  {\bibfnamefont {G.}~\bibnamefont {Onida}}, \bibinfo {author} {\bibfnamefont
  {D.}~\bibnamefont {Hamann}},\ and\ \bibinfo {author} {\bibfnamefont
  {D.}~\bibnamefont {Allan}},\ }\bibfield  {title} {\bibinfo {title} {A brief
  introduction to the abinit software package},\ }\href
  {https://hal.archives-ouvertes.fr/hal-00020828} {\bibfield  {journal}
  {\bibinfo  {journal} {Zeitschrift f{\"u}r Kristallographie}\ }\textbf
  {\bibinfo {volume} {220}},\ \bibinfo {pages} {558} (\bibinfo {year}
  {2005})}\BibitemShut {NoStop}%
\bibitem [{\citenamefont {Troullier}\ and\ \citenamefont
  {Martins}(1991)}]{Troullier-Martins_Pseudpotentials}%
  \BibitemOpen
  \bibfield  {author} {\bibinfo {author} {\bibfnamefont {N.}~\bibnamefont
  {Troullier}}\ and\ \bibinfo {author} {\bibfnamefont {J.~L.}\ \bibnamefont
  {Martins}},\ }\bibfield  {title} {\bibinfo {title} {Efficient
  pseudopotentials for plane-wave calculations},\ }\href
  {https://doi.org/10.1103/PhysRevB.43.1993} {\bibfield  {journal} {\bibinfo
  {journal} {Phys. Rev. B}\ }\textbf {\bibinfo {volume} {43}},\ \bibinfo
  {pages} {1993} (\bibinfo {year} {1991})}\BibitemShut {NoStop}%
\bibitem [{\citenamefont {Levine}\ and\ \citenamefont
  {Allan}(1989)}]{LevinePRL89}%
  \BibitemOpen
  \bibfield  {author} {\bibinfo {author} {\bibfnamefont {Z.~H.}\ \bibnamefont
  {Levine}}\ and\ \bibinfo {author} {\bibfnamefont {D.~C.}\ \bibnamefont
  {Allan}},\ }\bibfield  {title} {\bibinfo {title} {Linear optical response in
  silicon and germanium including self-energy effects},\ }\href
  {https://doi.org/10.1103/PhysRevLett.63.1719} {\bibfield  {journal} {\bibinfo
   {journal} {Phys. Rev. Lett.}\ }\textbf {\bibinfo {volume} {63}},\ \bibinfo
  {pages} {1719} (\bibinfo {year} {1989})}\BibitemShut {NoStop}%
\bibitem [{\citenamefont {Blakemore}(1982)}]{BlakemoreJAppPh82}%
  \BibitemOpen
  \bibfield  {author} {\bibinfo {author} {\bibfnamefont {J.~S.}\ \bibnamefont
  {Blakemore}},\ }\bibfield  {title} {\bibinfo {title} {Semiconducting and
  other major properties of gallium arsenide},\ }\href
  {https://doi.org/10.1063/1.331665} {\bibfield  {journal} {\bibinfo  {journal}
  {Journal of Applied Physics}\ }\textbf {\bibinfo {volume} {53}},\ \bibinfo
  {pages} {R123} (\bibinfo {year} {1982})}\BibitemShut {NoStop}%
\bibitem [{\citenamefont {Momma}\ and\ \citenamefont
  {Izumi}(2011)}]{MommaJAC11}%
  \BibitemOpen
  \bibfield  {author} {\bibinfo {author} {\bibfnamefont {K.}~\bibnamefont
  {Momma}}\ and\ \bibinfo {author} {\bibfnamefont {F.}~\bibnamefont {Izumi}},\
  }\bibfield  {title} {\bibinfo {title} {{\it VESTA3} for three-dimensional
  visualization of crystal, volumetric and morphology data},\ }\href
  {https://doi.org/10.1107/S0021889811038970} {\bibfield  {journal} {\bibinfo
  {journal} {Journal of Applied Crystallography}\ }\textbf {\bibinfo {volume}
  {44}},\ \bibinfo {pages} {1272} (\bibinfo {year} {2011})}\BibitemShut
  {NoStop}%
\bibitem [{\citenamefont {Roessler}\ and\ \citenamefont
  {Walker}(1967)}]{RoesslerPhRev67}%
  \BibitemOpen
  \bibfield  {author} {\bibinfo {author} {\bibfnamefont {D.~M.}\ \bibnamefont
  {Roessler}}\ and\ \bibinfo {author} {\bibfnamefont {W.~C.}\ \bibnamefont
  {Walker}},\ }\bibfield  {title} {\bibinfo {title} {Electronic spectrum and
  ultraviolet optical properties of crystalline mgo},\ }\href
  {https://doi.org/10.1103/PhysRev.159.733} {\bibfield  {journal} {\bibinfo
  {journal} {Phys. Rev.}\ }\textbf {\bibinfo {volume} {159}},\ \bibinfo {pages}
  {733} (\bibinfo {year} {1967})}\BibitemShut {NoStop}%
\bibitem [{\citenamefont {Jackson}(1998)}]{Jackson}%
  \BibitemOpen
  \bibfield  {author} {\bibinfo {author} {\bibfnamefont {J.~D.}\ \bibnamefont
  {Jackson}},\ }\href@noop {} {\emph {\bibinfo {title} {Classical
  Electrodynamics}}}\ (\bibinfo  {publisher} {Wiley},\ \bibinfo {address}
  {Hoboken},\ \bibinfo {year} {1998})\BibitemShut {NoStop}%
\bibitem [{\citenamefont {Landau}\ and\ \citenamefont
  {Lifshitz}(1960)}]{LandauElectrodynamics}%
  \BibitemOpen
  \bibfield  {author} {\bibinfo {author} {\bibfnamefont {L.}~\bibnamefont
  {Landau}}\ and\ \bibinfo {author} {\bibfnamefont {E.}~\bibnamefont
  {Lifshitz}},\ }\href@noop {} {\emph {\bibinfo {title} {Electrodynamics of
  continuous media}}},\ Teoreticheskaja fizika\ (\bibinfo  {publisher}
  {Pergamon Press},\ \bibinfo {year} {1960})\BibitemShut {NoStop}%
\end{thebibliography}%

\end{document}